\documentclass[a4paper,11pt]{article}
\usepackage{graphicx,amssymb,amstext,amsmath,mathabx}
\usepackage{color, yfonts, tcolorbox}
\usepackage{floatflt,rotating}
\usepackage{relsize}
\usepackage{exscale}

\numberwithin{equation}{section}

\definecolor{cbl}{rgb}{0,0,1}                

\newcommand{\bc}{\begin{center}}
\newcommand{\ec}{\end{center}}
\def\ba#1{\begin{array}{#1}\displaystyle}
\newcommand{\ea}{\end{array}}

\newcommand{\beq}{\begin{equation}}
\newcommand{\eeq}{\end{equation}}
\newcommand{\beqa}{\begin{eqnarray}}
\newcommand{\eeqa}{\end{eqnarray}}

\newcommand{\bi}{\begin{itemize}}
\newcommand{\ei}{\end{itemize}}

\newcommand{\bra}{\langle}
\newcommand{\ket}{\rangle}

\newcommand{\N}{{\mathbb{N}}}

\newcommand{\ep}{\epsilon}

\newcommand{\Tr}{{\rm Tr}}

\newcommand{\TT}{{\cal T}}

\newcommand{\bbc}{\mathsmaller{\bigcup}}

\begin{document}
\begin{titlepage}
\vspace{0.2cm}
\begin{center}

{\large{\bf{Entanglement Content of Quantum Particle Excitations II. \\ Disconnected Regions and Logarithmic Negativity}}}

\vspace{0.8cm} 
{\large \text{Olalla A. Castro-Alvaredo${}^{\heartsuit}$, Cecilia De Fazio{${}^{\clubsuit}$}, Benjamin Doyon{${}^{\diamondsuit}$}, and Istv\'an M. Sz\'ecs\'enyi{${}^{\spadesuit}$}}}

\vspace{0.8cm}
{\small
 ${}^{\heartsuit\, {\clubsuit}  \, \spadesuit}$ Department of Mathematics, City, University of London, 10 Northampton Square EC1V 0HB, UK\\
\vspace{0.2cm}
{${}^{\diamondsuit}$}Department of Mathematics, King's College London, Strand WC2R 2LS, UK}\\

\end{center}

\vspace{1cm}
In this paper we study the increment of the entanglement entropy and of the (replica) logarithmic negativity in a zero-density excited state of a free massive bosonic theory, compared to the ground state. This extends the work of two previous publications by the same authors. We consider the case of two disconnected regions and find that the change in the entanglement entropy depends only on the combined size of the regions and is independent of their connectivity. We subsequently generalize this result to any number of disconnected regions.  For the replica negativity we find that its increment is a  polynomial with integer coefficients depending only on the sizes of the two regions. The logarithmic negativity turns out to have a more complicated functional structure than its replica version, typically involving roots of polynomials on the sizes of the regions. We obtain our results by two methods already employed in previous work:  from a qubit picture and by computing four-point functions of branch point twist fields in finite volume. We  test our results against numerical simulations on a harmonic chain and find excellent agreement.
\medskip

\vfill

\noindent 
${}^{\heartsuit}$ o.castro-alvaredo@city.ac.uk\\
{${}^\clubsuit$} cecilia.de-fazio.2@city.ac.uk\\
{${}^{\diamondsuit}$} benjamin.doyon@kcl.ac.uk\\
{${}^{\spadesuit}$} istvan.szecsenyi@city.ac.uk\\

\hfill \today

\end{titlepage}


\section{Introduction}
In recent years there has been much progress in the understanding of entanglement measures in one-dimensional many body quantum systems (see e.g.~the review articles in \cite{EERev1,specialissue, EERev2}).  
These include critical systems, such as conformal field theories (CFTs) and critical quantum spin chains  \cite{CallanW94,HolzheyLW94,latorre1, Latorre2,Calabrese:2004eu,Calabrese:2005in}, as well as gapped systems such as massive quantum field theories (QFTs) (integrable or not) \cite{casini1, casini2, entropy, next, CH, nexttonext, review}, gapped quantum spin chains \cite{latorre3,Jin,Lambert,Keating, Weston,fabian,parity}  and lattice models \cite{peschel,rava1,rava2}. Much of this work has focussed on one particular measure of entanglement, the entanglement entropy \cite{bennet} and on a particular state, the ground state. More recently, the entanglement of different classes of excited states  \cite{ Vincenzo2, german1,german2, Vincenzo, the4ofus1, the4ofus2}, as measured by the entanglement entropy and other measures  of entanglement \cite{disco1, disco11, negativity1,negativity2,casini-dis}, has been studied.
{\begin{floatingfigure}[h]{7.5cm} 
 \begin{center} 
 \includegraphics[width=7cm]{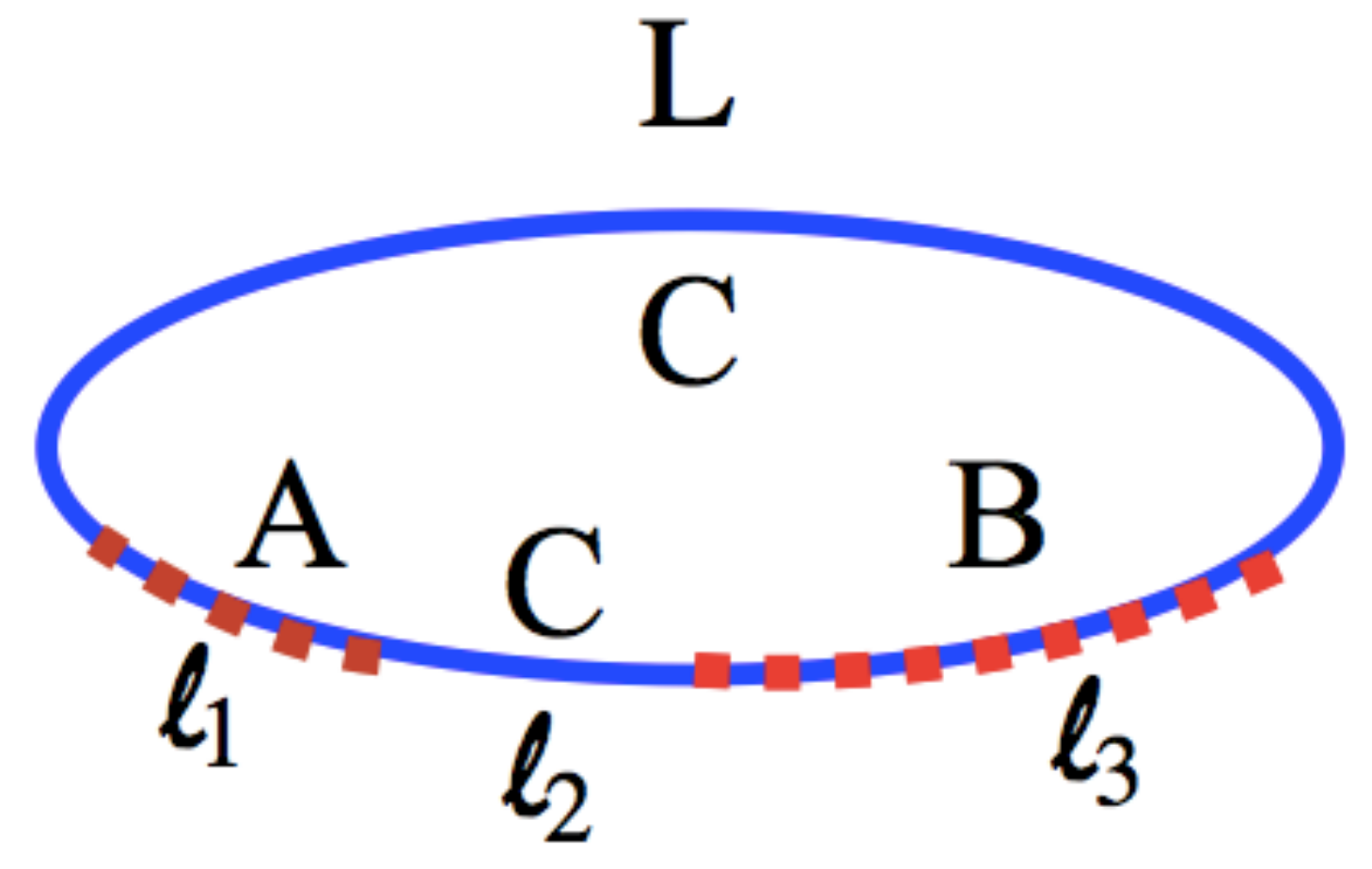} 
 \end{center} 
 \caption{Two disconnected regions in one dimension.\vspace{1cm}} 
 \label{typical} 
 \end{floatingfigure}}
 In two recent works \cite{the4ofus1,the4ofus2} we have investigated the entanglement entropy of gapped systems in zero-density excited states, where excitations have a (quasi-)particle interpretation. We found a very intuitive mathematical structure for the entropy differences between the ground state and such excited states. This structure allowed for simple closed formulae in particular limits. At the basis of our results laid the observation that the additional entanglement entropy due to such excitations is equal to the entanglement entropy of very simple states, involving a small number of qubits with amplitudes which represent the probabilities of finding particles in various system's regions. This idea of localized excitations and their one-to-one correspondence with increased entanglement has turned out to be surprisingly general. So far it has been shown to work for systems without interactions and in some instances of interacting models and particular excited states. We also found numerical evidence of the validity of our results for the harmonic chain and lattice (2 dimensions) and beyond the universal quantum field theory regime. In the present work we extend this intuition, with precise results for the free boson theory, to the entanglement entropy of disconnected regions and the  (replica) logarithmic negativity. 
In one dimension, the configuration that we will be considering throughout this manuscript is illustrated by Fig.~1. Here $L$ is the full length of the system and $\ell_1, \ell_3$ are the lengths of subsystems $A$ and $B$ respectively. The length of subsystem $C$ is then $L-\ell_1-\ell_3$, which includes the mid-region of length $\ell_2$.
 
The von Neumann and R\'enyi {\it entanglement entropies} are measures of the amount of quantum
entanglement, in a pure quantum state, between the degrees of
freedom associated to two sets of independent observables whose
union is complete on the Hilbert space. For a subsystem consisting of two disconnected regions as in Fig.~1, the Hilbert space may be seen as having the structure 
\beq
\mathcal{H}=\mathcal{H}_{A  {\bbc}  B} \otimes \mathcal{H}_C\, ,
\eeq
Consider a bipartition where the two sets of observables correspond to the local observables in the two finite-size complementary regions, $A \bbc B$ and $C$. Let the system be in a state $|\Psi\ket_L$. Then the von Neumann entropy associated to region $A\bbc B$ is 
\beq
S_1^\Psi(\ell_1,\ell_3, L)=-\Tr_{A\bbc B}(\rho_{A\bbc B} \log \rho_{A \bbc B})\,,
\label{trace}
\eeq
where $\rho_{A \bbc B}=\Tr_C(|\Psi \rangle_L{}_L \langle \Psi|)$ is the reduced density matrix associated to subsystem $A \bbc B$. One may obtain the von Neumann entropy $S_1^\Psi(\ell_1,\ell_3,L)$ as a limiting case of the sequence of  $n$th R\'enyi entropies  defined as
 \beq 
 S_n^\Psi(\ell_1,\ell_3,L)=\frac{\log \Tr_{A\bbc B}\rho_{A\bbc B}^n }{1-n}\,,
 \label{renyi}
 \eeq 
thanks to the property 
\beq 
\lim_{n\rightarrow 1} S_n^\Psi(\ell_1,\ell_3,L)=S_1^\Psi(\ell_1,\ell_3,L)\,.
\label{rtrick}
\eeq
The {\it logarithmic negativity} of the same state $|\Psi\ket_L$ \cite{Eisert, ZHSL, Eisert2, ple, erratumple, VW} is defined as:
 \beq
 \mathcal{E}^\Psi=\log \mathrm{Tr}_{A\bbc B} \left| \rho_{A\bbc B}^{T_B}  \right|\,.
 \label{neg1}
 \eeq
 The geometry is the same as in Fig.~1 but the density matrix $\rho_{A\bbc B}^{T_B} $ differs from $\rho_{A\bbc B}$, thanks to the operation $T_B$, which represents partial transposition on subsystem $B$.
 Crucially, after such transposition, the matrix  $\rho_{A\bbc B}^{T_B} $ is no longer guaranteed to be positive-definite: specifically, it may have positive and negative eigenvalues. The operation $ \mathrm{Tr}_{A\bbc B}|\cdot|$ involved in (\ref{neg1}) represents the ``trace norm" of  $\rho_{A\bbc B}^{T_B}$, namely the sum of the absolute values of its eigenvalues.
 
 Similarly as for the von Neumann entropy, it is possible to define the logarithmic negativity as a limit of a more general set of quantities depending on an additional parameter $n$ as:
 \beq
 \mathcal{E}^\Psi_n(\ell_1,\ell_3,L):=\log \mathrm{Tr}_{A\bbc B} \left( \rho_{A\bbc B}^{T_B}  \right)^n\,.
 \eeq
 such that 
 \beq
 \mathcal{E}^\Psi(\ell_1,\ell_3,L):=\lim_{n\rightarrow 1} \mathcal{E}^\Psi_n(\ell_1,\ell_3,L), \quad \mathrm{from\,\, {\it n}\,\, even}\,. 
 \eeq
The meaning of the last equation is that one sets $n=2m$ for $m\in\N$, and then analytically continues in $m$ towards $m=1/2$ (under appropriate conditions in the complex $m$ plane making this analytic continuation unique). This definition was first proposed in \cite{negativity1,negativity2}. From now on we will call these functions {\it replica logarithmic negativities}.
 
Note that the (replica) logarithmic negativity is a measure of the entanglement between the non-complementary regions $A$ and $B$ whereas the von Neumann and R\'enyi entropies defined earlier are measures of the entanglement between the two complementary regions $A\bbc B$ and $C$. Therefore they are distinct measures capturing different information about the state.  

From the definitions above it is clear that in order to compute these measures of entanglement we need to have access to the reduced density matrix $\rho_{A \bbc B}$ and its partially transposed version, or at least to their eigenvalues. This may be achieved numerically for some systems, but if the system admits a quantum field theory description, then is also accessible analytically by relating the $n$th powers  $\rho_{A \bbc B}^n$ and  $(\rho^{T_B}_{A \bbc B})^n$ to specific correlators of local fields in quantum field theory \cite{Calabrese:2004eu,Calabrese:2005in,entropy}. In \cite{entropy} these fields were termed {\it branch point twist fields} and characterized as symmetry fields in an $n$-copy version of the model under study. Branch point twist fields have since been systematically employed in the computation of measures of entanglement in massive QFTs and also in CFT. The entanglement measures considered in this paper, in the setup of Fig.~\ref{typical},  were studied systematically in the ground state of CFT in \cite{disco1,disco11,negativity1,negativity2,negativity3} and also partially in massive integrable QFT \cite{ourneg,freeboson}.

In the present work we want to extend our study to zero-density excited states. More precisely, we want to study the {\it increment} of the entanglement entropies of two disconnected regions and of the replica logarithmic negativities in a zero-density excited state, compared to the ground state. We define the new variables, 
 \beq
 r_i:=\frac{\ell_i}{L}\,,\quad \mathrm{for}\quad i=1,2,3, \quad \mathrm{and} \quad 
 r:=1-r_1-r_3\,.
 \label{r}
 \eeq
 In this paper we will consider the scaling limit, defined as
 \beq
 L,\ell_1,\ell_2,\ell_3 \rightarrow \infty\,, \qquad \mathrm{with}\qquad  r, r_1, r_2, r_3 \quad \mathrm{finite,}\,\qquad \mathrm{and} \qquad 0\leq r, r_1, r_2, r_3\leq 1\,. \label{scaling}
 \eeq
 In terms of branch point twist field correlators, the increment of the R\'enyi entropies of two disconnected regions is given by:
 \beqa
\Delta S_n^\Psi(r)&:=& S_n^{\Psi}(r_1,r_3)-S_n^0(r_1,r_3) \label{3}\\
&=&\lim_{L\rightarrow \infty}\frac{1}{1-n}\log \left[\frac{{}_L\bra \Psi|\TT(0)\tilde{\TT}(r_1 L)\TT((r_1+r_2) L)\tilde{\TT}((r_1+r_2+r_3) L)|\Psi\ket_L}{{}_L\bra 0|\TT(0)\tilde{\TT}(r_1 L)\TT((r_1+r_2) L)\tilde{\TT}((r_1+r_2+r_3) L)|0\ket_L}\right]\,,\nonumber\
 \eeqa 
where $S_n^0(r_1,r_3)$ are the entropies in the ground state. As the notation suggests, we will see later that $\Delta S_n^\Psi(r)$ is a function of $r$ {\it only}.
 The increments of the replica logarithmic negativities are given by:
 \beqa
\Delta \mathcal{E}_n^\Psi(r_1,r_3)&:=& \mathcal{E}_n^\Psi(r_1,r_3)-\mathcal{E}_n^0(r_1,r_3) \label{4}
\\
&=&\lim_{L\rightarrow \infty}\log \left[\frac{{}_L\bra \Psi|\TT(0)\tilde{\TT}(r_1 L)\tilde{\TT}((r_1+r_2) L){\TT}((r_1+r_2+r_3) L)|\Psi\ket_L}{{}_L\bra 0|\TT(0)\tilde{\TT}(r_1 L)\tilde{\TT}((r_1+r_2) L){\TT}((r_1+r_2+r_3) L)|0\ket_L}\right]\,, \nonumber
 \eeqa 
 and are functions of $r_1$ and $r_3$ {\it only}. As standard, $\TT$ is the branch point twist field, $\tilde{\TT}$ is its hermitian conjugate and $|0\ket_L$ is the ground state in the compactified space of Fig.~1. 
 As is well-known \cite{entropy} the branch point twist fields are symmetry fields associated with cyclic permutation symmetry in an $n$-copy version of the theory under consideration. From this definition it follows that $\TT=\tilde{\TT}$ if $n=2$, and we will see later that, indeed, the results for the replica  logarithmic negativity equal those for the R\'enyi entropy (up to a sign) for $n=2$.

When employing the branch point twist field technique, we will therefore be computing ratios of four-point functions in the infinite volume limit. Exact explicit formulae for the four-point functions above are generally inaccessible even in the ground state and/or CFT. Remarkably though, the increments computed here, in this particular scaling limit, turn out to be the logarithms of order $k\times n$ polynomials (where $k$ is the number of excitations) in the parameters $r_1, r_3$ (or just in $r$ for the entanglement entropy) with positive, integer coefficients.
 
 \medskip
 
 This paper is organized as follows: In section \ref{mainresults} we summarize the main results of the paper. In section \ref{qubit} we show how these results may be obtained from a qubit interpretation. That is, we show that the formulae for the increment of entropies (negativities) in excited states of QFT equal the entropies  (negativities) of a very special class of states involving a finite number of qubits. In section \ref{twist} we obtain the same  results from a branch point twist field computation in the massive free boson theory. In section 5 we present some numerical results for the harmonic chain. We conclude in section 6. In Appendix A we summarize some formulae for the R\'enyi entropies and replica negativities for particular values of $n$ and $k$. In Appendix B we derive the logarithmic negativities of states of two and three identical excitations from the exact diagonalization of the reduced and partially transposed density matrix. In Appendix C we extend our qubit results for the R\'enyi entropies of two disconnected regions to the case of any number of disconnected regions. In Appendix D we summarize some useful results for the residua of certain types of first, second and third order poles which play a role in our branch point twist field computation.
 
 \section{Summary of the Main Results}
 \label{mainresults}
 Consider a state consisting of a single particle excitation. Let us denote the entropy increments of such a state by $\Delta S_n^1(r)$ with $r$ as defined earlier (\ref{r}). We find that
\beq
\Delta{S}_n^{1}(r)=\frac{ \log(r^n+(1-r)^n)}{1-n}\,.
\label{ren}
\eeq
This is identical to the result found in \cite{the4ofus1,the4ofus2} for a connected region of length $r$ and indeed, further results for the von Neumann and other types of entropy, for excited states consisting of a higher number of excitations, identical or not, also take the same form as found in \cite{the4ofus1,the4ofus2}. We will not report all of these here again. We report however the formula for a $k$-particle excited state consisting of identical excitations, as it will feature later on. This is given by
\beq
\Delta{S}_n^{k}(r)=\frac{1}{1-n}\log\left(\sum_{p=0}^k  \left[{k\choose p}r^{p} (1-r)^{k-p}\right]^n
\right)\,.\label{ren2}
\eeq
This means that the entanglement entropy increase due to the presence of a discrete number of excitations  depends only upon the size of the regions $A$ and $B$ relative to that of the whole system, and {\it not on their connectivity}. Indeed, in Appendix C we show that the same conclusion can be drawn for a subsystem consisting of any number of disconnected regions.
 \begin{figure}[h!]
 \begin{center} 
 \includegraphics[width=7cm]{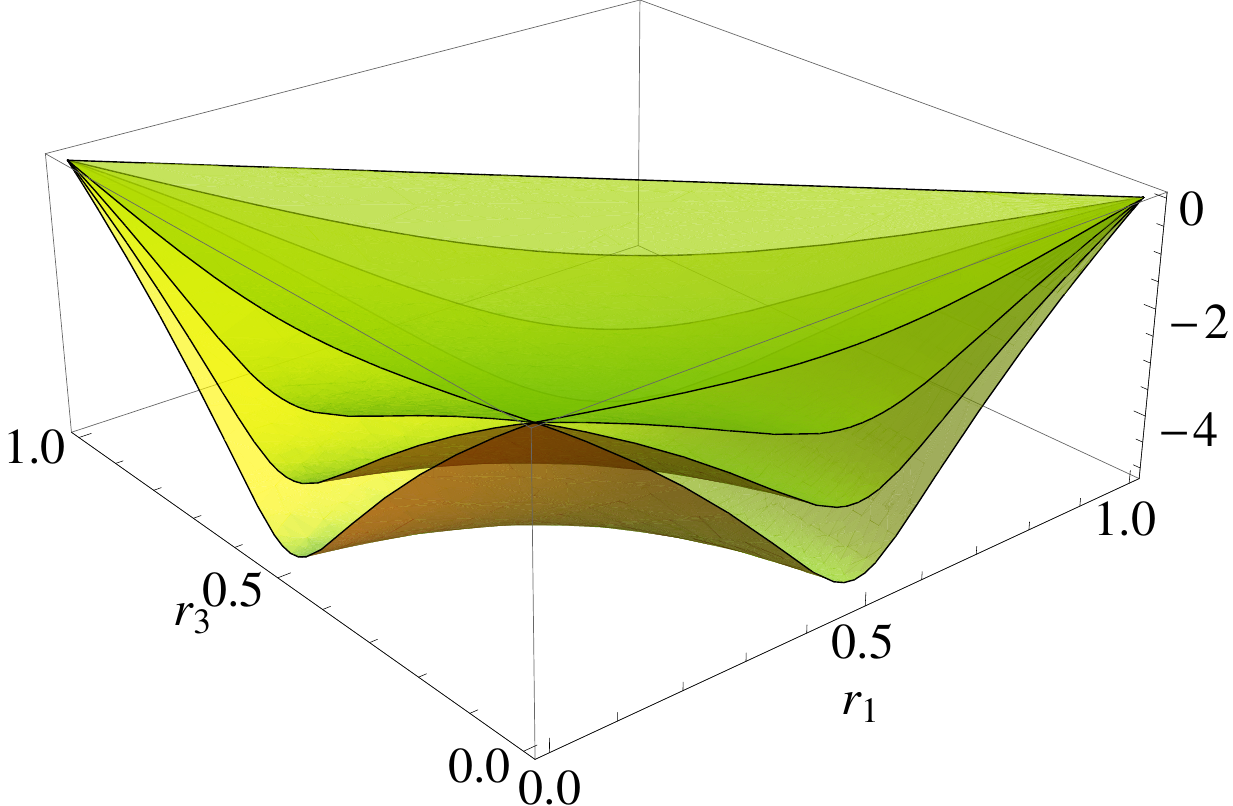} 
  \includegraphics[width=7cm]{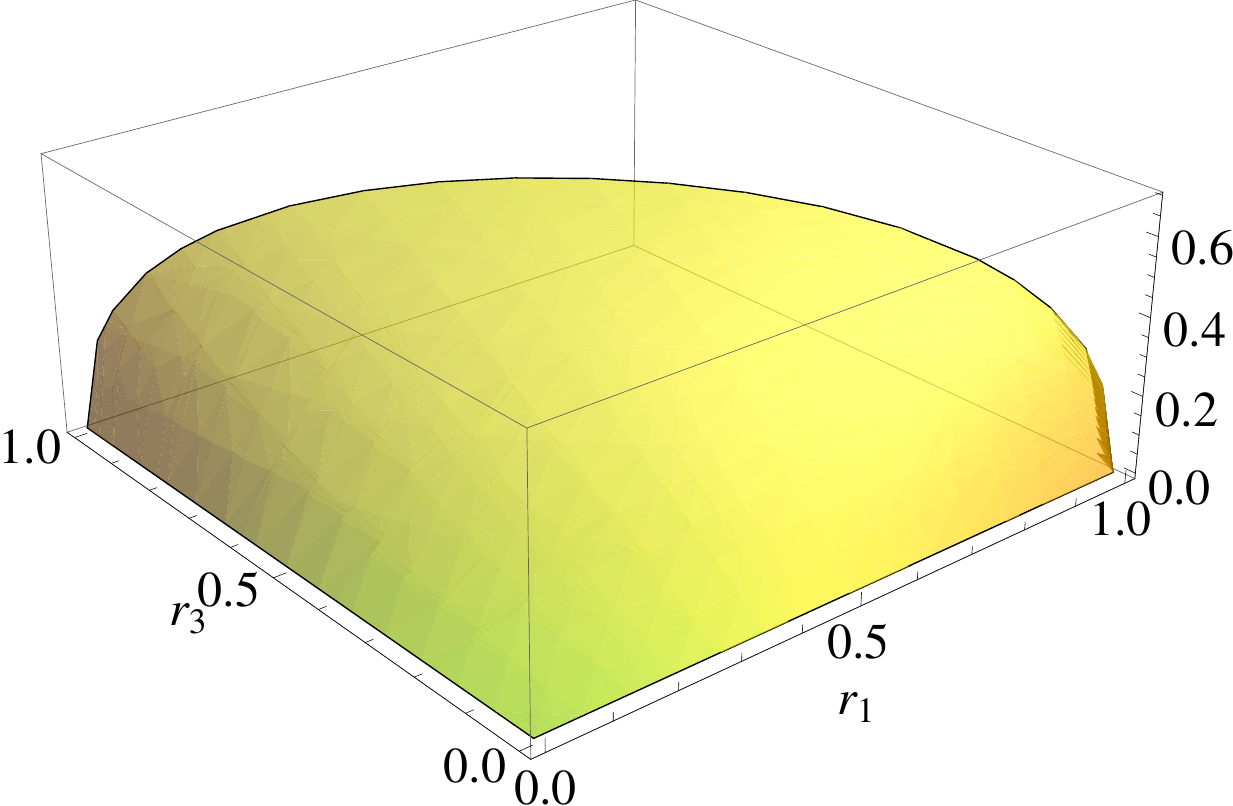} 
 \end{center} 
 \caption{Left: The function (\ref{de}) for $n=2,4, 6$ and $8$ (the higher the value of $n$, the more negative the value of the function).  Right: The function (\ref{ep1}). As we can see, there is 
 a change of curvature and sign when taking the limit $n\rightarrow 1$. A similar interesting analytic behaviour near $n=1$ was found in \cite{negativity2} for the compactified free boson. Recall that $0\leq r_1+r_3\leq 1$, which restricts the domain of definition of the functions shown. A cross-section of these surfaces is provided in Fig.\ref{negative333} (left).
 \vspace{1cm}} 
 \label{negative1} 
 \end{figure}
 
Results are more interesting for the (replica) logarithmic negativity.  First, we find that it is a function of $r_1, r_3$, where each parameter now enters independently. 
 For  a state consisting of  a single particle excitation the increment of the  replica logarithmic negativity is given by the simple expression:
 \beq
 \Delta \mathcal{E}_n^1(r_1,r_3)= \log\left(r_1^n + r_3^n+ \sum_{p=0}^{[\frac{n}{2}]} \frac{n}{n-p} \left(\begin{array}{c}
 n-p\\
 p\\
 \end{array}\right) \,
 r^{n-2p} r_1^p r_3^p\right)\,,
 \label{de}
 \eeq 
where $[.]$ denotes the integer part. The increment of the logarithmic negativity can be obtained by analytically continuing this expression from $n$ even to $n=1$. Evaluation of the sum above for $n=2m$ and $m\in \mathbb{N}$ gives
 \beq
 \Delta \mathcal{E}_{2m}^1(r_1,r_3)= \log\left(r_1^{2m} + r_3^{2m}+\left(\frac{\sqrt{r^2+4r_1 r_3}+r}{2}\right)^{2m}+\left(\frac{\sqrt{r^2+4r_1 r_3}-r}{2}\right)^{2m} \right)\,.
  \label{dem}
 \eeq 
so that the analytic continuation is simply,
  \beq
  \Delta \mathcal{E}^1(r_1,r_3):=\lim_{m\rightarrow \frac{1}{2}}  \Delta \mathcal{E}_{2m}^1(r_1,r_3)
= \log(r_1+r_3+ \sqrt{r^2+4 r_1 r_3})\,.
  \label{ep1}
 \eeq
 For a state consisting of $k$ distinct excitations (particles with distinct momenta), the result is simply $k$ times the above, just as for the R\'enyi entropies.
  The case of identical excitations (identical momenta) is more interesting. Consider an excited state of $k$ identical excitations. The increment  of the replica logarithmic negativity is given by:
  \beq
 \Delta \mathcal{E}_n^k(r_1,r_3)=\log\left(\sum_{p=-k}^{k} \sum_{q=\max(0,-np)}^{[\frac{n}{2}(k-p)]} \mathcal{A}_{p,q} \,r_1^{n p + q} r^{n(k-p)-2q} r_3^{q}\right)\,,
   \label{de2}
  \eeq
   where the coefficients $\mathcal{A}_{p,q}$ are defined as follows:
 \beq
\mathcal{A}_{p,q}=\sum_{\{k_1,\ldots,k_n\}\in \sigma_0^n(q)} \prod_{j=1}^n \frac{k!}{(p+k_j)!(k-p-k_{j+1}-k_j)! k_{j+1}!}\,,
\label{pro}
\eeq
and $\sigma_0^n(q)$ represents the set of integer partitions of $q$ into $n$ non-negative parts. Note that the coefficients are zero whenever any of the arguments of the factorials in the denominator becomes negative and this selects out the partitions that contribute to each coefficient for given values of $p$ and $q$. As should be, formula (\ref{de}) is the $k=1$ case of (\ref{de2}). Indeed the coefficients inside the sum (\ref{de}) are nothing but the number of partitions of $p$ into $n$ parts, $p$ of which are 1 and $n-p$ of which are 0, with the constraint that there are no consecutive 1s. 

Since the coefficients $\mathcal{A}_{p,q}$ are rather non-trivial, it is not easy to perform the sums in (\ref{de2}) explicitly and the analytic continuation leading to the logarithmic negativity is rather involved. We have not been able to obtain closed formulae for the logarithmic negativity for all $k$ but have obtained expressions for $k=1,2,3$. For $k=2$ it is given by
 \beq
\Delta \mathcal{E}^2(r_1,r_3)=\log\left(r_1^2+r_3^2 + 2 (r_1+r_3) \sqrt{r^2+2r_1r_3}+ \frac{r^2+r_1 r_3}{3}+\frac{2^{2/3} \sqrt[3]{\Delta}}{3}+\frac{2 {\sqrt[3]{2}} \Delta_0}{3\sqrt[3]{\Delta}}\right),\,
\label{ep2}
   \eeq
   where
   \beq
   \Delta:=\Delta_1+\sqrt{\Delta_1^2-4
   \Delta_0^3}\,,
   \label{delta}
   \eeq
   and
\beq
 \Delta_0=r^4+10r^2 r_1 r_3 + 7 r_1^2 r_3^2\,, \qquad
\Delta_1=-2 r^6-30 r^4 r_1 r_3-69 r^2 r_1^2 r_3^2+20 r_1^3 r_3^3\,. \label{d1}
\eeq
The case $k=3$ has been studied in Appendix B and the logarithmic negativity in this case is given by (\ref{E3par}). 
The explicit polynomials (\ref{de2}) for a few values of $n$ and $k$ are listed in the tables of Appendix A. 

 \begin{figure}[h]
 \begin{center} 
   \includegraphics[width=7cm]{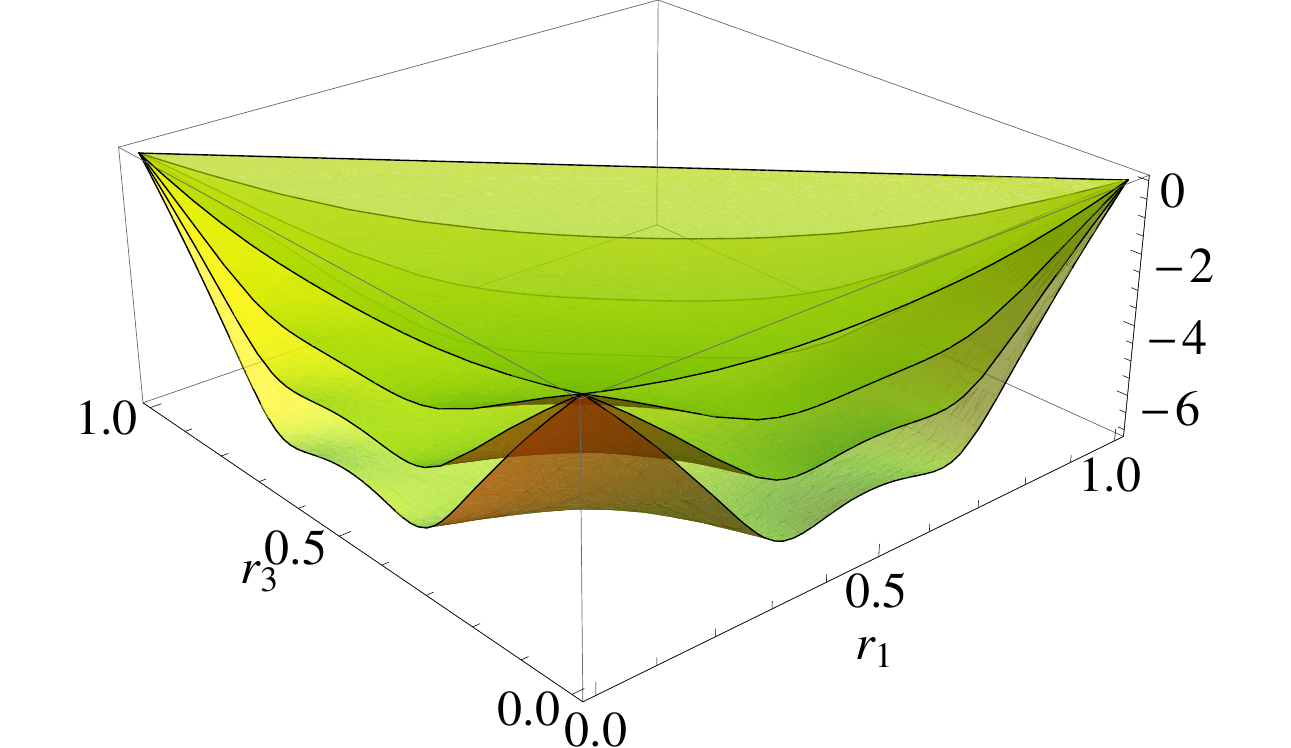} 
  \includegraphics[width=7cm]{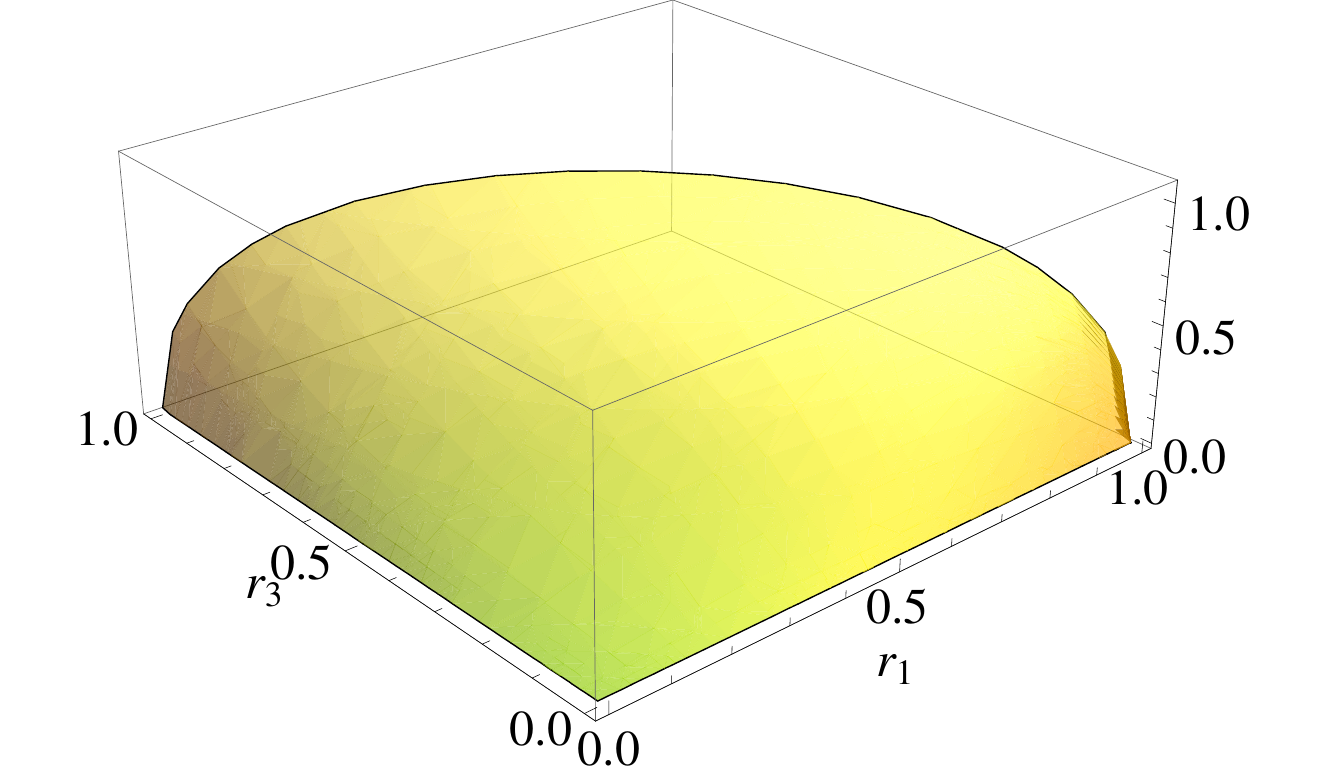} 
 \end{center} 
 \caption{Left: The function (\ref{de2}) for $n=2,4, 6$ and $8$ and $k=2$.  Right: The function (\ref{ep2}). Again, there is 
 a change of curvature and sign when taking the limit $n\rightarrow 1$.  Recall that $0\leq r_1+r_3\leq 1$, which restricts the domain of definition of the functions shown.  A cross section of these surfaces is provided in Fig.\ref{negative333} (right).
 \vspace{1cm}} 
 \label{negative2} 
 \end{figure}
 
For mixed excited states consisting of $k_i$ particles of identical momentum $p_i$ with $i=1,\ldots m$ and $p_i\neq p_j$ for $i\neq j$ we have that the R\'enyi entropies and the (replica) logarithmic negativities can be expressed in terms of the building blocks given above, namely:
 \beqa
\Delta S_n^{k_1,\dots,k_m} (r)&=&\sum_{q=1}^m \Delta S_n^{k_q}(r)\,, \\
\Delta \mathcal{E}_n^{k_1,\dots,k_m}(r_1,r_3)&=&\sum_{q=1}^m \Delta \mathcal{E}_n^{k_q}(r_1,r_3)\,.
\label{sumar}
\eeqa
 
Note that the functions (\ref{ep1}), (\ref{ep2}) give nice simple illustrations of the non-trivial nature of the analytic continuation from $n$ even to $n=1$. Although for integer $n$ the functions (\ref{de}) and (\ref{de2}) are simple polynomials in $r_1, r_3$, the logarithmic negativity involves roots which appear to be of increasing order as $k$ is increased.
 As explained in Appendix B, we have obtained the analytic continuation of \eqref{de2}, for $k=2$ giving (\ref{ep2}) and for $k=3$ giving (\ref{E3par}), by explicitly computing the eigenvalues of the partially transposed reduced density matrix obtained from a qubit interpretation.

\begin{figure}[h]
 \begin{center} 
   \includegraphics[width=7cm]{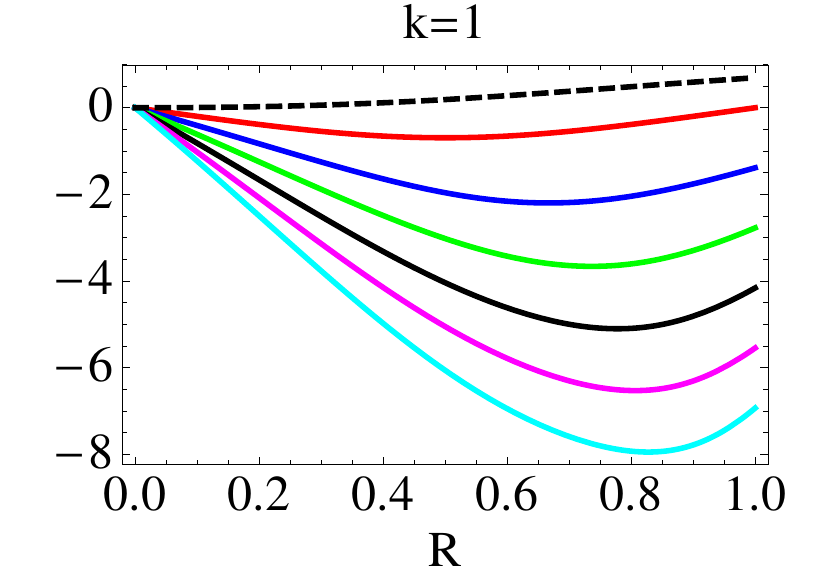} 
  \includegraphics[width=7cm]{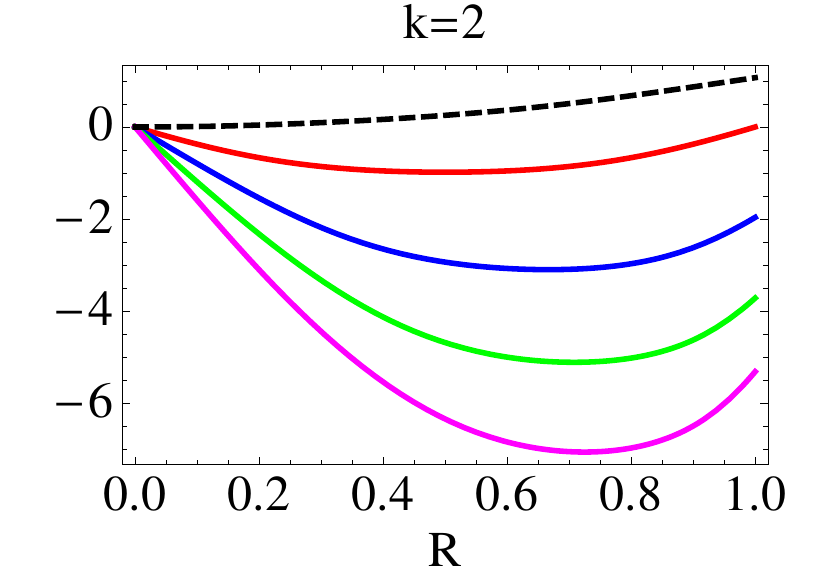} 
 \end{center} 
 \caption{Left: The functions (\ref{de}) for $n=2,4, 6,8,10$ and $12$ and the function (\ref{ep1}). Right: The functions (\ref{de2}) for $n=2,4, 6$ and $8$ and the function (\ref{ep2}). In all cases we consider $r_1=r_3=\frac{R}{2}$. The solid lines are the replica logarithmic negativities, which take more negative values for larger
 values of $n$. The dashed lines are the functions (\ref{ep1}), (\ref{ep2}). These figures provide a cross-section view of Figs.~\ref{negative1} and \ref{negative2} and illustrate the gradual change in curvature as $n\rightarrow 1$.
  \vspace{1cm}} 
 \label{negative333} 
 \end{figure}

The marked difference between functions (\ref{ep1}), (\ref{ep2}) and (\ref{de}), (\ref{de2}) can also be seen in Figs.~\ref{negative1} and \ref{negative2} where they are shown to exhibit opposite sign and curvature. The gradual change in curvature as $n$ is reduced is particularly clear from the cross-section views given in Fig.~\ref{negative333}.  
 
The expression (\ref{de2}) with (\ref{pro}) suggests a combinatorial interpretation for our results and indeed such interpretation can be made concrete.
 A one-to-one relationship between (\ref{de2}) and the generating  function of the number of simply-connected graphs with $2nk$ nodes, $nk$ edges and certain connectivity conditions will be proven in \cite{the4ofus4}.

 Let us for now present two approaches whereby these results can be derived: from an interpretation in terms of qubits, and from a direct calculation within massive QFT. 
 
 \section{Qubit Interpretation}
 \label{qubit}
 It was shown in \cite{the4ofus1,the4ofus2} that exact expressions for the increment of entanglement entropy in a zero-density excited state and a bipartition involving only simply-connected domains could be derived from a so-called ``qubit interpretation". This interpretation was first suggested by the result (\ref{ren}) and the realization that it is identical to the R\'enyi entropy of a two-qubit state of the form 
 $|\Psi_{\mathrm{qb}}\ket=\sqrt{r}|01\ket + \sqrt{1-r} |10\ket$: the entanglement entropy increase depends only on whether the particle is localized in region $A\bbc B$ with probability $r$, or on region $C$, with probability $1-r$. The strength of this picture lays on the fact that it can be extended naturally to states consisting of multiple excitations. Their increase of entanglement entropy is also the entropy of multi-qubit states, with coefficients which also have a simple probabilistic interpretation. For a single connected region, this interpretation has been numerically shown to hold even in the two-dimensional harmonic lattice \cite{the4ofus2}.

This interpretation worked so well for  the entanglement entropy of simply-connected domains and appears so natural that one may expect it to still hold for disconnected regions and for the logarithmic negativity. Clearly, it suggests that results should be independent of the topology of the regions. This is in agreement with our first main formula (\ref{ren}) (and its multiple-particle generalisation) and holds as well in the case of a region composed of any number of disconnected parts (see Appendix C). As we see below, this interpretation also allows us to obtain the result \eqref{de2} for the the replica logarithmic negativities. 
 \subsection{Single Particle Excitation}
 \label{single}
  Let us consider the simplest case, that of an excited state of a single excitation. We can simply write the following three qubit state:
\beq
|\Psi^{(1)}_{\rm qb}\ket=\sqrt{r_1}|100\ket+\sqrt{r}|010\ket+ \sqrt{r_3}|001\ket\,.
\label{Psi1}
\eeq
where
the first qubit represents the presence (1) or not (0) of a particle in region $A$, the second qubit represents the same for region $C$, and the final qubit likewise for region $B$. Tracing over the mid-qubit
we have that 
\beqa
\rho_{A \bbc B}&=&\mathrm{Tr}_C\left( |\Psi_{\rm qb}\ket \bra \Psi_{\rm qb}|\right)\nonumber\\
&=&
r_1 |1 0\ket \bra 1 0|+r|00\ket \bra 0 0|+r_3 |0 1\ket \bra 0 1|  + \sqrt{r_1 r_3}\left(|1 0\ket \bra 0 1 |+ |0 1\ket \bra 1 0| \right)\,,
\label{matrix1}
\eeqa
and 
\beqa
\rho_{A \bbc B}^{T_\mathrm{B}}=
r_1 |1 0\ket \bra 1 0|+r|00\ket \bra 0 0|+r_3 |0 1\ket \bra 0 1| + \sqrt{r_1 r_3}\left(|1 1\ket \bra 0 0 |+ |0 0\ket \bra 1 1| \right)\,.
\label{matrix2}
\eeqa
In matrix form we have
\beq
\rho_{A \bbc B}=\left(
\begin{array}{c|cccc}
&1 1 & 1 0 & 0 1& 0 0\\
\hline
1 1 &0&0&0& 0\\
1 0 &0&r_1&\sqrt{r_1 r_3}&0\\
0 1 &0&\sqrt{r_1 r_3}&r_3&0\\
0 0 &0&0&0&r
\end{array}
\right), \quad 
\rho_{A \bbc B}^{T_{B}}=\left(
\begin{array}{c|cccc}
&1 1 & 1 0 & 0 1& 0 0\\
\hline
1 1 &0&0&0& \sqrt{r_1 r_3}\\
1 0 &0&r_1&0&0\\
0 1 &0&0&r_3&0\\
0 0 &\sqrt{r_1 r_3}&0&0&r
\end{array}
\right)\,,
\eeq
where the first row and first column refer to the states involved in (\ref{matrix1})-(\ref{matrix2}). 
The eigenvalues of  $\rho_{A \bbc B}$ are:
\beq
\lambda_1=0\,, \quad \lambda_2=0\,, \quad \lambda_3=1-r\,, \quad \lambda_4=r\,,
\eeq 
and those of $\rho_{A \bbc B}^{T_{ B}}$
\beq
\lambda^t_1=r_1\,, \quad \lambda^t_2=r_3\,, \quad \lambda^t_3=\frac{r+\sqrt{r^2+4 r_1 r_3}}{2}\,, \quad \lambda^t_4=\frac{r-\sqrt{r^2+4 r_1 r_3}}{2}\,,
\eeq
Note that in the latter case, the last eigenvalue is clearly negative. This means that, for a one-particle excitation we have:
 \beq
  S_n^1(r)=\frac{\log(r^n+(1-r)^n)}{1-n},
 \label{ds}
 \eeq 
which is what we expected, and 
 \beq
  \mathcal{E}_n^1(r_1,r_3)= \log\left(r_1^n + r_3^n+ \frac{(r+\sqrt{r^2+4 r_1 r_3})^n}{2^n}+ \frac{(r-\sqrt{r^2+4 r_1 r_3})^n}{2^n}\right).
 \label{dede}
 \eeq
Interestingly, for $n$ integer, even or odd, there is no square-root dependence of the polynomial above (the square-roots always cancel).
Indeed, it can be equivalently written as given in (\ref{de}).
However, the logarithmic negativity itself involves the square root in (\ref{dede}). From the eigenvalues above this gives
 \beq
   \mathcal{E}^1(r_1,r_3)=\log\left(\sum_{i=1}^4 |\lambda_i^t| \right)=\log(r_1+r_3+ \sqrt{r^2+4 r_1 r_3}),
 \eeq
 which is the result (\ref{ep1}) reported earlier. 
As we can see, (\ref{ds}) and (\ref{dede}) are in general rather different functions. However, it is easy to show that the polynomials inside the logarithm coincide for $n=2$. 

 \subsection{ Multiple Particle Excitations}

The approach developed above, where we compute the eigenvalues of the reduced (partially transposed) density matrix, has been applied also to the cases of two  and three identical excitations, leading to the results (\ref{ep2}) and (\ref{E3par}). The explicit computation is reported in Appendix B. For higher number of excitations,  this exact diagonalization approach becomes quickly intractable. Here, we instead present an approach based on evaluating the replica trace more generally. It directly leads to the results (\ref{ren}) and \eqref{de2}. However, in the latter case, we do not know in general how to extract the analytic continuation to $n=1$ from even values, necessary for the logarithmic negativity.

\subsubsection{Distinct Excitations}

Having considered the particular case above, it is worth noting that the qubit states associated with multiple distinct excitations are by construction factorized states. For instance for two distinct excitations we have the state:
 \beqa
|\Psi^{(1,1)}_{\rm qb}\ket&=&{r_1}|1100000\ket+r|001100\ket+r_3|000011\ket+\sqrt{r_1r_3}\left(|100001\ket+|010010\ket\right)\nonumber\\
&+&\sqrt{r_1r}\left(|100100\ket+|011000\ket\right)+ \sqrt{r r_3}\left(|000110\ket+|001000\ket\right)\nonumber\\
&=& |\Psi^{(1)}_{\rm qb}\ket \otimes |\Psi^{(1)}_{\rm qb}\ket
\eeqa
where the first two qubits live in region $A$, the next two in region $C$ and the final two in region $B$ and $ |\Psi^{(1)}_{\rm qb}\ket$ is the state (\ref{Psi1}). Thus the trace is also factorized and can be written as
\beqa
\mathrm{Tr}_C\left( |\Psi^{(1,1)}_{\rm qb}\ket \bra \Psi^{(1,1)}_{\rm qb}|\right)=\left[\mathrm{Tr}_C\left( |\Psi^{(1)}_{\rm qb}\ket \bra \Psi^{(1)}_{\rm qb}|\right)\right]^2.\eeqa
From this factorization it is easy to see that both the R\'enyi entropies and the logarithmic negativity will be twice their value for a single excitation. This generalizes to $k$-particle states consisting of distinct excitations. 

\subsubsection{Identical Excitations}

Consider instead a $k$-particle state consisting of $k$ identical excitations. Its associated qubit state can be written as:
\beq
 |\Psi^{(k)}_{\rm qb}\ket = \sum_{\{k_1,k_2, k_3\} \in \sigma^3_0(k)} c_{k_1,k_2, k_3} \,|k_1k_2 k_3 \ket\,,
\eeq
where $\sigma^3_0(k)$ represents the set of integer partitions of $k$ into three non-negative parts,
\beq
c_{k_1,k_2, k_3}:=\sqrt{\frac{k! r_1^{k_1} r^{k_2} r_3^{k_3}}{k_1! k_2! k_3!}} \delta_{k_1+k_2+k_3,k},\,
\label{cdef}
\eeq
and we recall once more the definitions (\ref{r}). 

The state may be interpreted as follows: each state $|k_1 k_2 k_3\ket$ is a state of $k_1$ excitations in region $A$, $k_2$ excitations in region $C$ and $k_3$ excitations in region $B$. The square of the corresponding coefficient $c_{k_1,k_2, k_3}^2$ is the associated probability that
this configuration occurs if we were to place randomly and independently, with uniform
distribution, $k$ particles on the interval $[0, 1]$ covered by three non-intersecting subintervals of
lengths $r_1, r$ and $r_3$.

Provided that all vectors $|k_1 k_2 k_3\ket$ are normalized to one, then the vector $ |\Psi^{(k)}_{\rm qb}\ket $ is also a unit vector,
\beq
\bra \Psi^{(k)}_{\rm qb}|\Psi^{(k)}_{\rm qb}\ket  \sum_{\{k_1,k_2, k_3\} \in \sigma^3_0(k)}  c^2_{k_1,k_2,k_3}=(r_1+r+r_3)^k=1.
\eeq
From this expression it is then possible to explicitly construct the reduced density matrix and its partially transposed version as:
\beqa
\bra k_1 k_3|\rho_{A \bbc B}|k'_1 k'_3\ket&=&\sum_{k_2 \in \mathbb{N}_0}  
c_{k_1 k_2 k_3} c_{k'_1 k_2 k'_3},\,\\
\bra k_1 k_3|\rho^{T_B}_{A \bbc B}|k'_1 k'_3\ket&=&\sum_{k_2 \in \mathbb{N}_0}  
c_{k_1 k_2 k'_3} c_{k'_1 k_2 k_3}.\,
\eeqa
Here the sums run over all non-negative integers, and whenever the constraint $k_1+k_2+k_3=k$ is violated, the corresponding coefficient $c_{k_1,k_2,k_3}$ is zero by definition. From this constraint we know in fact that $0\leq k_2 \leq k$ so we could have restricted the summation range $k_2 \in \mathbb{N}_0$ much more. However we will write it like this for now, for simplicity, and discuss the summation ranges more precisely at the end of our calculation. With these results we can now evaluate the matrix elements of the $n$-th powers of the reduced density matrices above. These are given by:
\beqa
\bra k^{1}_1 k^{1}_3|\rho^n_{A \bbc B}|k^{n+1}_1 k^{n+1}_3\ket&=& 
\sum_{k_1^s, k_3^s \in \N_0; s=2,\ldots,n \atop k_2^r \in \N_0; r \in I_n}
  \prod_{j=1}^n c_{k^{j}_1 k_2^{j} k_3^{j}} c_{k^{j+1}_1 k^{j}_2 k^{j+1}_3},\,\\
\bra k^{1}_1 k^{1}_3|\left(\rho^{T_B}_{A \bbc B}\right)^n|k^{n+1}_1 k^{n+1}_3\ket&=&  \sum_{k_1^s, k_3^s \in \N_0; s=2,\ldots,n \atop k_2^r \in \N_0; r \in I_n}	 \prod_{j=1}^n c_{k^{j}_1 k_2^{j} k_3^{j+1}} c_{k^{j+1}_1 k^{j}_2 k^{j}_3}\,,
\label{trace2}
\eeqa
where $I_n:=\{1,\ldots, n\}$.
Finally, we are interested in the R\'enyi entropies and the replica logarithmic negativities, which means we need to take the trace over $A \bbc  B$ of the matrices above. This gives the following results:
\beqa
S_n^k(r_1,r_3)&=&\frac{1}{1-n}\log\left( \sum_{\{k_i^{s} \in \mathbb{N}_0; i\in I_3; s\in I_n\}}\prod_{j=1}^n c_{k^{j}_1 k_2^{j} k_3^{j}} c_{k^{j+1}_1 k^{j}_2 k^{j+1}_3}\right)\,,\\
\mathcal{E}_n^k(r_1,r_3)&=& \log\left( \sum_{\{k_i^{s} \in \mathbb{N}_0; i\in I_3; s\in I_n\}} \prod_{j=1}^n c_{k^{j}_1 k_2^{j} k_3^{j+1}} c_{k^{j+1}_1 k^{j}_2 k^{j}_3}\right)\,,
\eeqa
 where we adopt the convention $k_i^{1} \equiv k_i^{n+1}$ for $i=1,2,3$.
We can write these formulae more explicitly by employing the definition (\ref{cdef}), giving
\beqa
S_n^k(r_1,r_3)&=&\frac{1}{1-n}\log\left( \sum_{\{k_i^{s} \in \mathbb{N}_0; i\in I_3; s\in I_n\}}  \prod_{j=1}^n \frac{k!  r_1^{k_1^{j}} r^{k_2^{j}}r_3^{k_3^{j}}}{k_1^{j}! k^{j}_2! k_3^{j}!}  \delta_{k_1^{j}+k_2^{j}+k_3^{j},k} \, \delta_{k_1^{j+1}+k_2^{j}+k_3^{j+1},k} 
\right),\,\\ \label{SSS}
\mathcal{E}_n^k(r_1,r_3)&=& \log\left( \sum_{\{k_i^{s} \in \mathbb{N}_0; i\in I_3; s\in I_n\}} \prod_{j=1}^n \frac{k!  r_1^{k_1^{j}} r^{k_2^{j}}r_3^{k_3^{j}}}{k_1^{j}! k^{j}_2! k_3^{j}!}  \delta_{k_1^{j}+k_2^{j}+k_3^{j+1},k} \, \delta_{k_1^{j+1}+k_2^{j}+k_3^{j},k} 
\right)\,.\label{EEE}
\eeqa 
\subsubsection{Results for R\'enyi Entropies}
\label{entropy2r}
We can now eliminate the delta-functions by implementing their constraints. Let us start with the R\'enyi entropies. We can substitute:
\beq
k_1^{j}=k-k_2^{j-1}-k_3^{j}\,,\quad \forall \, j\,,
\eeq
and this will eliminate the sums over $k_1^{j}$ with $j=1,\ldots,n$. We then have sums over $k_2^{j}$ and $k_3^{j}$ left but we can also eliminate one of these by implementing the second set of delta-functions together with the conditions above. This gives the constraints,
\beq
k_2^{j}=k_2^{j+1}=:p \quad \forall \, j\,.
\label{defp}
\eeq
Therefore, $k_1^{j}=k-p-k_3^{j}$. This means that the factor $\prod_j r_1^{k_1^{j}} r^{k_2^{j}}r_3^{k_3^{j}}$ in (\ref{SSS}) becomes 
\beq
\prod_{j=1}^n r_1^{k_1^{j}} r^{k_2^{j}}r_3^{k_3^{j}} = r_1^{n(k-p)-q} r^{np} r_3^{q}\,,
\eeq
where we defined $q:=\sum_{j=1}^n k_3^{j}$. This finally allows us to rewrite (\ref{SSS}) as
\beq
S_n^k(r_1,r_3)=\frac{1}{1-n}\log\left(\sum_{p=0}^k   \sum_{q=0}^{n(k-p)}\mathcal{Z}_{p,q}  r_1^{n(k-p)-q} r^{np} r_3^{q}
\right)\,,
\eeq
where
\beq
\mathcal{Z}_{p,q}= \sum_{\{k_1,\ldots,k_n \} \in \sigma_0^n(q)} \prod_{j=1}^n \frac{k!}{p! k_j! (k-p-k_j)!}\,,
\eeq
and $\sigma_0^n(q)$ represents the set of integer partitions of $q$ into $n$ non-negative parts. 
We have relabelled $k_3^j:=k_j$, and the range of the sums in $p$ and $q$ is determined by the condition of $\mathcal{Z}_{p,q}\neq 0$. From the definition (\ref{defp}), $0\leq p \leq k$. Regarding the values of $q$, we know that $q$ can not be negative (by definition) so $q\geq 0$. Its maximum value is obtained if $k-p-k_j=0$ for all $j$. This corresponds to $q=n(k-p)$. In fact the sum over $q$ can be rewritten as
\beq
  \sum_{q=0}^{n(k-p)}\mathcal{Z}_{p,q}  r_1^{n(k-p)-q} r_3^{q} = \prod_{j=1}^n \sum_{k_j=0}^{k-p}  \frac{k!}{p! k_j! (k-p-k_j)!} r_1^{k-p-k_j} r_3^{k_j} =  \left[ {k \choose p} (r_1+r_3)^{k-p}\right]^n\,,
\eeq
with $r_1+r_3=1-r$ so that
\beq
S_n^k(r_1,r_3)=:S_n^k(r)=\frac{1}{1-n}\log\left(\sum_{p=0}^k  \left[{k\choose p}r^{p} (1-r)^{k-p}\right]^n
\right),\,
\label{onlyr}
\eeq
as reported in (\ref{ren2}).
Therefore, the entanglement entropy depends {\it only} on the parameter $r$ and is given by exactly the same expression as found in \cite{the4ofus1,the4ofus2}. In other words, in the qubit picture, the entanglement entropy depends only on the overall size of regions and not on whether or not they are connected. This implies that the same result should also hold for more than two disconnected regions. Indeed this can be shown by similar methods. The proof is presented in Appendix C.
\subsubsection{Results for Replica Logarithmic Negativity}
\label{324}
A similar analysis can be carried out for the replica logarithmic negativity. Starting with (\ref{EEE}) the second delta function gives the condition:
\beq
k_1^{j}=k-k_2^{j-1}-k_3^{j-1},\quad \forall \, j\,,
\eeq
and this eliminates the sums over $k_1^{j}$ with $j=1,\ldots,n$. We then have sums over $k_2^{j}$ and $k_3^{j}$ left but we can also eliminate one of these by implementing the second set of delta-functions together with the conditions above. This gives the constraints,
\beq
k_2^{j}-k_2^{j-1}=k_3^{j-1}-k_3^{j+1},\quad \forall \, j\,.
\eeq
We may regard this equation as a first order difference equation for the sequence $k_2^{j}$. The solution to such an equation is the sum of the solution to its homogenous version (a constant) and a particular solution of the full equation which can be worked out by inspection to be $-k_3^{j}-k_3^{j+1}$. The general solution is then
\beq
k_2^{j}=\gamma-k_3^{j}-k_3^{j-1}\,,
\label{relk2}
\eeq
where $\gamma$ is an arbitrary constant. With this we also have that $k_1^{j}=k-\gamma+k_3^{j}$. We can now evaluate the product 
\beq
\prod_{j=1}^n r_1^{k_1^{j}} r^{k_2^{j}}r_3^{k_3^{j}} = r_1^{n p+q} r^{n(k-p)-2q} r_3^{q}\,,
\eeq
where $q:=\sum_{j=1}^n k_3^{j}$ and $p=k-\gamma$. Relabelling $k_3^{j+1}:=k_j$ we then find 
\beq
\mathcal{E}_n^k(r_1,r_3)=\frac{1}{1-n}\log\left(\sum_{p=-k}^k   \sum_{q=\max(0,-np)}^{{[\frac{n}{2}(k-p)]}}\mathcal{A}_{p,q}  r_1^{n p+q} r^{n(k-p)-2q} r_3^{q}
\right),\,
\eeq
where
\beq
\mathcal{A}_{p,q}= \sum_{\{k_1,\ldots,k_n \} \in \sigma_0^n(q)} \prod_{j=1}^n \frac{k!}{(p+k_j)! (k-p-k_j-k_{j+1})! k_j!},\,
\label{apq}
\eeq
as given in (\ref{de2}) and (\ref{pro}). The range of sums in $p$ and $q$ is fixed by selecting out those contributions for which $\mathcal{A}_{p,q}\neq 0$. This requires that the arguments of the factorials in the denominator remain non-negative, which in turn restricts the type of partitions that can contribute to the sum over $k_1, \ldots, k_n$. 

Consider the sum in $p$. The range of this sum can be determined easily from the relation (\ref{relk2}). This implies that $0\leq \gamma \leq 2k$. Together with the definition  of $p$ this gives $-k\leq p \leq k$. This guarantees that all arguments of the factorials in the denominator remain non-negative. 

The range of values of $q$ can also be determined as follows. The lower limit is easy to establish as whenever $p<0$ the partitions contributing to $\mathcal{A}_{p,q}$ must have $k_j\geq -p$. Thus, the smallest value of $q$ giving a non-vanishing contribution corresponds to taking all $k_j=-p$ for all $j$ which gives $q=-np$. On the other hand, if $p\geq 0$ then the smallest value $q$ can take is zero corresponding to all $k_j=0$. This fixes the lower bound to $\max(0,-np)$. Let us now consider the upper bound. Given a certain $p$, the largest value $q$ can take corresponds to having 
$k-p-k_j-k_{j+1}=0$ for all $j$, or $k_j+k_{j+1}=k-p$. Writing 
\beq
\sum_{j=1}^n (k_j+k_{j+1})=\sum_{j=1}^n (k-p)\,,
\eeq
we have obviously that the right hand side gives $n(k-p)$ whereas  the left hand side gives $2 \sum_{j=1}^n k_j=2q$. Therefore, for generic parity of $n$ and $k-p$, we obtain $q=[\frac{n(k-p)}{2}]$. 
This gives the range $\max(0,-np) \leq q \leq [\frac{n(k-p)}{2}]$.


\section{Computation from Branch Point Twist Fields}
\label{twist}

In this section we evaluate the leading volume contribution to the increment
of R\'enyi entropies of two disconnected regions
and the increment of the replica logarithmic negativity using branch point twist field methods in the {\it free massive boson theory}. This provides a formal derivation, from QFT methods, of the results presented in the previous sections, which were argued for based on the qubit picture.  We follow the same  strategy  as for the computation presented in \cite{the4ofus2}, generalized to the study of the four-point functions entering the definitions (\ref{3}) and (\ref{4}).  
All the techniques that we use in this section (branch point twist fields, finite volume form factors, doubling trick) have been exhaustively reviewed in \cite{the4ofus2} and we refer the reader to this paper for further details.  Here we will just enumerate the main ideas and techniques needed.

\subsection{Techniques and Main Formulae}

As usual in the context of branch point twist fields we will be working with a replica free boson theory, consisting of $n$ copies of the original model.  We then first implement the {\it doubling trick} \cite{doubling} on the replica free boson
theory to construct a replica free massive complex boson theory. The doubling introduces a new $U(1)$ internal symmetry on each replica. This makes it possible to diagonalize the action of the branch point twist fields on asymptotic states. Associated to these new $U(1)$ symmetries there are $n$ twist fields $\TT_p$ with $p=1,\ldots,n$ in terms of which the branch point twist field can be expressed as:
\begin{eqnarray}
\mathcal{T}  =  \prod_{p=1}^{n}\mathcal{T}_{p}\,,
\label{diagonal}
\end{eqnarray}
where $p$ labels the $U(1)$ sectors. The form factors of the $U(1)$ twist fields will constitute the building blocks of all our computations so we will recall the main formulae here. They can be computed by the standard techniques  \cite{KW,SmirnovBook,YZam}  and have been known for a long time \cite{SMJ, VEVFB}. The formulae that we will need here are:
\beq 
f^{n}_p(\theta_{12}):=\frac{{}_p\bra 0| \TT_p(0) \textfrak{a}_{p}^\dag(\theta_1) \textfrak{b}_{p}^\dag(\theta_2)|0\ket_p}{{}_p\bra 0| \TT_p |0\ket_p}=- \sin\frac{\pi p}{n} \frac{e^{\left(\frac{p}{n}-\frac{1}{2} \right)\theta_{12}}}{\cosh \frac{\theta_{12}}{2}}\,,\label{minus}
\eeq 
where $\theta_{12}:=\theta_1-\theta_2$. This is the normalized (by the vacuum expectation value) matrix element of a sector-$p$ $U(1)$ twist field between the vacuum in that sector and a two-particle state. The state consists of a complex free boson (with creation operator $\textfrak{a}_{p}^\dag(\theta_1)$ and rapidity $\theta_1$) and its charge-conjugate boson (with creation operator $\textfrak{b}_{p}^\dag(\theta_2)$ and rapidity $\theta_2$). The dependence on the sector is determined by the form factor equations, in particular the crossing relation $f^{n}_p(\theta+2\pi i)=e^{\frac{2\pi i p}{n}} f_p^n(\theta)$.
Since the theory is free, higher particle form factors can be obtained by simply employing Wick's theorem. 
 For the complex free boson they have the structure
\beqa 
\!\!\!\!\!\!F^{p,n}(\theta_1,\ldots,\theta_m; \beta_1,\ldots, \beta_m)&:=&
 {}_p\bra 0|\TT_p(0)\textfrak{a}_{p}^\dag(\theta_1)\cdots \textfrak{a}_{p}^\dag(\theta_m) \textfrak{b}_{p}^\dag(\beta_1)\cdots \textfrak{b}_{p}^\dag(\beta_m)|0\ket_p \nonumber\\
&=&
 {}_p\bra 0| \TT_p |0\ket_p \sum_{\sigma \in S_m} f_p^n(\theta_{\sigma(1)}-\beta_1)\cdots 
f_p^n(\theta_{\sigma(m)}-\beta_{m})\,, \label{suma}
\eeqa 
where $\sigma$ are all elements of the permutation group $S_m$ of $m$ symbols and each two-particle form factor can  be associated with a particle-antiparticle contraction.
 Note that any other form factors are zero (e.g. if the numbers of particles and antiparticles are different).

The second important ingredient in our calculation is understanding the structure of the Hilbert space of the theory in the replica doubled theory. That is, the description of the asymptotic states 
in each sector of the theory in a basis upon which the $U(1)$ twist fields act diagonally. The solution to this problem was discussed at length in \cite{the4ofus2} and can be schematically presented as follows. First,
\beq
a_j^\dagger(\theta) \mapsto \frac{1}{\sqrt{2}}((a_j^+)^\dagger(\theta)+(a_j^-)^\dag(\theta)), \quad j=1,\ldots,n
\label{brexit1}
\eeq
that is, the creation operator of a real boson on copy $j$ is mapped after doubling into a linear combination of creation operators on copy $j$ associated with a complex free boson (+) and its charge conjugate (-). As expected, the branch point twist field action on these operators is such that they are mapped to operators on copy $j+1$ of the theory. However, there exists a new basis $\{\textfrak{a}_{p}^\dag(\theta), \textfrak{b}_{p}^\dag(\theta)\}$ with $p=1,\ldots,n$ on which the branch point twist field acts diagonally, with each $U(1)$ field (\ref{diagonal}) acting on a particular sector. For example, in can be shown
\beq
(a_j^+)^\dag(\theta)=\frac{1}{\sqrt{n}} \sum_{p=1}^n e^{-\frac{2\pi i j p}{n}} \textfrak{a}_{p}^\dag(\theta)\,,
\eeq
where the operators $ \textfrak{a}_{p}^\dag(\theta),  \textfrak{b}_{p}^\dag(\theta)$ satisfy also a complex boson algebra. 
We will from now on perform computations on this basis, where we can employ the matrix elements of $U(1)$ twist fields seen above. Although this simplifies the computation of form factors enormously, the price to pay is a more complicated structure for the excited states which must now be also expressed in this new basis (see e.g. (\ref{415}) for an example). 

The third and final issue to consider is the finite volume extension of the above. In particular, rapidities will be quantized  in finite volume and the quantization conditions will depend upon the sector $p$ the corresponding creation/anhilation operator is acting on, and upon an index $\alpha$ that parametrizes the periodicity conditions for the fields $\Phi_j(x+L)=\Phi_{j+\alpha}(x)$, where $\Phi_j(x)$ is the complex bosonic field with associated creation  operators in (\ref{brexit1}). A set of (generalized) Bethe-Yang equations \cite{Bethe, YangYang} can be written as
\beq
h_{\alpha}^{p,\epsilon}\left(\theta\right)=2\pi J\, \qquad \mathrm{with} \qquad
h_{\alpha}^{p,\epsilon}\left(\theta\right)=mL\sinh\left(\theta\right)-2 \pi \epsilon \alpha\frac{p}{n}\,,\label{eq:quant_cond}
\eeq
 where $\epsilon=\pm$ is the $U(1)$ charge of the
particle and $J\in\mathbb{Z}$. The sum over a complete set of states in sector $p$ with quantization condition $\alpha$ can be written as
\begin{eqnarray}
\label{eq:complete_set_p}
\mathbf{I}_{\alpha}^{p} & = & \sum_{m^{\pm}=0}^{\infty}\frac{1}{m^{+}!m^{-}!}\sum_{\left\{ J^{\pm}\right\} }\prod_{i=1}^{m^{+}}\mathfrak{a}_{p}^{\dagger}\left({\theta_{i}^+}\right)\prod_{j=1}^{m^{-}}\mathfrak{b}_{p}^{\dagger}\left({\theta_{j}^-}\right)\left| 0\right \rangle \! \big. _{p;L}\, \big._{p;L}\! \left\langle 0\right|\prod_{j=1}^{m^{-}}\mathfrak{b}_{p}\left({\theta_{j}^-}\right)\prod_{i=1}^{m^{+}}\mathfrak{a}_{p}\left({\theta_{i}^+}\right)\,,
\end{eqnarray}
The summation runs over the integer sets $\left\{ J^{\pm}\right\} =\left\{ J_{1}^{+},\dots,J_{m^{+}}^{+},J_{1}^{-},\dots,J_{m^{-}}^{-}\right\} $
and the rapidities satisfy the quantization conditions (\ref{eq:quant_cond}) as:
\beqa
 h_{\alpha}^{p,\epsilon}\left({\theta_i^\pm}\right)=2 \pi J^\epsilon_i\,\qquad i=1,\ldots m^\epsilon\,.
 \eeqa
We can then define the complete sum over all sectors
\begin{eqnarray}
\mathbf{I}_{\alpha} =  \otimes_{p=1}^{n}\mathbf{I}_{\alpha}^{p}\,.
\end{eqnarray}
The branch point twist fields intertwine between the different quantization
sectors. Denoting the corresponding Hilbert space by $\mathcal{H}_{\alpha}$ we can write
\begin{eqnarray}
\mathcal{T}:\mathcal{H}_{\alpha}\to\mathcal{H}_{\alpha-1}\,, & \quad & \tilde{\mathcal{T}}:\mathcal{H}_{\alpha}\to\mathcal{H}_{\alpha+1}\,.\label{eq:twist_field_action_on_sectors}
\end{eqnarray}
The excited states $|\Psi\ket_L$ we consider in this calculation are elements of
the trivial quantization sector $\mathcal{H}_{0}\equiv \mathcal{H}_n$. This,  combined with the properties of the branch point twist fields means that the four point functions of interest may be spanned as
\beq
\,_{L}\left\langle \Psi\right|\mathcal{T}\left(0\right)\tilde{\mathcal{T}}\left(x_{1}\right)\mathcal{T}\left(x_{2}\right)\tilde{\mathcal{T}}\left(x_{3}\right)\left|\Psi\right\rangle _{L}  =  \,_{L}\left\langle \Psi\right|\mathcal{T}\left(0\right)\,\mathbf{I}_{1}\,\tilde{\mathcal{T}}\left(x_{1}\right)\,\mathbf{I}_{0}\,\mathcal{T}\left(x_{2}\right)\,\mathbf{I}_{1}\,\tilde{\mathcal{T}}\left(x_{3}\right)\left|\Psi\right\rangle _{L}\,,
\eeq
\beq
\,_{L}\left\langle \Psi\right|\mathcal{T}\left(0\right)\tilde{\mathcal{T}}\,\left(x_{1}\right)\tilde{\mathcal{T}}\left(x_{2}\right)\mathcal{T}\left(x_{3}\right)\left|\Psi\right\rangle _{L} =  \,_{L}\left\langle \Psi\right|\mathcal{T}\left(0\right)\,\mathbf{I}_{1}\tilde{\mathcal{T}}\,\left(x_{1}\right)\,\mathbf{I}_{0}\,\tilde{\mathcal{T}}\left(x_{2}\right)\,\mathbf{I}_{-1}\,\mathcal{T}\left(x_{3}\right)\left|\Psi\right\rangle _{L}\,,
\eeq
where $x_{1,2,3}$ denote the positions of the branch point twist fields which is related to the original lengths in Fig.~1 as
\beq
 x_{1}:=\ell_{1}\,, \quad x_{2}:=\ell_{1}+\ell_{2}\,,\quad \mathrm{and} \quad x_{3}:=\ell_{1}+\ell_{2}+\ell_{3}\,.
 \eeq
Employing (\ref{diagonal})  the four-point functions then factorize into products of contributions from every sector, involving the finite volume matrix elements of $U(1)$ twist fields. These matrix elements are related to infinite volume form factors (\ref{suma}) in a simple well-known fashion \cite{PT1,PT2} which involves dividing each infinite volume form factor by the square root of the product of the associated particle energies (in general, for interacting theories, this would be the density of states) of the excitations involved, and employing the crossing property of form factors. This can be summarized as
\beqa
&&  {}_{p;L}\bra 0|\prod_{i_1=1}^s \textfrak{a}_p(\theta_{i_1})\prod_{i_2=1}^q \textfrak{b}_p(\beta_{i_2})\TT_p(0)\prod_{i_4=1}^{q'} \textfrak{b}^\dagger_p(\beta'_{i_4})\prod_{i_3=1}^{s'} \textfrak{a}_p^\dagger(\theta'_{i_3})|0\ket_{p;L} = \label{ftofL}\\
&& \quad  \frac{F^{p,n}(\theta'_1,\ldots,\theta'_{s'},\beta_1+i\pi,\ldots, \beta_q+i\pi; \beta'_1,\ldots, \beta'_{q'},\theta_1+i\pi,\ldots, \theta_s + i\pi)}{\sqrt{\prod_{i_1=1}^{s} L E(\theta'_{i_1})\prod_{i_2=1}^{q} L E(\beta'_{i_2})\prod_{i_3=1}^{s'} L E(\theta_{i_3}) \prod_{i_4=1}^{q'} L E(\beta'_{i_4})}} \,,
\nonumber
\eeqa
up to exponentially decaying corrections $O(e^{-\mu L})$ and $E(\theta)=m \cosh \theta$.

An important feature of the two-particle form factors (\ref{minus}) is their behaviour near kinematic poles
\beq 
f_p^n(\theta\pm i\pi) \underset{\theta\approx 0}{=}\pm i\frac{1-e^{\pm\frac{ i 2 \pi p}{n}}}{ \theta}\,.
\label{z12}
\eeq 

\subsection{Four Point Functions in Single-Particle Excited States}
Let us focus on the calculation for a single-particle excited
state denoted by $|1\ket_L$. If the excitation has rapidity $\theta$, the state in the $U(1)$ basis has the form 
\begin{eqnarray}
\left|1\right\rangle _{L} & = & \sum_{\left\{ N^{\pm}\right\} }C_{n}\left(\left\{ N^{\pm}\right\} \right)\prod_{p=1}^{n}\left[\mathfrak{a}_{p}^{\dagger}\left(\theta\right)\right]^{N_{p}^{+}}\left[\mathfrak{b}_{p}^{\dagger}\left(\theta\right)\right]^{N_{p}^{-}}\left|0\right\rangle _{L}\,,
\label{415}
\end{eqnarray}
where $C_{n}\left(\left\{ N^{\pm}\right\} \right)$ are the (known) coefficients
containing all the phase factors of the basis transformations (see subsection 4.1.3 in \cite{the4ofus2} for an explicit example). The
rapidity is the solution of the quantization condition (\ref{eq:quant_cond})
\begin{equation}
h_{0}^{p,\epsilon}\left(\theta\right)=mL\sinh\left(\theta\right)=2 \pi  I^0\,.
\end{equation}
The summation runs over the integer
sets $\left\{ N^{\pm}\right\} =\left\{ N_{1}^{+},N_{1}^{-},\dots,N_{n}^{+},N_{n}^{-}\right\} $
subject to the condition 
\begin{eqnarray}
\sum_{p=1}^{n}\sum_{\epsilon=\pm}N_{p}^{\epsilon} & = & n\,.
\end{eqnarray}
Following previous considerations, the four point functions  are 
\begin{eqnarray}
\,_{L}\left\langle 1\right|\mathcal{T}\left(0\right)\tilde{\mathcal{T}}\left(x_{1}\right)\mathcal{T}\left(x_{2}\right)\tilde{\mathcal{T}}\left(x_{3}\right)\left|1\right\rangle _{L} & = & \sum_{\left\{ N^{\pm}\right\} }\sum_{\left\{ \tilde{N}^{\pm}\right\} }\left[C_{n}\left(\left\{ N^{\pm}\right\} \right)\right]^{*}C_{n}\left(\left\{ \tilde{N}^{\pm}\right\} \right)\nonumber \\
 &  & \times\prod_{p=1}^{n}\mathcal{F}_{p}\left(N_{p}^{\pm},\tilde{N}_{p}^{\pm}\right)\,,
 \label{419}
 \eeqa
 \beqa
\,_{L}\left\langle 1\right|\mathcal{T}\left(0\right)\tilde{\mathcal{T}}\left(x_{1}\right)\tilde{\mathcal{T}}\left(x_{2}\right)\mathcal{T}\left(x_{3}\right)\left|1\right\rangle _{L} & = & \sum_{\left\{ N^{\pm}\right\} }\sum_{\left\{ \tilde{N}^{\pm}\right\} }\left[C_{n}\left(\left\{ N^{\pm}\right\} \right)\right]^{*}C_{n}\left(\left\{ \tilde{N}^{\pm}\right\} \right)\nonumber \\
 &  & \times\prod_{p=1}^{n}\tilde{\mathcal{F}}_{p}\left(N_{p}^{\pm},\tilde{N}_{p}^{\pm}\right)\,,
 \label{420}
\end{eqnarray}
with the different sector contributions
\begin{eqnarray}
\mathcal{F}_{p}\left(N_{p}^{\pm},\tilde{N}_{p}^{\pm}\right) & = & \,_{p;L}\left\langle 0\right|\left[\mathfrak{a}_{p}\left(\theta\right)\right]^{N_{p}^{+}}\left[\mathfrak{b}_{p}\left(\theta\right)\right]^{N_{p}^{-}}\mathcal{T}_{p}\left(0\right)\,\mathbf{I}_{1}^{p}\,\tilde{\mathcal{T}}_{p}\left(x_{1}\right)\,\mathbf{I}_{0}^{p} \label{eq:EE_cont_sector_p} \\
 &  & \times\mathcal{T}_{p}\left(x_{2}\right)\,\mathbf{I}_{1}^{p}\,\tilde{\mathcal{T}}_{p}\left(x_{3}\right)\left[\mathfrak{a}_{p}^{\dagger}\left(\theta\right)\right]^{\tilde{N}_{p}^{+}}\left[\mathfrak{b}_{p}^{\dagger}\left(\theta\right)\right]^{\tilde{N}_{p}^{-}}\left|0\right\rangle _{p;L}\,,\nonumber\\
\tilde{\mathcal{F}}_{p}\left(N_{p}^{\pm},\tilde{N}_{p}^{\pm}\right) & = & \,_{p;L}\left\langle 0\right|\left[\mathfrak{a}_{p}\left(\theta\right)\right]^{N_{p}^{+}}\left[\mathfrak{b}_{p}\left(\theta\right)\right]^{N_{p}^{-}}\mathcal{T}_{p}\left(0\right)\,\mathbf{I}_{1}^{p}\,\tilde{\mathcal{T}}_{p}\left(x_{1}\right)\,\mathbf{I}_{0}^{p} \label{eq:LogNeg_cont_sector_p}\\
 &  & \times\tilde{\mathcal{T}}_{p}\left(x_{2}\right)\,\mathbf{I}_{-1}^{p}\,\mathcal{T}_{p}\left(x_{3}\right)\left[\mathfrak{a}_{p}^{\dagger}\left(\theta\right)\right]^{\tilde{N}_{p}^{+}}\left[\mathfrak{b}_{p}^{\dagger}\left(\theta\right)\right]^{\tilde{N}_{p}^{-}}\left|0\right\rangle _{p;L}\,. \nonumber
\end{eqnarray}
In sector $n$ the $U(1)$ twist fields coincide
with the identity operator, hence the contributions from this sector are
just the normalization of the state 
\begin{eqnarray}
\mathcal{F}_{n}\left(N_{n}^{\pm},\tilde{N}_{n}^{\pm}\right) & = & \tilde{\mathcal{F}}_{n}\left(N_{n}^{\pm},\tilde{N}_{n}^{\pm}\right)=N_{n}^{+}!N_{n}^{-}!\delta_{N_{n}^{+},\tilde{N}_{n}^{+}}\delta_{N_{n}^{-},\tilde{N}_{n}^{-}}\,.
\end{eqnarray}
The strategy of the calculation from this point on is to evaluate the
matrix elements of the twist fields using the finite volume form factor
formula (\ref{ftofL}), turn the summation
for quantum numbers into contour integrals, and by contour manipulation
extract the leading volume  contribution. First, we focus
on the increment of the R\'enyi entropy of two
disconnected regions. We will later see that the logarithmic negativity can be computed in a very similar way.


\subsection{Entanglement Entropy of two Disconnected Regions}
\label{sec:EE_2regions_ff}

Let us compute the ratio of correlators (\ref{3})  for a single-particle excited state. We need to evaluate (\ref{419}) with (\ref{eq:EE_cont_sector_p}). We will start by focussing on the evaluation of (\ref{eq:EE_cont_sector_p}) through the introduction of the complete sums ${\bf I}^p_{0}$ and  ${\bf I}^p_{1}$. 
We first introduce some
notations. We denote the rapidities of the complete sets of states,
starting from left to right, by $\beta_i^{\epsilon},\alpha_i^{\epsilon},\tilde{\beta}_i^{\epsilon}$
and their numbers by $m^{\epsilon},k^{\epsilon},\tilde{m}^{\epsilon}$,
respectively.  As usual $\epsilon=\pm$. The rapidities satisfy the following Bethe-Yang quantization
conditions (\ref{eq:quant_cond})
\begin{eqnarray}
h_{1}^{p,\epsilon}\left(\beta_{i}^{\epsilon}\right) & = & mL\sinh\left(\beta_{i}^{\epsilon}\right)-2\pi\epsilon\frac{p}{n}=2\pi J_{i}^{\epsilon}\,,\label{by1}\\
h_{0}^{p,\epsilon}\left(\alpha_{i}^{\epsilon}\right) & = & mL\sinh\left(\alpha_{i}^{\epsilon}\right)=2\pi I_{i}^{\epsilon}\,,\label{by2}\\
h_{1}^{p,\epsilon}\left(\tilde{\beta}_{i}^{\epsilon}\right) & = & mL\sinh\left(\tilde{\beta}_{i}^{\epsilon}\right)-2\pi\epsilon\frac{p}{n}=2\pi\tilde{J}_{i}^{\epsilon} \,. \label{by3}
\end{eqnarray}
We introduce also the following notation for the various sets of rapidities 
\begin{eqnarray}
\left\{ \theta\right\} ^{N} & = & \underbrace{\left\{ \theta,\dots,\theta\right\} }_{N}\,,\nonumber\\
\left\{ \beta^{\epsilon}\right\} _{m} & = & \left\{ \beta_{1}^{\epsilon},\beta_{2}^{\epsilon},\dots,\beta_{m}^{\epsilon}\right\} \,,\nonumber\\
\{ \hat{\beta}^{\epsilon}\} _{m} & = & \left\{ \beta_{1}^{\epsilon}+i\pi,\beta_{2}^{\epsilon}+i\pi,\dots,\beta_{m}^{\epsilon}+i\pi\right\} \,,
\end{eqnarray}
and similarly for the $\alpha_i^{\epsilon}$ and $\tilde{\beta}_i^{\epsilon}$ rapidities. Using
the finite volume form factor formula (\ref{ftofL}) and these notations, (\ref{eq:EE_cont_sector_p})
takes the form 
\begin{eqnarray}
\mathcal{F}_{p}\left(N_{p}^{\pm},\tilde{N}_{p}^{\pm}\right) & = & \prod_{\epsilon=\pm}\sum_{m^{\epsilon}=0}^{\infty}\frac{1}{m^{\epsilon}!}\sum_{\left\{ J^{\epsilon}\right\} }\sum_{k^{\epsilon}=0}^{\infty}\frac{1}{k^{\epsilon}!}\sum_{\left\{ I^{\varepsilon}\right\} }\sum_{\tilde{m}^{\epsilon}=0}^{\infty}\frac{1}{\tilde{m}^{\epsilon}!}\sum_{\left\{ \tilde{J}^{\epsilon}\right\} }  \frac{e^{-ix_{3}\tilde{N}_p^{\epsilon}P\left(\theta\right)}}{\left[\sqrt{LE\left(\theta\right)}\right]^{N_{p}^{\epsilon}+\tilde{N}_{p}^{\epsilon}}}\nonumber\\
 &  & \times\frac{e^{ix_{1}\sum_{i=1}^{m^{\epsilon}}P\left(\beta_{i}^{\epsilon}\right)}e^{i\left(x_{2}-x_{1}\right)\sum_{i=1}^{k^{\epsilon}}P\left(\alpha_{i}^{\epsilon}\right)}e^{i\left(x_{3}-x_{2}\right)\sum_{i=1}^{\tilde{m}^{\epsilon}}P\left(\tilde{\beta}_{i}^{\epsilon}\right)}}{\left[\prod_{i=1}^{m^{\epsilon}}LE\left(\beta_{i}^{\epsilon}\right)\right]\left[\prod_{i=1}^{k^{\epsilon}}LE\left(\alpha_{i}^{\epsilon}\right)\right]\left[\prod_{i=1}^{\tilde{m}^{\epsilon}}LE\left(\tilde{\beta}_{i}^{\epsilon}\right)\right]}\nonumber\\
 &  & \times F^{p,n}\left(\{ \hat{\theta}\} ^{N_{p}^{-}},\{ \beta^{+}\} _{m^{+}}; \{ \hat{\theta} \} ^{N_{p}^{+}}, \{ \beta^{-}\} _{m^{-}}\right)\nonumber\\
 &  & \times F^{n-p,n}\left(\{ \alpha^{+}\} _{k^{+}},\{ \hat{\beta}^{-}\} _{m^{-}};\{ \alpha^{-}\} _{k^{-}},\{ \hat{\beta}^{+}\} _{m^{+}}\right)\nonumber\\
 &  & \times F^{p,n}\left(\{ \hat{\alpha}^{-}\} _{k^{-}},\{ \tilde{\beta}^{+}\} _{\tilde{m}^{+}};\{ \hat{\alpha}^{+}\} _{k^{+}},\{ \tilde{\beta}^{-}\} _{\tilde{m}^{-}}\right)\nonumber\\
 &  & \times F^{n-p,n}\left(\{ \theta\} ^{\tilde{N}_{p}^{+}},\{\hat{\tilde{\beta}}^{-}\} _{\tilde{m}^{-}};\{ \theta\} ^{\tilde{N}_{p}^{-}},\{ \hat{\tilde{\beta}}^{+}\} _{\tilde{m}^{+}}\right)\,,
  \label{big}
\end{eqnarray}
where $P(\theta)=m \sinh \theta$ are the particle momenta. We have also used the symmetry property of the form factors (since we are dealing with free bosons, the form factors are symmetric in all rapidities).
We stress that at this stage all the rapidities in the formula are solutions
of the appropriate Bethe-Yang equations. This will not be the case  anymore when we rewrite our formulae in terms of contour integrals.
We will start by transforming  the $\{\beta^\epsilon\}$ set by writing
\begin{eqnarray}
\sum_{J_{i}^{\epsilon}\in\mathbb{Z}}\frac{f\left(\beta_{i}^{\epsilon},\dots\right)}{LE\left(\beta_{i}^{\epsilon}\right)} & = & \int_{\mathcal{C}_{p}^{\epsilon}}\frac{\mathrm{d}\beta_{i}^{\epsilon}}{2\pi}\frac{f\left(\beta_{i}^{\epsilon},\dots\right)}{e^{ih_{1}^{p,\epsilon}\left(\beta_{i}^{\epsilon}\right)}-1}\,,
\end{eqnarray}
where we now consider the function $h_{1}^{p,\epsilon}\left(\beta_{i}^{\epsilon}\right)$  as the function of $\beta_i^\epsilon$ defined in (\ref{by1}) so as to ensure that the integrand has a pole exactly when the Bethe-Yang equation is satisfied.
The function $f$ on the left hand side of the equation is to be taken from the $J_i^\ep$ summand in (\ref{big}), 
and the contour is the sum of small contours around the solutions of the
Bethe-Yang quantization condition (\ref{eq:quant_cond}) $h_{1}^{p,\epsilon}\left(\beta_i^{\epsilon}\right)=2\pi J_i^{\epsilon}$, (note that these all lie along the real line)
\begin{eqnarray}
\mathcal{C}_{p}^{\epsilon} & = & \sum_{J_i^\epsilon \in\mathbb{Z}}\mathcal{C}_{\beta_i^{\epsilon}}\,.
\end{eqnarray}
We deform the contour into a contour encircling the real axis with positive
orientation $\mathcal{C}_{\leftrightarrows}$ and subtract the residua
of the form factors' kinematic poles. The form factors (\ref{suma}) have kinematic
poles, when either $\beta_i^{\epsilon}=\theta$
in the first form factor or $\beta_i^{\epsilon}=\alpha_j^{\epsilon}$ for some $j$, 
in the second form factor. However, if the quantum number $I_{j}^{\epsilon}=I^{0}$,
the quantum number of the excited state, then $\alpha_{j}^{\epsilon}=\theta$
and  a second order pole can result from the combination of the kinematic
singularities of the two form factors. This leads to a different residue,
hence we should proceed with care. 

In order to correctly account for the different contributions,
we separate the quantum number sums into a group that coincide with
$I^{0}$ and the rest
\begin{eqnarray}\label{kM}
\sum_{k^{\epsilon}=0}^{\infty}\frac{1}{k^{\epsilon}!}\sum_{\left\{ I^{\epsilon}\right\} }f\left(\left\{ \alpha^{\epsilon}\right\} _{k^{\epsilon}},\dots\right) & \to & \sum_{M^{\epsilon}=0}^{\infty}\frac{1}{M^{\epsilon}!}\sum_{k^{\epsilon}=0}^{\infty}\frac{1}{k^{\epsilon}!}\sum_{{\left\{ I^{\epsilon}\right\} \not\ni I^{0}}}f\left(\left\{ \alpha^{\epsilon}\right\} _{k^{\epsilon}},\left\{ \theta\right\} ^{M^{\epsilon}},\dots\right)\,,
\end{eqnarray}
where we used the symmetry property of the form factors and the invariance 
of the integral under relabelling of variables $\{\alpha^\epsilon\}_{k^\epsilon}$. With these considerations, the contour after the deformation
is \begin{eqnarray}
\mathcal{C}_{p}^{\epsilon} =  \mathcal{C}_{\leftrightarrows}-\mathcal{C}_{\theta}-\sum_{j=1}^{k^{\epsilon}}\mathcal{C}_{\alpha_{j}^\epsilon}\,,
\label{contour1}
\end{eqnarray}
where $\alpha_j^\epsilon \neq \theta$, and the multi-contour for all $\beta_i^{\epsilon}$ rapidities is
\begin{eqnarray}
\left[\mathcal{C}_{p}^{\epsilon}\right]^{m^{\epsilon}} =  \underbrace{\mathcal{C}_{p}^{\epsilon}\times\mathcal{C}_{p}^{\epsilon}\times\cdots\times\mathcal{C}_{p}^{\epsilon}}_{m^{\epsilon}}\,.
\label{contour2}
\end{eqnarray}
We can expand the integration multi-contour by substituting (\ref{contour1}) into (\ref{contour2}) and employing the same symmetries of the integrand mentioned above.
As a result there are several terms of the multi-contour
that give the same residue and we can write
\begin{eqnarray}
\left[\mathcal{C}_{p}^{\epsilon}\right]^{m^{\epsilon}} & \sim & \sum_{m_{\leftrightarrows}^{\epsilon},m_{\theta}^{\epsilon},\left\{ \gamma^{\epsilon}\right\} _{m_{\alpha}^{\epsilon}}\subseteq\left\{ \alpha^{\epsilon}\right\} _{k^{\epsilon}}}\left(-1\right)^{m_{\alpha}^{\epsilon}+m_{\theta}^{\epsilon}} G\left( m_{\leftrightarrows}^{\epsilon},m_{\theta}^{\epsilon}, m_{\alpha}^{\epsilon}\right) \nonumber \\
 & & \times \left[\mathcal{C}_{\leftrightarrows}\right]^{m_{\leftrightarrows}^{\epsilon}}\times\left[\prod_{j=1}^{m_{\alpha}^{\epsilon}}\mathcal{C}_{\gamma_{j}^{\epsilon}}\right]\times\left[\mathcal{C}_{\theta}\right]^{m_{\theta}^{\epsilon}}\,,
 \label{gammas}
\end{eqnarray}
where $G\left( m_{\leftrightarrows}^{\epsilon},m_{\theta}^{\epsilon}, m_{\alpha}^{\epsilon}\right)$ is a combinatorial factor arising from the power expansion, $m_\theta^\epsilon$ is the number of contours about the rapidities $\theta$, and $m_\alpha^\epsilon$ is the number of rapidities in the set $\{\gamma^\epsilon\}_{m_\alpha^\epsilon}$ which is the subset of $\{\alpha^\epsilon\}_{k^\epsilon}$, itself being involved in a particular term of the expansion \eqref{kM}. The nonnegative integer summation indices are constrained by 
\begin{eqnarray}
m^{\epsilon}  =  m_{\leftrightarrows}^{\epsilon}+m_{\theta}^{\epsilon}+m_{\alpha}^{\epsilon}\,,
\end{eqnarray}
Note that although, in general, the order in which the integrals over the various contours are performed matters, in the expansion (\ref{gammas}) we can obviate this by employing the fact that all such orderings are equivalent under relabeling of rapidities and that in this case all such relabelings are equivalent due to the symmetries of the free boson form factors. 

Let us focus on the residua of the contour $\mathcal{C}_{\theta}$ (that is the poles with $\beta_i^\epsilon=\theta$).
As mentioned before, if $M^{\epsilon}\neq0$, the product of the two
first form factors produces both second and first oder poles. We denote by $m_{\theta,1/2}^{\epsilon}$
the number of first order poles arising from the first/second form factor, and by $m_{\theta,d}^{\epsilon}$ the number of second
order poles. They satisfy the condition 
\begin{eqnarray}
m_{\theta}^{\epsilon}  =  m_{\theta,1}^{\epsilon}+m_{\theta,2}^{\epsilon}+m_{\theta,d}^{\epsilon}\,.
\end{eqnarray}
Evaluating the residua of these poles produces several types of factors. In the phase factors, the rapidity is trivially replaced by $\theta$ or $\gamma^{\epsilon}_{i}$ (recall that the rapidities $\{\gamma^\epsilon\}_{m_\alpha^\epsilon}$ are a subset of $\{\alpha^\epsilon\}_{k^\epsilon}$ as defined in (\ref{gammas})). 
There are constant factors coming from the combination of the form factor residua and the denominators, simplified by using the Bethe-Yang equations (\ref{by1}) and (\ref{by2}). These factors contain derivatives in the case of higher order poles. 
Moreover, in case of first order poles, the contracted rapidity $\theta$ or $\gamma^{\epsilon}_{j}$ appears in the non-singular form factor. In the case of $\mathcal{C}_\theta$, this also means taking the regular part of that form factor, defined as the part that does not contain any two-particle form factors depending on $\beta^{\epsilon}_i-\theta$. The singular terms thus omitted are already accounted for in the double order pole residua. The form factors minus such omitted terms are what we later denote by an additional subscript ``reg" (see Eq.~(\ref{monster})).
There are combinatorial factors coming from counting all possible contractions of $\beta_i^{\epsilon}$ with $\theta$ or $\gamma^{\epsilon}_{j}$ rapidities inside a given form factor, that we discuss in detail later.
The residua of the singular two-particle form factor(s) are given by
\begin{eqnarray}
\mathcal{R}_{\theta,1}^{\epsilon}=1\,,&\quad&\mathcal{R}_{\theta,2}^{\epsilon}=e^{2\pi i\frac{\epsilon p}{n}}\,, \nonumber \\
\mathcal{R}_{\alpha}^{\epsilon}=e^{2\pi i\frac{\epsilon p}{n}}\,, &\quad&\mathcal{R}_{\theta,d}^{\epsilon}=LE\left(\theta\right)g_{\epsilon p}^{n}\left(r_{1}\right)\,,
\end{eqnarray}
where $\mathcal{R}_{\theta,1}^{\epsilon}, \mathcal{R}_{\theta,2}^{\epsilon}$ and $\mathcal{R}_{\alpha}^{\epsilon}$ correspond to first order poles and $\mathcal{R}_{\theta,d}^{\epsilon}$ corresponds to second order poles. $r_{1}=\frac{x_{1}}{L}=\frac{\ell_{1}}{L}$ and  
\beq
g_{\epsilon p}^{n}(r) = 1-r+r e^{\frac{2 \pi i \epsilon p}{n}}\,,\label{g_function}
\eeq
the function that already played an important role in the computation of the entanglement entropies of one interval \cite{the4ofus2}. 

The same consideration is valid for the $\tilde{\beta}^{\epsilon}$
rapidities. $\tilde{m}_{\theta,1/2}^{\epsilon}$ denotes
the number of first order poles resulting from the third/fourth form factor
having a singularity, and $\tilde{m}_{\theta,d}^{\epsilon}$ denotes the number of second
order poles. The residua of the singular two-particle form factor(s) are given by
\begin{eqnarray}
\tilde{\mathcal{R}}_{\theta,1}^{\epsilon}=1\,,&\quad&\tilde{\mathcal{R}}_{\theta,2}^{\epsilon}=e^{2\pi i\frac{\epsilon p}{n}}\,, \nonumber\\
\tilde{\mathcal{R}}_{\alpha}^{\epsilon}=1\,, &\quad&\tilde{\mathcal{R}}_{\theta,d}^{\epsilon}=LE\left(\theta\right)g_{\epsilon p}^{n}\left(r_{3}\right)\,,
\end{eqnarray}
where $r_{3}=\frac{x_{3}-x_{2}}{L}=\frac{\ell_{3}}{L}$. We stress that, as earlier, only the second order poles contribute a factor proportional to the volume $LE\left(\theta\right)$.

Successive computation of the residua associated with rapidities $\beta_i^\epsilon$ about any of the rapidities $\{\gamma^\epsilon\}_{m_\alpha^\epsilon}$ and $\tilde{\beta}_i^\epsilon$ about rapidities  $\{\tilde{\gamma}^\epsilon\}_{\tilde{m}_\alpha^\epsilon}$ results in a sum of contributions where a  ``reshufling" of the rapidities  $\{\alpha^{\epsilon}\}_{k^{\epsilon}}$ has taken place. These will now appear in the first or second and in the third or fourth form factors, in various combinations. To reflect this, we introduce the following partition of the rapidities
\begin{eqnarray}
\left\{ \alpha^{\epsilon}\right\} _{k^{\epsilon}} =  \left\{ \alpha_{13}^{\epsilon}\right\} _{k_{13}^{\epsilon}}\bbc\left\{ \alpha_{14}^{\epsilon}\right\} _{k_{14}^{\epsilon}}\bbc\left\{ \alpha_{23}^{\epsilon}\right\} _{k_{23}^{\epsilon}}\bbc\left\{ \alpha_{24}^{\epsilon}\right\} _{k_{24}^{\epsilon}}\,,
\end{eqnarray}
where $\{ \alpha_{ij}^{\epsilon}\} _{k_{ij}^{\epsilon}}$
denotes the subset of rapidities in  $\{\alpha^\epsilon\}_{k^\epsilon}$ that feature as arguments of the $i$th
and $j$th form factors after the residue evaluation. $k_{ij}^{\epsilon}$
are the cardinalities of these sets satisfying
\begin{eqnarray}
k^{\epsilon}=k_{13}^{\epsilon}+k_{14}^{\epsilon}+k_{23}^{\epsilon}+k_{24}^{\epsilon}\,,\qquad
m_{\alpha}^{\epsilon}  =  k_{13}^{\epsilon}+k_{14}^{\epsilon}\,,\qquad
\tilde{m}_{\alpha}^{\epsilon} =  k_{14}^{\epsilon}+k_{24}^{\epsilon}\,.
\end{eqnarray}
With this partitioning, a generic sum over $k^{\epsilon}$ turns into
\begin{eqnarray}
\sum_{k^{\epsilon}=0}^{\infty}\frac{1}{k^{\epsilon}!}\sum_{{\left\{ I^{\epsilon}\right\} \not\ni I^{0}}}f\left(\left\{ \alpha^{\epsilon}\right\} _{k^{\epsilon}},\dots\right)  & = & \prod_{i=1,2}\prod_{j=3,4}\sum_{k_{ij}^{\epsilon}  =  0}^{\infty}\frac{1}{k_{ij}^{\epsilon}!}\sum_{{\left\{ I_{ij}^{\epsilon}\right\} \not\ni I^{0}}} \nonumber\\
 & & \times f\left(\left\{ \alpha_{13}^{\epsilon}\right\} _{k_{13}^{\epsilon}},\left\{ \alpha_{14}^{\epsilon}\right\} _{k_{14}^{\epsilon}},\left\{ \alpha_{23}^{\epsilon}\right\} _{k_{23}^{\epsilon}},\left\{ \alpha_{24}^{\epsilon}\right\} _{k_{24}^{\epsilon}},\dots\right)\,,
\end{eqnarray}
where also
$
\{I^\epsilon\}_{k^\epsilon}= \left\{I_{13}^{\epsilon}\right\} _{k_{13}^{\epsilon}}\bbc\left\{I_{14}^{\epsilon}\right\} _{k_{14}^{\epsilon}}\bbc\left\{I_{23}^{\epsilon}\right\} _{k_{23}^{\epsilon}}\bbc\left\{ I_{24}^{\epsilon}\right\} _{k_{24}^{\epsilon}}\,.
$
The new form of (\ref{eq:twist_field_action_on_sectors}) after all these manipulations is given by
\begin{eqnarray}
 &  & \mathcal{F}_{p}\left(N_{p}^{\pm},\tilde{N}_{p}^{\pm}\right) 
 =  \Biggl(\prod_{\epsilon=\pm}\prod_{i=1,2}\prod_{j=3,4}\sum_{k_{ij}^{\epsilon}=0}^{\infty}\frac{1}{k_{ij}^{\epsilon}!}\sum_{{\left\{ I_{ij}^{\epsilon}\right\} \not\ni I^{0}}}\sum_{M^{\epsilon}=0}^{\infty}\frac{1}{M^{\epsilon}!}\sum_{s=1,2,d}\sum_{m_{\theta,s}^{\epsilon},\tilde{m}_{\theta,s}^{\epsilon}}' \nonumber\\
 &  & \times\sum_{m_{\leftrightarrows}^{\epsilon}=0}^{\infty}\frac{1}{m_{\leftrightarrows}^{\epsilon}!}\prod_{i=1}^{m_{\leftrightarrows}^{\epsilon}}\int_{\mathcal{C}_{\leftrightarrows}}\frac{\mathrm{d}\beta_{i}^{\epsilon}}{2\pi}\sum_{\tilde{m}_{\leftrightarrows}^{\epsilon}=0}^{\infty}\frac{1}{\tilde{m}_{\leftrightarrows}^{\epsilon}!}\prod_{i=1}^{\tilde{m}_{\leftrightarrows}^{\epsilon}}\int_{\mathcal{C}_{\leftrightarrows}}\frac{\mathrm{d}\tilde{\beta}_{i}^{\epsilon}}{2\pi} \, \mathcal{A}(\{\alpha^\epsilon\}_{k^\epsilon},\{\beta^\epsilon\}_{m_{\leftrightarrows}^{\epsilon}}, \{\tilde{\beta^\epsilon}\}_{\tilde{m}_{\leftrightarrows}^{\epsilon}} )\nonumber\\
 &  & \times\mathcal{B}( \left\{ \alpha_{13}^{\epsilon}\right\} _{k_{13}^{\epsilon}},\left\{ \alpha_{14}^{\epsilon}\right\} _{k_{14}^{\epsilon}},\left\{ \alpha_{23}^{\epsilon}\right\} _{k_{23}^{\epsilon}},\left\{ \alpha_{24}^{\epsilon}\right\} _{k_{24}^{\epsilon}})\, \mathcal{D}(\theta,m_\theta^\epsilon, M^\epsilon, N_p^\epsilon,\tilde{N}_p^\epsilon)\nonumber\\
  &  & \times \mathcal{R}(m_{\theta,1}^{\epsilon},m_{\theta,2}^{\epsilon},m_{\theta,d}^{\epsilon}, m_{\alpha}^{\epsilon},\tilde{m}_{\theta,1}^{\epsilon},\tilde{m}_{\theta,2}^{\epsilon},\tilde{m}_{\theta,d}^{\epsilon}, \tilde{m}_{\alpha}^{\epsilon})\nonumber\\
 &  & \times F_{\mathrm{reg}}^{p,n}\left(\{ \hat{\theta}\} ^{N_{p}^{-}-m_{\theta,1}^{-}-m_{\theta,d}^{-}},\{ \theta\} ^{m_{\theta,2}^{+}},\{ \alpha_{13}^{+}\} _{k_{13}^{+}},\{ \alpha_{14}^{+}\} _{k_{14}^{+}},\{ \beta^{+}\} _{m_{\leftrightarrows}^{+}}\right.\nonumber\\
 &  & \left.\quad\qquad;\{ \hat{\theta}\} ^{N_{p}^{+}-m_{\theta,1}^{+}-m_{\theta,d}^{+}},\{ \theta\} ^{m_{\theta,2}^{-}},\{ \alpha_{13}^{-}\} _{k_{13}^{-}},\{ \alpha_{14}^{-}\} _{k_{14}^{-}},\{ \beta^{-}\} _{m_{\leftrightarrows}^{-}}\right)\nonumber\\
 &  & \times F_{\mathrm{reg}}^{n-p,n}\left(\{\theta\} ^{M^{+}-m_{\theta,2}^{+}-m_{\theta,d}^{+}},\{ \hat{\theta}\} ^{m_{\theta,1}^{-}},\{ \alpha_{23}^{+}\} _{k_{23}^{+}},\{ \alpha_{24}^{+}\} _{k_{24}^{+}},\{ \hat{\beta}^{-}\} _{m_{\leftrightarrows}^{-}}\right.\nonumber\\
 &  & \left.\quad\qquad;\{ \theta\} ^{M^{-}-m_{\theta,2}^{-}-m_{\theta,d}^{-}},\{ \hat{\theta}\} ^{m_{\theta,1}^{+}},\{ \alpha_{23}^{-}\} _{k_{23}^{-}},\{ \alpha_{24}^{-}\} _{k_{24}^{-}},\{ \hat{\beta}^{+}\} _{m_{\leftrightarrows}^{+}}\right)\nonumber\\
 &  & \times F_{\mathrm{reg}}^{p,n}\left(\{ \hat{\theta}\} ^{M^{-}-\tilde{m}_{\theta,1}^{-}-\tilde{m}_{\theta,d}^{-}},\{ \theta\} ^{\tilde{m}_{\theta,2}^{+}},\{ \hat{\alpha}_{13}^{-}\} _{k_{13}^{-}},\{ \hat{\alpha}_{23}^{-}\} _{k_{23}^{-}},\{ \tilde{\beta}^{+}\} _{\tilde{m}_{\leftrightarrows}^{+}}\right.\nonumber\\
 &  & \left.\quad\qquad;\{ \hat{\theta}\} ^{M^{+}-\tilde{m}_{\theta,1}^{+}-\tilde{m}_{\theta,d}^{+}},\{ \theta\} ^{\tilde{m}_{\theta,2}^{-}},\{ \hat{\alpha}_{13}^{+}\} _{k_{13}^{+}},\{ \hat{\alpha}_{23}^{+}\} _{k_{23}^{+}},\{ \tilde{\beta}^{-}\} _{\tilde{m}_{\leftrightarrows}^{-}}\right)\nonumber\\
 &  & \times F_{\mathrm{reg}}^{n-p,n}\left(\{\theta\} ^{\tilde{N}_{p}^{+}-\tilde{m}_{\theta,2}^{+}-\tilde{m}_{\theta,d}^{+}},\{ \hat{\theta}\} ^{\tilde{m}_{\theta,1}^{-}},\{ \hat{\alpha}_{14}^{-}\} _{k_{14}^{-}},\{ \hat{\alpha}_{24}^{-}\} _{k_{24}^{-}},\{ \hat{\tilde{\beta}}^{-}\} _{\tilde{m}_{\leftrightarrows}^{-}}\right.\nonumber\\
 &  & \left.\quad\qquad;\{ \theta\} ^{\tilde{N}_{p}^{-}-\tilde{m}_{\theta,2}^{-}-\tilde{m}_{\theta,d}^{-}},\{ \hat{\theta}\} ^{\tilde{m}_{\theta,1}^{+}},\{ \hat{\alpha}_{14}^{+}\} _{k_{14}^{+}},\{\hat{\alpha}_{24}^{+}\} _{k_{24}^{+}},\{ \hat{\tilde{\beta}}^{+}\} _{\tilde{m}_{\leftrightarrows}^{+}}\right)\,,
 \label{monster}
\end{eqnarray}
with
\beq
 \mathcal{A}(\{\alpha^\epsilon\}_{k^\epsilon},\{\beta^\epsilon\}_{m_{\leftrightarrows}^{\epsilon}}, \{\tilde{\beta^\epsilon}\}_{\tilde{m}_{\leftrightarrows}^{\epsilon}} ):=\frac{e^{ix_{1}\sum_{i=1}^{m_{\leftrightarrows}^{\epsilon}}P\left(\beta_{i}^{\epsilon}\right)}e^{i\left(x_{2}-x_{1}\right)\sum_{i=1}^{k^{\epsilon}}P\left(\alpha_{i}^{\epsilon}\right)}e^{i\left(x_{3}-x_{2}\right)\sum_{i=1}^{\tilde{m}_{\leftrightarrows}^{\epsilon}}P\left(\tilde{\beta}_{i}^{\epsilon}\right)}}{\left[\prod_{i=1}^{m_{\leftrightarrows}^{\epsilon}}h_{1}^{p,\epsilon}\left(\beta_{i}^{\epsilon}\right)\right]\left[\prod_{i=1}^{\tilde{m}_{\leftrightarrows}^{\epsilon}}h_{1}^{p,\epsilon}\left(\tilde{\beta}_{i}^{\epsilon}\right)\right]}\,,
\eeq
\beqa
&&\mathcal{B}( \left\{ \alpha_{13}^{\epsilon}\right\} _{k_{13}^{\epsilon}},\left\{ \alpha_{14}^{\epsilon}\right\} _{k_{14}^{\epsilon}},\left\{ \alpha_{23}^{\epsilon}\right\} _{k_{23}^{\epsilon}},\left\{ \alpha_{24}^{\epsilon}\right\} _{k_{24}^{\epsilon}}):=\nonumber\\
&& \frac{e^{ix_{2}\sum_{i=1}^{k_{13}^{\epsilon}}P\left(\alpha_{13,i}^{\epsilon}\right)}}{\prod_{i=1}^{k_{13}^{\epsilon}}LE\left(\alpha_{13,i}^{\epsilon}\right)}\frac{e^{i\left(x_{2}-x_{1}\right)\sum_{i=1}^{k_{23}^{\epsilon}}P\left(\alpha_{23,i}^{\epsilon}\right)}}{\prod_{i=1}^{k_{23}^{\epsilon}}LE\left(\alpha_{23,i}^{\epsilon}\right)}\frac{e^{ix_{3}\sum_{i=1}^{k_{14}^{\epsilon}}P\left(\alpha_{14,i}^{\epsilon}\right)}}{\prod_{i=1}^{k_{14}^{\epsilon}}LE\left(\alpha_{14,i}^{\epsilon}\right)}\frac{e^{i\left(x_{3}-x_{1}\right)\sum_{i=1}^{k_{24}^{\epsilon}}P\left(\alpha_{24,i}^{\epsilon}\right)}}{\prod_{i=1}^{k_{24}^{\epsilon}}LE\left(\alpha_{24,i}^{\epsilon}\right)}\,,
\eeqa
\beqa
\mathcal{D}(\theta,m_\theta^\epsilon, M^\epsilon, N_p^\epsilon,\tilde{N}_p^\epsilon):= \frac{e^{ix_{1} m_{\theta}^{\epsilon} P\left(\theta\right)}e^{i\left(x_{2}-x_{1}\right)M^{\epsilon}P\left(\theta\right)}e^{i\left(x_{3}-x_{2}\right)\tilde{m}_{\theta}^{\epsilon}P\left(\theta\right)}e^{-ix_{3}\tilde{N}_p^{\epsilon}P\left(\theta\right)}}{\left[LE\left(\theta\right)\right]^{\frac{N_{p}^{\epsilon}}{2}+\frac{\tilde{N}_{p}^{\epsilon}}{2}+M^{\epsilon}}}\,,
\eeqa
and, finally
\beqa
&&\mathcal{R}(m_{\theta,1}^{\epsilon},m_{\theta,2}^{\epsilon},m_{\theta,d}^{\epsilon}, m_{\alpha}^{\epsilon},\tilde{m}_{\theta,1}^{\epsilon},\tilde{m}_{\theta,2}^{\epsilon},\tilde{m}_{\theta,d}^{\epsilon}, \tilde{m}_{\alpha}^{\epsilon}):=\nonumber\\
&&  \tilde{G}^{\epsilon}\left(m_{\theta,1}^{\epsilon},m_{\theta,2}^{\epsilon},m_{\theta,d}^{\epsilon}, m_{\alpha}^{\epsilon},\tilde{m}_{\theta,1}^{\epsilon},\tilde{m}_{\theta,2}^{\epsilon},\tilde{m}_{\theta,d}^{\epsilon}, \tilde{m}_{\alpha}^{\epsilon}\right) \label{RG}\\
&&\times  \left[\mathcal{R}_{\theta,1}^{\epsilon}\right]^{m_{\theta,1}^{\epsilon}}
\left[\mathcal{R}_{\theta,2}^{\epsilon}\right]^{m_{\theta,2}^{\epsilon}}
\left[\mathcal{R}_{\theta,d}^{\epsilon}\right]^{m_{\theta,d}^{\epsilon}}
\left[\mathcal{R}_{\alpha}^{\epsilon}\right]^{m_{\alpha}^{\epsilon}}
\left[\tilde{\mathcal{R}}_{\theta,1}^{\epsilon}\right]^{\tilde{m}_{\theta,1}^{\epsilon}}
\left[\tilde{\mathcal{R}}_{\theta,2}^{\epsilon}\right]^{\tilde{m}_{\theta,2}^{\epsilon}}
\left[\tilde{\mathcal{R}}_{\theta,d}^{\epsilon}\right]^{\tilde{m}_{\theta,d}^{\epsilon}}
\left[\tilde{\mathcal{R}}_{\alpha}^{\epsilon}\right]^{\tilde{m}_{\alpha}^{\epsilon}}\,,\nonumber
\eeqa
where the $``\mathrm{reg}"$ subscript of the form factors indicates they do
not contain any contractions within the sets of $\theta$ rapidities (e.g.~as explained earlier). $\tilde{G}^{\epsilon}$ in (\ref{RG}) contains the combinatorial factors from the summation over contours and from the combinatorics of the residua. The prime over  the summation of $m_{\theta,i}^{\epsilon},\tilde{m}_{\theta,i}^{\epsilon}$ for $i=1,2,d$
indicates certain
constraints due to the limited total number of $\theta$ rapidities that can take part in the residue calculation, namely
\begin{eqnarray}
m_{\theta,d}^{\epsilon}\le\min\left(N_{p}^{\epsilon},M^{\epsilon}\right)\,, & \quad & \tilde{m}_{\theta,d}^{\epsilon}\le\min\left(\tilde{N}_{p}^{\epsilon},M^{\epsilon}\right)\,,\nonumber\\
m_{\theta,1}^{\epsilon}+m_{\theta,d}^{\epsilon}\le N_{p}^{\epsilon}\,, & \quad & \tilde{m}_{\theta,1}^{\epsilon}+\tilde{m}_{\theta,d}^{\epsilon}\le M^{\epsilon}\,,\nonumber\\
m_{\theta,2}^{\epsilon}+m_{\theta,d}^{\epsilon}\le M^{\epsilon}\,, & \quad & \tilde{m}_{\theta,2}^{\epsilon}+\tilde{m}_{\theta,d}^{\epsilon}\le\tilde{N}_{p}^{\epsilon}\,.\label{eq:constraints_m}
\end{eqnarray}
The next step of the calculation is to transform the quantum number
sums for $\left\{ I^{\epsilon}\right\} $ into contour integrals as
well
\begin{eqnarray}
\sum_{I_{i}^{\epsilon}\neq I^{0}}\frac{f\left(\alpha_{i}^{\epsilon},\dots\right)}{LE\left(\alpha_{i}^{\epsilon}\right)} & = & \int_{\mathcal{C}_{0}^{\epsilon}}\frac{\mathrm{d}\alpha_{i}^{\epsilon}}{2\pi}\frac{f\left(\alpha_{i}^{\epsilon},\dots\right)}{e^{ih_{0}^{p,\epsilon}\left(\alpha_{i}^{\epsilon}\right)}-1}\,,
\label{inte}
\end{eqnarray}
where the contour is a combination of small contours 
\begin{eqnarray}
\mathcal{C}_{0}^{\epsilon} & = & \sum_{I\neq I^{0}}\mathcal{C}_{\alpha_{i}^{\epsilon}}
\end{eqnarray}
about the Bethe-Yang solutions (\ref{by2}).
As before, we deform the contour into one encircling the real axis $\tilde{\mathcal{C}}_{\leftrightarrows}$,
however, we need to subtract the residua of poles at $\alpha_{i}^{\epsilon}={\theta}$,
since these poles are not included in the original contour. That is
\begin{eqnarray}
\mathcal{C}_{0}^{\epsilon} & = & \tilde{\mathcal{C}}_{\leftrightarrows}-\mathcal{C}_{\theta}\,.
\end{eqnarray}
It is important to note that $\tilde{\mathcal{C}}_{\leftrightarrows}$
must be chosen such as to run closer to the real axis than the contour $\mathcal{C}_{\leftrightarrows}$,
for the $\left\{\beta^\epsilon\right\}_{m_{\leftrightarrows}^{\epsilon}}$ and $\{\tilde{\beta}^\epsilon\}_{\tilde{m}_{\leftrightarrows}^{\epsilon}}$ rapidities,
to avoid capturing undesired residua. 

The denominator $e^{ih_{0}^{p,\epsilon}\left(\alpha_{i}^{\epsilon}\right)}-1$  of (\ref{inte}) is singular at $\alpha_i^\epsilon=\theta$
and the form factors (which would be part of the function $f(\alpha_i^\epsilon,\ldots)$ in the numerator) can also have kinematic singularities at this point. Since any given $\alpha_{ij}^{\epsilon}$ rapidity appears in two form factors it follows then that we can have first, second, and third order poles. Using the residue formulae in Appendix \ref{Ape}, it is clear that only third order poles can generate contributions that are proportional to $mL$. This is the reason why we will only need to consider third order poles in order to obtain the large-volume leading contribution to the entanglement entropies.
To find the leading contribution, let us denote by $s^\epsilon_{ij}$ the number of third order poles evaluated by rapidities belonging to the set $\{ \alpha_{ij}^{\epsilon}\} _{k_{ij}^{\epsilon}}$.
These numbers are constrained by the number of $\alpha_{ij}^{\epsilon}$ and $\theta$
rapidities. Clearly $s_{ij}^{\epsilon}\leq k_{ij}^{\epsilon}$. In addition, we have the following less trivial constraints
\begin{eqnarray}
s_{13}^{\epsilon}+s_{14}^{\epsilon}\le N_{p}^{\epsilon}-m_{\theta,d}^{\epsilon}-m_{\theta,1}^{\epsilon}\,, & \quad & s_{23}^{\epsilon}+s_{24}^{\epsilon}\le m_{\theta,1}^{\epsilon}\,,\nonumber\\
s_{14}^{\epsilon}+s_{24}^{\epsilon}\le\tilde{N}_{p}^{\epsilon}-\tilde{m}_{\theta,d}^{\epsilon}-\tilde{m}_{\theta,2}^{\epsilon}\,, & \quad & s_{13}^{\epsilon}+s_{23}^{\epsilon}\le\tilde{m}_{\theta,2}^{\epsilon}\,,\nonumber\\
s_{13}^{\epsilon}+s_{14}^{\epsilon}+s_{23}^{\epsilon}+s_{24}^{\epsilon} & \le & \min\left(N_{p}^{\epsilon}-m_{\theta,d}^{\epsilon},\tilde{N}_{p}^{\epsilon}-\tilde{m}_{\theta,d}^{\epsilon}\right)\,.\label{eq:constraints_sij}
\end{eqnarray}
After evaluation of all residua of possible third order poles, the volume dependence of the whole expression is $\left[LE\left(\theta\right)\right]^{\Delta}$
with 
\begin{eqnarray}
\Delta & = & \sum_{\epsilon=\pm}\left(m_{\theta,d}^{\epsilon}+\tilde{m}_{\theta,d}^{\epsilon}+s_{13}^{\epsilon}+s_{14}^{\epsilon}+s_{23}^{\epsilon}+s_{24}^{\epsilon}-\frac{N_{p}^{\epsilon}}{2}-\frac{\tilde{N}_{p}^{\epsilon}}{2}-M^{\epsilon}\right)\,.
\end{eqnarray}
We aim to maximize $\Delta$ to extract the leading large-volume contribution
of the four-point function. We may rearrange the expression as
\begin{eqnarray}
\Delta & = & \sum_{\epsilon=\pm}\left(\frac{s_{13}^{\epsilon}+s_{14}^{\epsilon}+s_{23}^{\epsilon}+s_{24}^{\epsilon}-\left(N_{p}^{\epsilon}-m_{\theta,d}^{\epsilon}\right)}{2}+\frac{m_{\theta,d}^{\epsilon}-M^{\epsilon}}{2}\right. \nonumber\\
 &  & \left.+\frac{s_{13}^{\epsilon}+s_{14}^{\epsilon}+s_{23}^{\epsilon}+s_{24}^{\epsilon}-\left(\tilde{N}_{p}^{\epsilon}-\tilde{m}_{\theta,d}^{\epsilon}\right)}{2}+\frac{\tilde{m}_{\theta,d}^{\epsilon}-M^{\epsilon}}{2}\right)\,,
\end{eqnarray}
 due
to the constraints (\ref{eq:constraints_m}) and (\ref{eq:constraints_sij}) each fraction inside the sum is less or equal than zero.
Therefore, the maximum of $\Delta$ is achieved when all inequalities are saturated, namely
\begin{eqnarray}
\label{eq:maximum_solution}
N_{p}^{\epsilon}=\tilde{N}_{p}^{\epsilon}\,, & \quad & m_{\theta,d}^{\epsilon}=\tilde{m}_{\theta,d}^{\epsilon}=M^{\epsilon}\,,\nonumber\\
m_{\theta,2}^{\epsilon}=\tilde{m}_{\theta,1}^{\epsilon}=0\,, & \quad & s_{13}^{\epsilon}=\tilde{m}_{\theta,2}^{\epsilon}-s_{23}^{\epsilon}\,,\nonumber\\
s_{24}^{\epsilon}=m_{\theta,1}^{\epsilon}-s_{23}^{\epsilon}\,, & \quad & s_{14}^{\epsilon}=N_{p}^{\epsilon}-M^{\epsilon}-m_{\theta,1}^{\epsilon}-\tilde{m}_{\theta,2}^{\epsilon}+s_{23}^{\epsilon}\,,
\end{eqnarray}
where $M^{\epsilon},m_{\theta,1}^{\epsilon},\tilde{m}_{\theta,2}^{\epsilon}$
and $s_{23}^{\epsilon}$ are still free parameters within the range
\begin{eqnarray}
\!\!\!\!\!\!\!0\le M^{\epsilon}\le N_{p}^{\epsilon}\,, & \quad & 0\le m_{\theta,1}^{\epsilon}\le N_{p}^{\epsilon}-M^{\epsilon}\,,\nonumber\\
\!\!\!\!\!\!\!0\le\tilde{m}_{\theta,2}^{\epsilon}\le N_{p}^{\epsilon}-M^{\epsilon}\,, & \quad & \max(M^{\epsilon}+m_{\theta,1}^{\epsilon}+\tilde{m}_{\theta,2}^{\epsilon}-N_{p}^{\epsilon},0)\le s_{23}^{\epsilon}\le\min\left(m_{\theta,1}^{\epsilon},\tilde{m}_{\theta,2}^{\epsilon}\right)\,.
\label{indep}
\end{eqnarray}
The maximum power of the volume is  then $\Delta=0$. This corresponds to the situation when all dependency on the $\theta$ rapidities has been cancelled by the evaluation of residua. 
Hence there are no other poles to consider other than third order ones (as these will produce subleading contributions).
Using the results of Appendix \ref{Ape}, the residua from third order poles are (up to factors depending on rapidities other than $\theta$)
\begin{eqnarray}
\mathcal{R}_{\alpha_{13}}^{\epsilon} & = & LE\left(\theta\right) \tilde{g}\left(r_{1}+r_{2}\right)\left[1-e^{-2\pi i\frac{\epsilon p}{n}}\right]^{2}\,,\nonumber\\
\mathcal{R}_{\alpha_{14}}^{\epsilon} & = & LE\left(\theta\right) \tilde{g}\left(r_{1}+r_{2}+r_{3}\right)\left[1-e^{2\pi i\frac{\epsilon p}{n}}\right]\left[1-e^{-2\pi i\frac{\epsilon p}{n}}\right]\,,\nonumber\\
\mathcal{R}_{\alpha_{23}}^{\epsilon} & = & LE\left(\theta\right) \tilde{g}\left(r_{2}\right)\left[1-e^{2\pi i\frac{\epsilon p}{n}}\right]\left[1-e^{-2\pi i\frac{\epsilon p}{n}}\right]\,\nonumber,\\
\mathcal{R}_{\alpha_{24}}^{\epsilon}  & = & LE\left(\theta\right) \tilde{g}\left(r_{2}+r_{3}\right)\left[1-e^{2\pi i\frac{\epsilon p}{n}}\right]^{2}\,, \label{residuos}
\end{eqnarray}
where
\begin{eqnarray}
\tilde{g}\left(r\right)  =  \frac{6r\left(r-1\right)+1}{12}\,,
\label{gtilde}
\end{eqnarray}
and $r_{2}=\frac{x_{2}-x_{1}}{L}=\frac{\ell_{2}}{L}$.  The origin of the function (\ref{gtilde}) is explained in Appendix D. 
After relabelling $k_{ij}^\epsilon -s_{ij}^\epsilon\to k_{ij}^\epsilon$ and using \eqref{eq:maximum_solution}, the combinatorial factor arising from the evaluation of the residua,
 that is from counting the number of ways of picking the rapidities for the residua and the different contractions, has the simple form 
\begin{equation}
\frac{\left(N^\epsilon_p !\right)^2}{s_{13}^\epsilon! s_{14}^\epsilon! s_{23}^\epsilon! s_{34}^\epsilon!}\,.
\end{equation}
Examining the remaining integrals and form factors (after extracting the residua and their combinatorics), one realizes what is left is exactly the vacuum four-point function, that is
\begin{eqnarray}
\frac{\lim\limits_{L\to\infty}\mathcal{F}_{p}\left(N_{p}^{\pm},{N}_{p}^{\pm}\right) }{\,_{p;L}\left\langle 0\right|\mathcal{T}_{p}\left(0\right)\tilde{\mathcal{T}}_{p}\left(x_{1}\right)\mathcal{T}_{p}\left(x_{2}\right)\tilde{\mathcal{T}}_{p}\left(x_{3}\right)\left|0\right\rangle _{p;L}}= \prod_{\epsilon=\pm}N_p^{\epsilon}!\,\Omega_{\epsilon p}^{n}\left(N^{\epsilon}_p,r_{1},r_{2},r_{3}\right)+\mathcal{O}\left(L^{-1}\right)\,,
\end{eqnarray}
with 
\begin{eqnarray}
 &  & \Omega_{\epsilon p}^{n}\left(N,r_{1},r_{2},r_{3}\right)\nonumber\\
 & = & N!\sum_{M=0}^{N}\frac{\left(-1\right)^{N-M}}{M!}\left[1-e^{2\pi i\frac{\epsilon p}{n}}\right]^{2\left(N-M\right)}\left[g_{\epsilon p}^{n}\left(r_{1}\right)g_{\epsilon p}^{n}\left(r_{3}\right)\right]^{M}\left[\tilde{g}\left(r_{1}+r_{2}+r_{3}\right)\right]^{N-M}\nonumber\\
 &  & \times\sum_{m=0}^{N-m}\left(-1\right)^{m}\left[\frac{\tilde{g}\left(r_{2}+r_{3}\right)}{\tilde{g}\left(r_{1}+r_{2}+r_{3}\right)}\right]^{m}
 \sum_{\tilde{m}=0}^{N-m}\left(-1\right)^{\tilde{m}}\left[\frac{\tilde{g}\left(r_{1}+r_{2}\right)}{\tilde{g}\left(r_{1}+r_{2}+r_{3}\right)}\right]^{\tilde{m}}\nonumber\\
 &  & \times\sum_{s=\max(M-N+m+\tilde{m},0)}^{\min\left(m,\tilde{m}\right)}\frac{\left[\tilde{g}\left(r_{1}+r_{2}+r_{3}\right)\tilde{g}\left(r_{2}\right)\right]^{s}\left[\tilde{g}\left(r_{1}+r_{2}\right)\tilde{g}\left(r_{2}+r_{3}\right)\right]^{-s}}{s!\left(m-s\right)!\left(\tilde{m}-s\right)!\left(N-M-m-\tilde{m}+s\right)!}\,.
\end{eqnarray}
Here the number of sums has been greatly reduced thanks to the constraints (\ref{eq:maximum_solution}). The four remaining sums correspond to the four independent variables in (\ref{indep})
which can been relabelled for convenience ($s_{13} \mapsto s$, $M^\epsilon \mapsto M, m_{\theta,1}^\epsilon \mapsto m, \tilde{m}_{\theta,1}^\epsilon \mapsto \tilde{m}$ and $N_p^\epsilon \mapsto N$).

It is easy to show that the sums in $m,\tilde{m}$ and $s$ are nothing but the multinomial expansion of $\frac{1}{\left(N-M\right)!}\left[1-\frak{A}-\frak{B}+\frak{ABC}\right]^{N-M}$ 
with 
\begin{eqnarray}
\frak{A}  = \frac{\tilde{g}\left(r_{2}+r_{3}\right)}{\tilde{g}\left(r_{1}+r_{2}+r_{3}\right)}\,,\qquad
\frak{B} =  \frac{\tilde{g}\left(r_{1}+r_{2}\right)}{\tilde{g}\left(r_{1}+r_{2}+r_{3}\right)}\,,\qquad
\frak{C} =  \frac{\tilde{g}\left(r_{1}+r_{2}+r_{3}\right)\tilde{g}\left(r_{2}\right)}{\tilde{g}\left(r_{1}+r_{2}\right)\tilde{g}\left(r_{2}+r_{3}\right)}\,,
\end{eqnarray}
The last three sums are then simply
\begin{eqnarray}
\frac{1}{\left(N-M\right)!}\left[1-\frak{A}-\frak{B}+\frak{ABC}\right]^{N-M}  =  \frac{1}{\left(N-M\right)!}\left[\frac{r_{1}r_{3}}{\tilde{g}\left(r_{1}+r_{2}+r_{3}\right)}\right]^{N-M}\,,
\label{multinomial}
\end{eqnarray}
and 
\begin{eqnarray}
\Omega_{\epsilon p}^{n}\left(N,r_{1},r_{2},r_{3}\right) =  \left[g_{\epsilon p}^{n}\left(r_{1}\right)g_{\epsilon p}^{n}\left(r_{3}\right)-\left[1-e^{2\pi i\frac{\epsilon p}{n}}\right]^{2}r_{1}r_{3}\right]^{N}=\left[g_{\epsilon p}^{n}\left(r_{1}+r_{3}\right)\right]^{N}\,.
\end{eqnarray}
Note that, from (\ref{multinomial}) we see that the dependence on the parameter $r_2$ drops out. Thus the results do not depend on the distance between regions $A$ and $B$ as long as $0\leq r_2 \leq 1$  in the scaling limit. This leads to the following final result for the four-point function 
\begin{eqnarray}
\!\!\!\!\lim_{L\rightarrow \infty}  \frac{\,_{L}\left\langle 1\right|\mathcal{T}\left(0\right)\tilde{\mathcal{T}}\left(x_{1}\right)\mathcal{T}\left(x_{2}\right)\tilde{\mathcal{T}}\left(x_{3}\right)\left|1\right\rangle _{L}}{\,_{L}\left\langle 0\right|\mathcal{T}\left(0\right)\tilde{\mathcal{T}}\left(x_{1}\right)\mathcal{T}\left(x_{2}\right)\tilde{\mathcal{T}}\left(x_{3}\right)\left|0\right\rangle _{L}}
 =  \sum_{\left\{ N^{\pm}\right\} }\left|C_{n}\left(\{ N^{\pm}\} \right)\right|^2\prod_{p=1}^{n}\prod_{\epsilon=\pm}N_{p}^{\epsilon}!\left[g_{\epsilon p}^{n}\left(r_{1}+r_{3}\right)\right]^{N_p^\epsilon}\,.
 \label{Cn1}
\end{eqnarray}
Recall that
this is exactly the same result as for the single region entanglement
entropy \cite{the4ofus2} with region length $r_1+r_3$. Therefore, as observed earlier for the qubit states, the increment of the entanglement entropy due to the presence of one excitation
is independent of the connectivity of the region under consideration. This can be shown more generally both with qubits (see Appendix C) and using form factors. 


\subsection{Replica Logarithmic Negativity}
\label{sec:LogNeg_ff}

To calculate the leading order large-volume contribution to \eqref{eq:LogNeg_cont_sector_p} we have to go through exactly same steps as presented in the previous section. For the logarithmic negativity the last two form factors are ``exchanged" hence the quantization condition in the complete set of states inserted between the last two fields is now different. This only affects the values of some residua. The changed residua are
\begin{eqnarray}
\tilde{\mathcal{R}}_{\theta,2}^{\epsilon}&=&e^{-2\pi i\frac{\epsilon p}{n}}\,,\nonumber \\
\tilde{\mathcal{R}}_{\theta,d}^{\epsilon}&=&LE\left(\theta\right)g_{-\epsilon p}^{n}\left(r_{3}\right)\,,\nonumber\\
\mathcal{R}_{\alpha_{13}}^{\epsilon} & = & LE\left(\theta\right) \tilde{g}\left(r_{1}+r_{2}\right)\left[1-e^{2\pi i\frac{\epsilon p}{n}}\right]\left[1-e^{-2\pi i\frac{\epsilon p}{n}}\right]\,,\nonumber\\
\mathcal{R}_{\alpha_{14}}^{\epsilon} & = & LE\left(\theta\right) \tilde{g}\left(r_{1}+r_{2}+r_{3}\right)\left[1-e^{-2\pi i\frac{\epsilon p}{n}}\right]^2\,,\nonumber\\
\mathcal{R}_{\alpha_{23}}^{\epsilon} & = & LE\left(\theta\right) \tilde{g}\left(r_{2}\right)\left[1-e^{2\pi i\frac{\epsilon p}{n}}\right]^2\,,\nonumber\\
\mathcal{R}_{\alpha_{24}}^{\epsilon}  & = & LE\left(\theta\right) \tilde{g}\left(r_{2}+r_{3}\right)\left[1-e^{2\pi i\frac{\epsilon p}{n}}\right]\left[1-e^{-2\pi i\frac{\epsilon p}{n}}\right]\,.
\end{eqnarray}
The final result for \eqref{eq:LogNeg_cont_sector_p} is
\begin{eqnarray}
\frac{\lim_{L\to\infty}\mathcal{\tilde{F}}_{p}\left(N_{p}^{\pm},{N}_{p}^{\pm}\right)}{\,_{p;L}\left\langle 0\right|\mathcal{T}_{p}\left(0\right)\tilde{\mathcal{T}}_{p}\left(x_{1}\right)\tilde{\mathcal{T}}_{p}\left(x_{2}\right)\mathcal{T}_{p}\left(x_{3}\right)\left|0\right\rangle _{p;L} }=  
 \left(\prod_{\epsilon=\pm}N_p^{\epsilon}!\,\tilde{\Omega}_{\epsilon p}^{n}\left(N_p^{\epsilon},r_{1},r_{2},r_{3}\right)\right)+\mathcal{O}\left(L^{-1}\right)\,,
\end{eqnarray}
where
\begin{eqnarray}
\tilde{\Omega}_{\epsilon p}^{n}\left(N,r_{1},r_{2},r_{3}\right) & = & \left[g_{\epsilon p}^{n}\left(r_{1}\right)g_{-\epsilon p}^{n}\left(r_{3}\right)-\left[1-e^{2\pi i\frac{\epsilon p}{n}}\right]\left[1-e^{-2\pi i\frac{\epsilon p}{n}}\right]r_{1}r_{3}\right]^{N} \nonumber\\
&=&\left[\hat{g}_{\epsilon p}^{n}\left(r_{1},r_{3}\right)\right]^{N}\,,
\end{eqnarray}
with the function 
\begin{equation}
\hat{g}_{\epsilon p}^{n}\left(r_{1},r_{3}\right)= 1-r_1 -r_3+r_1 e^{2\pi i\frac{\epsilon p}{n}}+r_3 e^{-2\pi i\frac{\epsilon p}{n}}\,.
\end{equation}
As for the R\'enyi entropies we find that  the function $\tilde{\Omega}_{\epsilon p}^{n}\left(N,r_{1},r_{2},r_{3}\right)$ does not ultimately depend on the parameter $r_2$.
The final result for the logarithmic negativity for the one-particle state becomes
\begin{eqnarray}
\lim_{L \rightarrow \infty}  \frac{\,_{L}\left\langle 1\right|\mathcal{T}\left(0\right)\tilde{\mathcal{T}}\left(x_{1}\right)\tilde{\mathcal{T}}_{p}\left(x_{2}\right)\mathcal{T}_{p}\left(x_{3}\right)\left|1\right\rangle _{L}}{\,_{L}\left\langle 0\right|\mathcal{T}\left(0\right)\tilde{\mathcal{T}}\left(x_{1}\right)\tilde{\mathcal{T}}_{p}\left(x_{2}\right)\mathcal{T}_{p}\left(x_{3}\right)\left|0\right\rangle _{L}}
  = \sum_{\left\{ N^{\pm}\right\} }\left|C_{n}\left(\{ N^{\pm}\} \right)\right|^{2}\prod_{p=1}^{n}\prod_{\epsilon=\pm}N_{p}^{\epsilon}!\left[\hat{g}_{\epsilon p}^{n}\left(r_{1},r_{3}\right)\right]^{N_{p}^{\epsilon}}\,.
  \label{Cn2}
\end{eqnarray}


\subsection{Multi-Particle States}
\label{sec:multi_part}

Following the notations introduced in \cite{the4ofus2}, a general multi-particle state in the $U(1)$ basis has the form
\beq
\left|k_1,\dots,k_m\right\rangle_L= \left(\prod_{q=1}^m \sum_{\left\{N^{q,\pm}\right\}} \frac{D^{k_q}_n\left(\left\{ N^{q,\pm}\right\}\right)}{\sqrt{k_q!}^n}  \prod_{p=1}^n\left[\mathfrak{a}_p^\dagger (\theta_q)\right]^{N^{q,+}_{p}} \left[\mathfrak{b}_p^\dagger (\theta_q)\right]^{N^{q,-}_{p}}  \right)\left| 0 \right\rangle_L\,, \label{eq:general_state}
\eeq
where $k_q$ is the multiplicity of rapidity $\theta_q$ in the multi-particle state. Each rapidity $\theta_q$ with $q=1,\ldots, m$ is a solution of the Bethe-Yang equation \eqref{by2} with  quantum number $I_{0,q}$. All such quantum numbers are distinct for different values of $q$.  $D_n^{k_q}\left(\left\{ N^{q,\pm}\right\}\right)$ are the numerical coefficients involved in expressing a $k_q$-particle state with coinciding rapidities in the $U(1)$ basis (they are related to the $C_n$ coefficients in (\ref{Cn1})-(\ref{Cn2}) and shown in Eq.~(4.43) of \cite{the4ofus2}). The integer sums obey the selection rules
\beq
\sum_{\epsilon=\pm}\sum_{p=1}^n N^{q,\epsilon}_p=n k_q\,, \quad \forall q\,. \label{eq:integer_selection_rule}
\eeq
The four-point functions factorize into the product of $U(1)$ sector contributions as before in \eqref{419} and \eqref{420}, but now they depend on all the rapidities 
\begin{eqnarray}
\mathcal{F}_{p}\left( \{ N_{p}^{\pm} \},\{\tilde{N}_{p}^{\pm}\}\right)  & =&  \,_{p;L}\left\langle 0\right|\left(\prod_{q=1}^m\left[\mathfrak{a}_{p}\left(\theta_q\right)\right]^{N_{p}^{q,+}}\left[\mathfrak{b}_{p}\left(\theta_q\right)\right]^{N_{p}^{q,-}}\right)\mathcal{T}_{p}\left(0\right)\,\mathbf{I}_{1}^{p}\,\tilde{\mathcal{T}}_{p}\left(x_{1}\right)\,\mathbf{I}_{0}^{p}  \nonumber\\
 &  & \times\mathcal{T}_{p}\left(x_{2}\right)\,\mathbf{I}_{1}^{p}\,\tilde{\mathcal{T}}_{p}\left(x_{3}\right)\left(\prod_{q=1}^m \left[\mathfrak{a}_{p}^{\dagger}\left(\theta_q\right)\right]^{\tilde{N}_{p}^{q,+}}\left[\mathfrak{b}_{p}^{\dagger}\left(\theta_q\right)\right]^{\tilde{N}_{p}^{q,-}} \right) \left|0\right\rangle _{p;L}\,,
 \eeqa
 and
 \beqa
\tilde{\mathcal{F}}_{p}\left(\{ N_{p}^{\pm} \},\{\tilde{N}_{p}^{\pm}\}\right) & = & \,_{p;L}\left\langle 0\right|\left(\prod_{q=1}^m\left[\mathfrak{a}_{p}\left(\theta_q\right)\right]^{N_{p}^{q,+}}\left[\mathfrak{b}_{p}\left(\theta_q\right)\right]^{N_{p}^{q,-}}\right)\mathcal{T}_{p}\left(0\right)\,\mathbf{I}_{1}^{p}\,\tilde{\mathcal{T}}_{p}\left(x_{1}\right)\,\mathbf{I}_{0}^{p} \nonumber\\
 &  & \times\tilde{\mathcal{T}}_{p}\left(x_{2}\right)\,\mathbf{I}_{-1}^{p}\,\mathcal{T}_{p}\left(x_{3}\right) \left(\prod_{q=1}^m \left[\mathfrak{a}_{p}^{\dagger}\left(\theta_q\right)\right]^{\tilde{N}_{p}^{q,+}}\left[\mathfrak{b}_{p}^{\dagger}\left(\theta_q\right)\right]^{\tilde{N}_{p}^{q,-}} \right)\left|0\right\rangle _{p;L}\,.
\end{eqnarray}
It is straightforward to generalize the calculation for these sector contributions along the lines of sections \ref{sec:EE_2regions_ff} and \ref{sec:LogNeg_ff} and to extract the leading contributions
\begin{eqnarray}
\lim_{L \rightarrow \infty}  \frac{\mathcal{F}_{p}\left(\{ N_{p}^{\pm} \},\{N_{p}^{\pm}\}\right)}{\,_{p;L}\left\langle 0\right|\mathcal{T}_{p}\left(0\right)\tilde{\mathcal{T}}_{p}\left(x_{1}\right)\mathcal{T}_{p}\left(x_{2}\right)\tilde{\mathcal{T}}_{p}\left(x_{3}\right)\left|0\right\rangle _{p;L} }  & = & \prod_{q=1}^m \prod_{\epsilon=\pm} N_p^{q,\epsilon}! \left[ g^n_{\epsilon p}(r_1+r_3)\right]^{N_p^{q,\epsilon}} \,, \label{r1pr3}\\
\lim_{L \rightarrow \infty}  \frac{\tilde{\mathcal{F}}_{p}\left( \{ N_{p}^{\pm}\},\{N_{p}^{\pm}\}\right)}{\,_{p;L}\left\langle 0\right|\mathcal{T}_{p}\left(0\right)\tilde{\mathcal{T}}_{p}\left(x_{1}\right)\tilde{\mathcal{T}}_{p}\left(x_{2}\right)\mathcal{T}_{p}\left(x_{3}\right)\left|0\right\rangle _{p;L} } & = & \prod_{q=1}^m \prod_{\epsilon=\pm} N_p^{q,\epsilon}! \left[ \hat{g}^n_{\epsilon p}(r_1,r_3)\right]^{N_p^{q,\epsilon}} \,.
\end{eqnarray}
Due to the product form of the sector contributions, the form of the general multi-particle state \eqref{eq:general_state} and the logarithm in the definition \eqref{3} and \eqref{4}, the increment of the entanglement entropy with two separate regions and of the replica negativity of a multi-particle state take the form
\beqa
\Delta S_n^{k_1,\dots,k_m} (r)&=&\sum_{q=1}^m \Delta S_n^{k_q}(r)\,, \\
\Delta \mathcal{E}_n^{k_1,\dots,k_m}(r_1,r_3)&=&\sum_{q=1}^m \Delta \mathcal{E}_n^{k_q}(r_1,r_3)\,, 
\eeqa
where
\beqa
\Delta S_n^{k}(r)&=&\frac{1}{1-n}\log\left(\sum_{\left\{N^{\pm}\right\}} \frac{\left|D^{k}_n\left(\left\{ N^{\pm}\right\}\right)\right|^2}{(k!)^n} \prod_{p=1}^n\prod_{\epsilon=\pm}N_p^{\epsilon}! \left[ g^n_{\epsilon p}(r)\right]^{N_p^{\epsilon}}\right)\,, \\
\Delta \mathcal{E}_n^{k}(r_1,r_3)&=&\log\left(\sum_{\left\{N^{\pm}\right\}} \frac{\left|D^{k}_n\left(\left\{ N^{\pm}\right\}\right)\right|^2}{(k!)^n} \prod_{p=1}^n\prod_{\epsilon=\pm}N_p^{\epsilon}! \left[ \hat{g}^n_{\epsilon p}(r_1,r_3)\right]^{N_p^{\epsilon}}\right)\,, 
\eeqa
are the increments for a $k$-particle state of identical particles and the integer sums still obey \eqref{eq:integer_selection_rule}. Recall that $r_1+r_3=1-r$ and (\ref{r1pr3}) is invariant under $r \mapsto 1-r$. 
We cross-checked these final results against those obtained from the qubit picture for several multi-particle states and found perfect agreement. 


\section{Numerical Results}

In this section we present some numerical results for the R\'enyi entropies and replica logarithmic negativities of a harmonic chain. Recall that this is a discrete theory whose continuum limit is the massive free boson. Therefore we expect that for appropriate parameter choices we should find agreement with our predictions of the previous sections. The dispersion relation is 
\beq
E(p)=\sqrt{m^2+4 \Delta x^{-2}\sin^2 \frac{\Delta x\, p}{2}}\,,
\eeq
where $\Delta x$ is the lattice spacing, $m$ is the mass and $p$ is the momentum of the excitation. The QFT regime corresponds to $\Delta x \cdot p \ll 1$ where we recover the relativistic dispersion relation. The numerical procedure was explained in detail in Appendix A of \cite{the4ofus2} and is based on exact diagonalization. In all the plots we have chosen:
\beq
\Delta x=0.02, \quad m=1, \quad \mathrm{and} \quad  L=20\,.
\eeq
As explained in \cite{the4ofus1}, the analytical predictions are expected to be valid in the ``quasiparticle regime", where the lengths of all connected regions (the regions $A$ and $B$, and the connected components of $C$) are large as compared to either the correlation length $m^{-1}$, or the De Broglie wavelengths $\frac{2\pi}{p}$ of all excitations (where $p$ is the momentum). Below,  all momenta are chosen to be well within the QFT regime.
\begin{figure}[h!]
\begin{center} 
    \includegraphics[width=7.5cm]{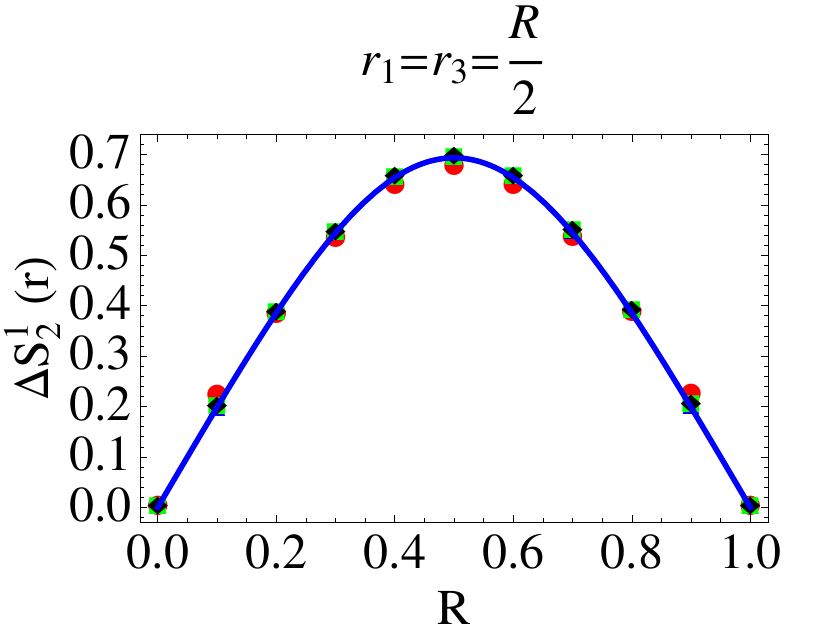} 
        \includegraphics[width=7.5cm]{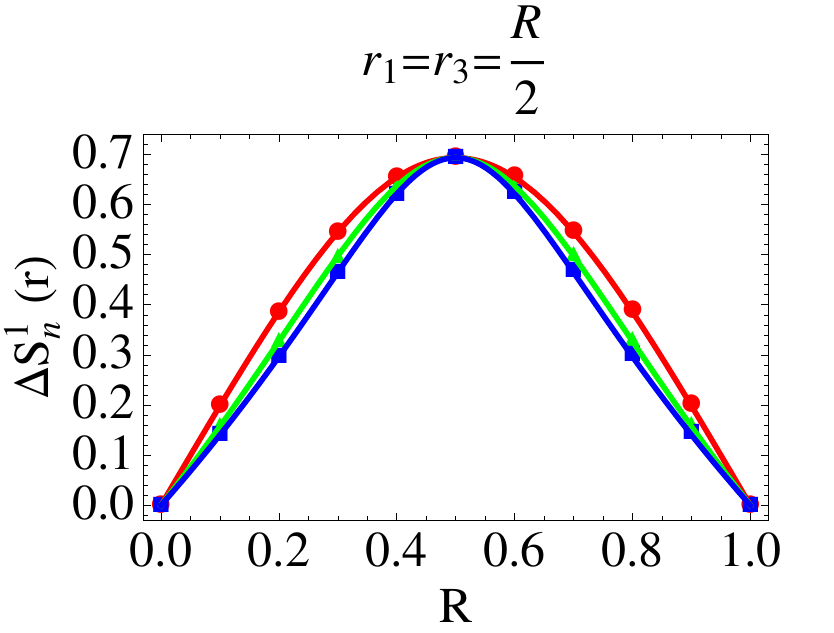} 
      \end{center} 
 \caption{Left: Numerical values of the 2nd R\'enyi entropy for  $r_1=r_3=\frac{R}{2}$ as a function of $r=1-R$ and a single excitation. These are given for four different choices of the momentum of the excitation: $p=0$ (circles), $p=\pi$ (triangles) and $p=2\pi$ (squares) and $4\pi$ (rombi).  Right: Numerical values of the 2nd (circles), 3rd (triangles) and 4th (squares) R\'enyi entropies for $p=4\pi$ (single excitation) compared to analytic formula (continuous curves). } 
 \label{cuspy} 
  \end{figure}
\subsection{R\'enyi Entropies}
We have evaluated numerically the R\'enyi entropies for a subsystem consisting of two disconnected regions of the same length $r_1=r_3$. The subsystem is $A\bbc B$ with $A = [0,RL/2]$ and $B=[L/2,L/2+RL/2]$, and the values of $R$ are taken from $0$ to $1$ by increments of $0.1$. The total subsystem's length is therefore $\ell_1+\ell_3 = RL$, ranging from $0$ to $L$.
\begin{figure}[h!]
\begin{center} 
    \includegraphics[width=7.5cm]{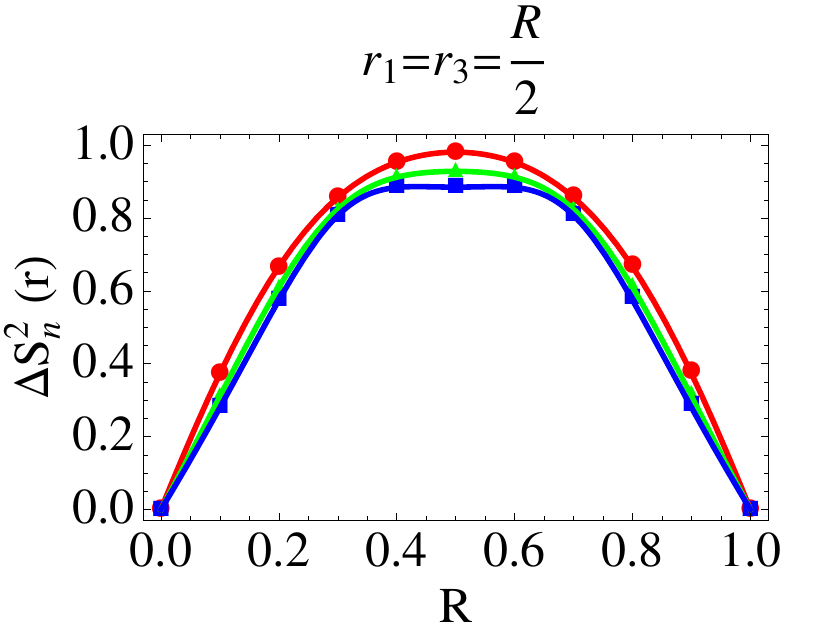} 
     \includegraphics[width=7.5cm]{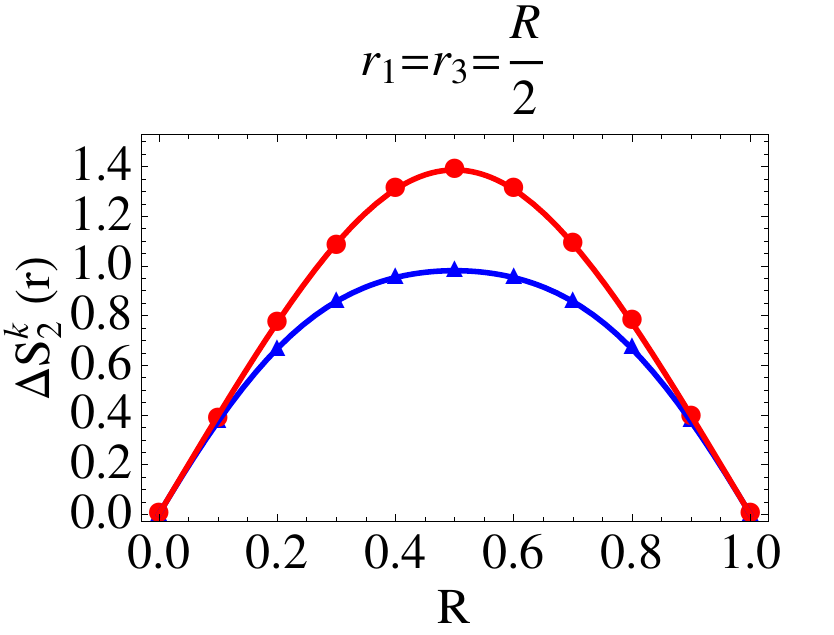} 
      \end{center} 
 \caption{Left: Numerical values of the 2nd (circles), 3rd (triangles) and 4th (squares) R\'enyi entropies for a state of two identical excitations with momenta for $p_1=p_2=4\pi$. Right: Numerical values of the 2nd R\'enyi entropy of a state of two identical excitations (circles) and two distinct excitation (triangles). The momenta are $p_1=p_2=4\pi$ for identical excitations ($k=2$) and $p_1=3\pi$, $p_2=4\pi$ for two distinct excitations. In all cases agreement with the analytic results (continuous curves) is excellent.} 
 \label{cuspy2} 
  \end{figure}

In Fig.~\ref{cuspy} we present results for the R\'enyi entropies of one excitation. The figure on the left explores the dependence on the momentum of the excitation; this dependence is shown to be very small, and the numerical results to agree very well with the analytical prediction. At $p=0$ the disagreement is slightly larger because in this case only the correlation length $m^{-1}=1$ controls the accuracy. The relative disagreement is however higher for the small value $R=0.1$ (omitting the trivial value $R=0$) than it is at $R=0.5$, where correlation length effects should be smallest. The entanglement is higher than predicted at the small value $R=0.1$, where the regions $A$ and $B$ are small, presumably because one observes additional entanglement due to the correlations within distances comparable to the correlation length; while it is smaller at $R=0.5$, presumably because one looses entanglement due to correlations between the regions $A$ and $B$ which are now adjacent. We observe that the numerical results for $p=\pi$ (which is equal to $10\cdot 2\pi/L$, giving a De Broglie's wavelength of $2$), $2\pi$ (De Broglie's wavelength of $1$) and $4\pi$ (De Broglie's wavelength of $0.5$) are virtually indistinguishable from each other and from the analytic curve (\ref{ren}) here presented as the solid curve. The figure on the right in Fig.~\ref{cuspy} shows very good agreement with the analytic predictions for three values of $n$ and relatively high momentum (De Broglie's wavelength of $0.5$). Similarly good agreement is found for states consisting of two identical (or not) excitations, whose R\'enyi entropies are presented in Fig.~\ref{cuspy2}.
  \subsection{Replica Logarithmic Negativity}
  We have evaluated numerically the replica logarithmic negativity between regions $A$ and $B$ with various ratios of lengths. We have chosen $A = [0, fRL]$ and $B=[fL,fL+(1-f)RL]$, where the ratio of lengths $|A|/|B| = f/(1-f)$ is controlled by the parameter $f$ taking values $1/2$, $2/5$ and $1/3$, giving $r_1=r_3=R/2$ ($f=1/2$), $r_1=2R/5,\,r_3=3R/5$ ($f=2/5$), and $r_1=R/3,\,r_3=2R/3$ ($f=1/3$) respectively. The overall scale is again $R$, ranging from $0$ to $1$ by increments of $0.1$. 
   \begin{figure}[h!]
\begin{center} 
    \includegraphics[width=7.5cm]{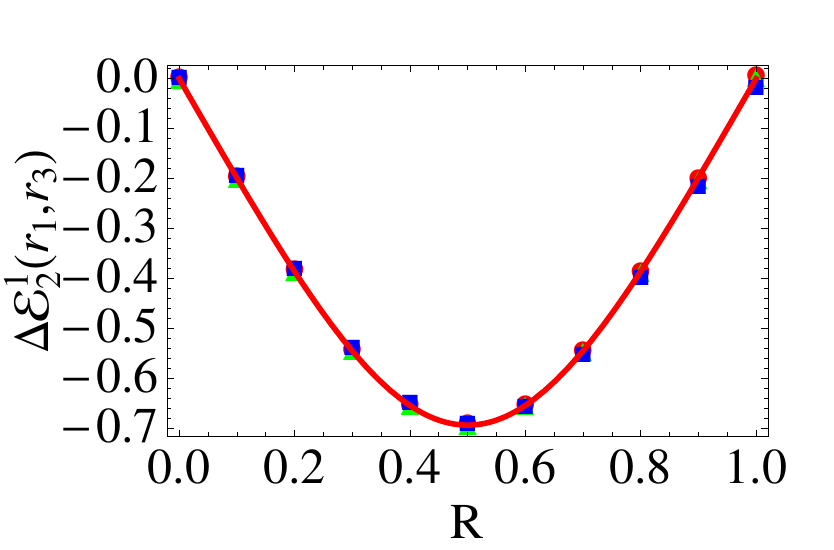} 
      \includegraphics[width=7.5cm]{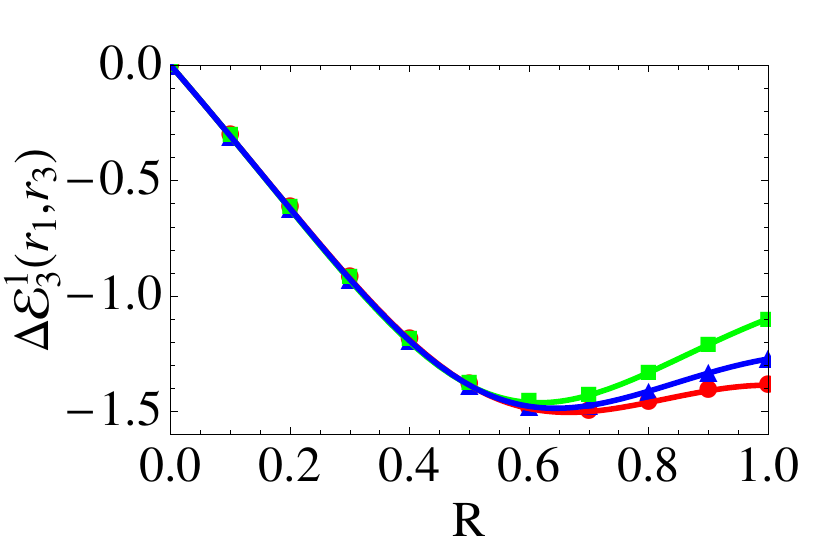}
      \end{center} 
 \caption{Results for $k=1$ and $p=4{\pi}$ with $n=2$ (left) and $n=3$ (right). For each case we present three curves corresponding to $r_1=r_3=\frac{R}{2}$ (circles), $r_1=\frac{2R}{5}$ and $r_3=\frac{3R}{5}$ (triangles), and $r_1=\frac{R}{3}$, $r_3=\frac{2R}{3}$ (squares). For $n=2$ all three sets of data are identical, since in this case the negativity only depends on the sum $r_1+r_3=R$ and equals minus the 2nd R\'enyi entropy.
 The symbols are numerical data and the curves are obtained  from the formula (\ref{de}).} 
 \label{negnr1r3dif1} 
  \end{figure}
 \begin{figure}[h!]
\begin{center} 
         \includegraphics[width=7.5cm]{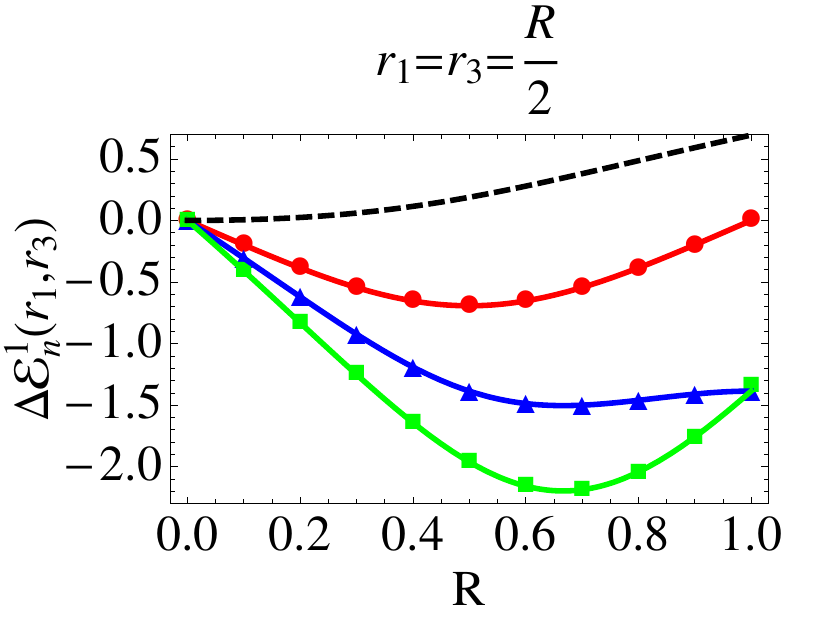}
           \includegraphics[width=7.5cm]{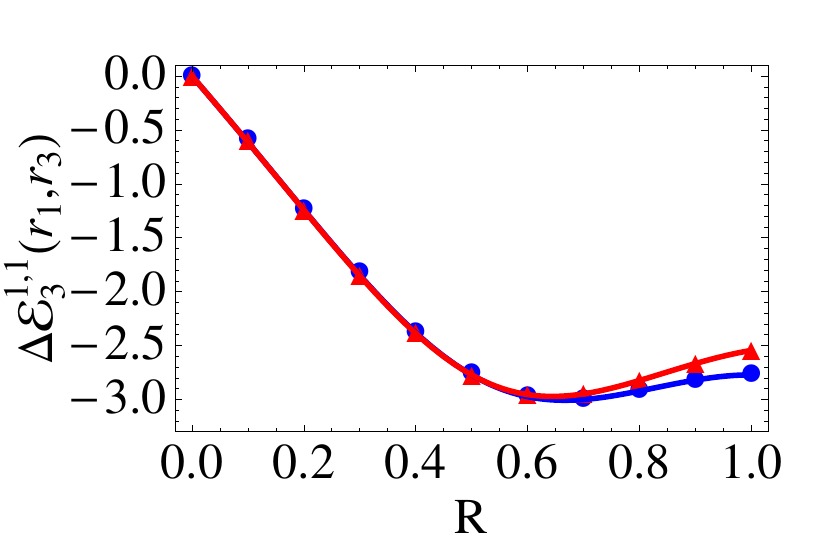}
      \end{center} 
 \caption{Left: Results for $k=1$ and $p=4{\pi}$ with $n=2$ (dots), $n=3$ (triangles), and $n=4$ (squares). The dashed curve is the logarithmic negativity, showing once again the gradual change in curvature as $n\rightarrow 1$. The symbols are numerical data and the solid curves are obtained  from the formula (\ref{de}). The dashed curve is obtained from (\ref{ep1}). Right: The replica logarithmic negativity for $n=3$ and a state of two distinct excitations of momenta $p_1=3{\pi}$ and $p_2=4{\pi}$. The circles give data for $r_1=r_3=\frac{R}{2}$ and the triangles correspond to $r_1=\frac{2R}{5}$ and $r_3=\frac{3R}{5}$. Agreement with the analytic curves is again excellent. These correspond to twice the replica logarithmic negativity of a single excitation. } 
 \label{negnr1r3dif2} 
  \end{figure}

\begin{figure}[h!]
\begin{center} 
         \includegraphics[width=7.5cm]{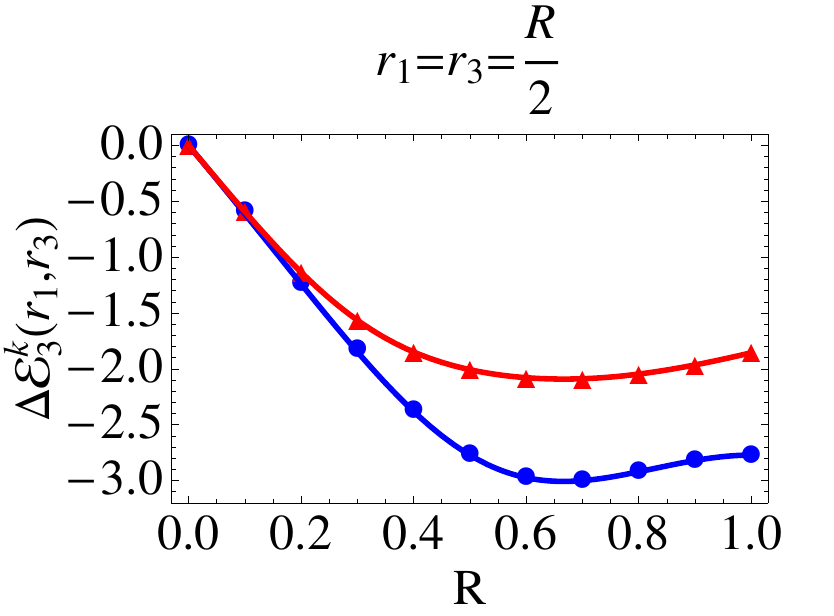}
         \includegraphics[width=7.5cm]{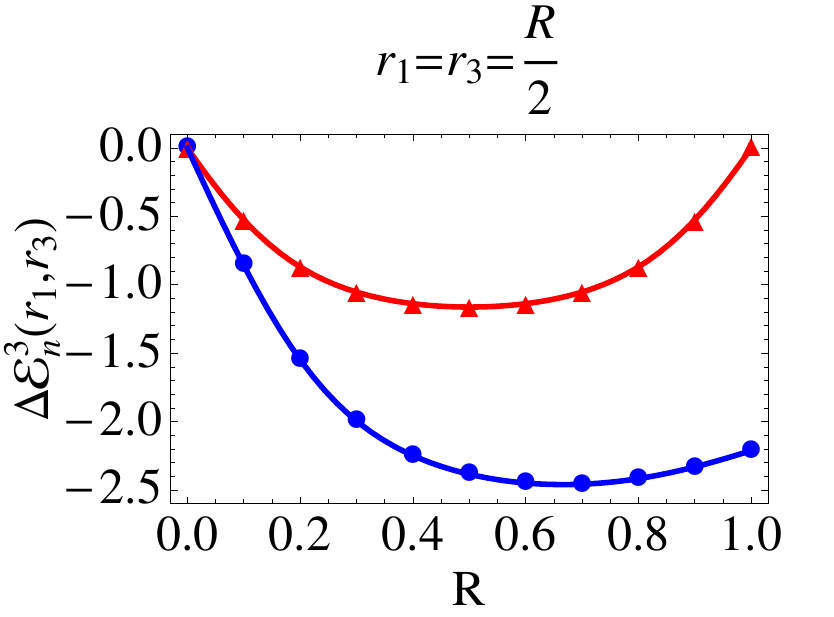}
      \end{center} 
 \caption{Left:  $n=3$ for a state of two equal particles $k=2$ with $p=4{\pi}$ (triangles) and for a state of two distinct particles  with momenta $p_1=3{\pi}$ and $p_2=4\pi$ (circles). The solid lines are the analytic formulae for $\Delta \mathcal{E}_3^2(r_1,r_2)$ (for equal momenta) and $2\Delta \mathcal{E}_3^1(r_1,r_2)$ for distinct momenta. Right: $k=3$ and $n=2$ (triangles), $n=3$ (circles) for a state of three identical excitations with momenta $p=4{\pi}$. Solid lines are the analytic formulae (\ref{de2}).} 
 \label{negnr1r3dif22} 
  \end{figure}

The regions' lengths are therefore $\ell_1 = fRL$ and $\ell_3 = (1-f)RL$, whose sum is $RL$ ranging from 0 to $L$. We have used different values of $n$ and different types of excited state (with distinct and equal excitations). Here we concentrate on higher momenta, which are still well within the QFT regime but also well within the quasiparticle regime, indeed finding excellent agreement with our analytic predictions.  In Figs.~\ref{negnr1r3dif1} and \ref{negnr1r3dif2} (left) we present results for $k=1$ and $n=2,3$ and 4 and different ratios $r_1/r_3$. Fig.~\ref{negnr1r3dif2} (right) and Fig.~\ref{negnr1r3dif22} explore some configurations for states with $k=2, 3$ and states of equal/distinct excitations. In particular we find confirmation of the statement (\ref{sumar}) that the replica logarithmic negativity of a state of $k$ distinct excitations is $k$ times the replica logarithmic negativity of the state of one excitation. 

\section{Conclusion}
In two previous publications \cite{the4ofus1,the4ofus2} we developed a methodology for the evaluation of the R\'enyi entropies (and related quantities) in zero-density excited states of one-dimensional free massive quantum field theory. Three methods were used: a finite-volume form factor approach based on the use of branch point twist fields and $U(1)$ twist fields, a qubit picture showing that the same excess entanglement can be obtained from simple qubit states, and, finally, a numerical approach based on exact diagonalization for a harmonic chain. 

These methods have been employed again in the current work to extend results to the R\'enyi entropies of any number of disconnected regions and to the replica logarithmic negativity, a measure of entanglement for non-complementary regions. Our results demonstrate that all three approaches can still be successfully employed to evaluate the increment experienced by these more mathematically complex quantities in zero-density excited states. In the current paper we have focussed on the case of the free boson. It would be interesting to extend our results to the massive Majorana fermion. 

We find that the increment of the R\'enyi, von Neumann, and single copy entropies of multiple disconnected regions of lengths $\ell_i$ takes exactly the same form as for a single connected region of length $\ell$, up to   the replacement $\ell \mapsto \sum_i \ell_i$. More precisely, in the scaling limit (\ref{scaling}) the increment of the R\'enyi and related entropies is independent of the connectivity of the regions. It depends only upon their overall length.

For the replica logarithmic negativities we find that their increment is the logarithm of a symmetric polynomial of the variables $r_1$ and $r_3$. 
By employing a qubit picture we have obtained closed formulae for all such polynomials (\ref{de2}). The numerical coefficients involved are given by (\ref{pro}) and point towards a combinatorial interpretation of the result. Indeed, we show in \cite{the4ofus4} that such polynomials are partitions functions for counting of particular families of graphs. This is also true for the R\'enyi entropies which may be recovered from the replica logarithmic negativity by taking either $r_1$ or $r_3$ to zero. 

From the qubit interpretation of our results we are also able to generalize our conclusions to any number of disconnected regions and find that both the increments of R\'enyi entropies and logarithmic negativities do not depend on the connectivity of the regions, just on the overall size of each subsystem.

This work, together with \cite{the4ofus1, the4ofus2} and the follow up paper \cite{the4ofus4}, provides a complete understanding of the most popular measures of entanglement in zero-density excited states of free one-dimensional massive quantum field theory. Some of the results have also been shown to hold for interacting models \cite{the4ofus2}, and all results hold as well in free higher dimensional theories \cite{the4ofus1, the4ofus4}, the most general proof being given in \cite{the4ofus4}. The main challenge now is to consider interacting models more generally and to show under which conditions  (if any) these results still are valid in the presence of interactions. We expect to study this problem in a future work. 

\paragraph{Acknowledgments:} Olalla A. Castro-Alvaredo, Benjamin Doyon, and Istv\'an M. Sz\'ecs\'enyi are grateful to EPSRC for funding through the standard proposal {\it ``Entanglement Measures, Twist Fields, and Partition Functions in Quantum Field Theory"} under reference numbers EP/P006108/1 and EP/P006132/1. Cecilia De Fazio gratefully acknowledges funding from the School of Mathematics, Computer Science and Engineering of City, University of London through a PhD Studentship. Benjamin Doyon acknowledges the hospitality of the \'Ecole Normale Sup\'erieure de Lyon, France, where parts of this work were done during an invited professorship.

\appendix

\section{Logarithmic Negativity for Identical Particles}
\label{apA}
The tables below give examples of the sort of polynomials that are obtained for the replica logarithmic negativities $\mathcal{E}_n^k(r_1,r_3)$ for a state of $k$ identical particles. Recall that, as previously, $r:=1-r_1-r_3$ and $r_1$, $r_3$ are the scaled versions of regions $A$ and $B$ sizes in Fig.~1. As indicated, the polynomials for $n=2$ can be rewritten entirely in terms of $r$ only. They are the same as those entering the expressions for the second R\'enyi entropies $\Delta S_2^k(r)$ obtained in \cite{the4ofus1,the4ofus2}. This is because for $n=2$ we have $\TT=\tilde{\TT}$ and so the two ratios of four-point functions in (\ref{3}) and (\ref{4}) are identical. 
\begin{table}[h!]
\begin{center}
\begin{tabular}{|c||l|}
\hline
$n=2$& $r^4+4 r_1^2 r^2+4 r_3^2 r^2+8 r_1 r_3 r^2+r_1^4+r_3^4+4 r_1 r_3^3+6 r_1^2 r_3^2+4 r_1^3 r_3$\\
& $=r^4+4r^2(1-r)^2+(1-r)^4$\\\hline
$n=3$& $r^6+12 r_1 r_3 r^4+8 r_1^3 r^3+8 r_3^3 r^3+27 r_1^2 r_3^2 r^2+12 r_1 r_3^4 r+12 r_1^4 r_3 r+r_1^6+r_3^6+8 r_1^3 r_3^3$ \\ \hline
$n=4$& $r^8+16 r_1 r_3 r^6+16 r_1^4 r^4+16 r_3^4 r^4+68 r_1^2 r_3^2 r^4+32 r_1 r_3^5 r^2+80 r_1^3 r_3^3 r^2+32 r_1^5 r_3 r^2$\\ 
&  $+r_1^8+r_3^8+8 r_1^2 r_3^6+18 r_1^4 r_3^4+8 r_1^6 r_3^2$ \\ \hline
$n=5$& $r^{10}+20 r_1 r_3 r^8+125 r_1^2 r_3^2 r^6+32 r_1^5 r^5+32 r_3^5 r^5+280 r_1^3 r_3^3 r^4+80 r_1 r_3^6 r^3+80 r_1^6 r_3 r^3$\\ 
&  $+205 r_1^4 r_3^4 r^2+40 r_1^2 r_3^7 r+40 r_1^7 r_3^2
   r+r_1^{10}+r_3^{10}+32 r_1^5 r_3^5$ \\ \hline
$n=6$& $r^{12}+ r_1^{12}+24 r_1 r_3 r^{10}+198 r_1^2 r_3^2 r^8+64 r_1^6 r^6+64 r_3^6 r^6+680 r_1^3 r_3^3 r^6+192 r_1 r_3^7 r^4+969 r_1^4 r_3^4 r^4$\\ 
&  $+192 r_1^7 r_3 r^4+144 r_1^2 r_3^8 r^2+504 r_1^5 r_3^5 r^2+144 r_1^8 r_3^2 r^2+r_3^{12}+16 r_1^3 r_3^9+66 r_1^6 r_3^6+16 r_1^9 r_3^3$ \\ 
\hline
\end{tabular}
\end{center}
\caption{The replica negativities $\exp(\mathcal{E}_n^2(r_1,r_3))$ for $k=2$.}
\end{table}
 \begin{table}[h!]
\begin{center}
\begin{tabular}{|c||l|}
\hline
$n=2$& $r^6+9 r_1^2 r^4+9 r_3^2 r^4+18 r_1 r_3 r^4+9 r_1^4 r^2+9 r_3^4 r^2+36 r_1 r_3^3 r^2+54 r_1^2 r_3^2 r^2$\\
& $+36 r_1^3 r_3 r^2+r_1^6+r_3^6+6 r_1 r_3^5+15 r_1^2 r_3^4+20 r_1^3
   r_3^3+15 r_1^4 r_3^2+6 r_1^5 r_3$\\
   & $= r^6 + 9r^2(1-r)^4+9r^4(1-r)^2 +(1-r)^6 $\\ \hline
$n=3$& $r^9+27 r_1 r_3 r^7+27 r_1^3 r^6+27 r_3^3 r^6+189 r_1^2 r_3^2 r^5+162 r_1 r_3^4 r^4+162 r_1^4 r_3 r^4$\\
& $+27 r_1^6 r^3+27 r_3^6 r^3+381 r_1^3 r_3^3 r^3+189 r_1^2 r_3^5 r^2+189r_1^5 r_3^2 r^2+27 r_1 r_3^7 r$\\ & $+162 r_1^4 r_3^4 r+27 r_1^7 r_3 r+r_1^9+r_3^9+27 r_1^3 r_3^6+27 r_1^6 r_3^3$ \\ \hline
$n=4$& $r^{12}+36 r_1 r_3 r^{10}+81 r_1^4 r^8+81 r_3^4 r^8+414 r_1^2 r_3^2 r^8+648 r_1 r_3^5 r^6+1840 r_1^3 r_3^3 r^6$
\\ & $ +648 r_1^5 r_3 r^6+81 r_1^8 r^4+81 r_3^8 r^4+1404 r_1^2 r_3^6 r^4+3114 r_1^4 r_3^4 r^4+1404 r_1^6 r_3^2 r^4$\\
& $+108 r_1 r_3^9 r^2+864 r_1^3 r_3^7 r^2+1656 r_1^5 r_3^5 r^2+864 r_1^7 r_3^3 r^2+108 r_1^9 r_3 r^2$\\ & $+
r_1^{12}+r_3^{12}+18 r_1^2
   r_3^{10}+99 r_1^4 r_3^8+164 r_1^6 r_3^6+99 r_1^8 r_3^4+18 r_1^{10} r_3^2$ \\ \hline
$n=5$& $r^{15}+45 r_1 r_3 r^{13}+720 r_1^2 r_3^2 r^{11}+243 r_1^5 r^{10}+243 r_3^5 r^{10}+5135 r_1^3 r_3^3 r^9$\\
& $+2430 r_1 r_3^6 r^8+2430 r_1^6 r_3 r^8+17100 r_1^4 r_3^4 r^7+7695 r_1^2  r_3^7 r^6+7695 r_1^7 r_3^2 r^6$\\
&$+243 r_1^{10} r^5+243 r_3^{10} r^5+25551 r_1^5 r_3^5 r^5+8910 r_1^3 r_3^8 r^4+8910 r_1^8 r_3^3 r^4$\\
&$+405 r_1 r_3^{11} r^3+14855 r_1^6 r_3^6 r^3+405 r_1^{11} r_3 r^3+3375 r_1^4 r_3^9 r^2+3375 r_1^9 r_3^4 r^2$\\\
& $+135 r_1^2 r_3^{12} r+2430 r_1^7 r_3^7 r+135 r_1^{12} r_3^2 r+r_1^{15}+r_3^{15}+243 r_1^5 r_3^{10}+243
   r_1^{10} r_3^5$ \\ \hline
$n=6$& $r^{18}+54 r_1 r_3 r^{16}+1107 r_1^2 r_3^2 r^{14}+729 r_1^6 r^{12}+729 r_3^6 r^{12}+11022 r_1^3 r_3^3 r^{12}$\\
& $+8748 r_1 r_3^7 r^{10}+57267 r_1^4 r_3^4 r^{10}+8748 r_1^7 r_3 r^{10}+36450 r_1^2 r_3^8 r^8+156330 r_1^5 r_3^5 r^8$\\
& $+36450 r_1^8 r_3^2 r^8+729 r_1^{12} r^6+729 r_3^{12} r^6+64152 r_1^3 r_3^9 r^6+214983 r_1^6 r_3^6 r^6+64152 r_1^9 r_3^3
   r^6$\\
   & $  +1458 r_1 r_3^{13} r^4+47385 r_1^4 r_3^{10} r^4+134190 r_1^7 r_3^7 r^4+47385 r_1^{10} r_3^4 r^4+1458 r_1^{13} r_3 r^4$\\
   & $+729 r_1^2 r_3^{14} r^2+12636 r_1^5 r_3^{11}
   r^2+31158 r_1^8 r_3^8 r^2+12636 r_1^{11} r_3^5 r^2+729 r_1^{14} r_3^2 r^2$\\
   & $+r_1^{18}+r_3^{18}+54 r_1^3 r_3^{15}+783 r_1^6 r_3^{12}+1460 r_1^9 r_3^9+783 r_1^{12} r_3^6+54
   r_1^{15} r_3^3$ \\
\hline
\end{tabular}
\end{center}
\caption{The replica negativities $\exp(\mathcal{E}_n^3(r_1,r_3))$ for $k=3$.}
\end{table}
\section{Eigenvalues of the Reduced (Partially Transposed) Density Matrix for $k=2$ and $k=3$.}
It is instructive to repeat the analysis of subsection~\ref{single} for the case of two and three identical excitations. 
\subsection{Qubit Computation for two Identical Particles}
\label{2particles}
In this case the state can be written simply as
\beqa
|\Psi^{(2)}_{\rm qb}\ket&=&{r_1}|200\ket + r |020\ket + r_3 |002\ket + \sqrt{2 r_1 r} |110\ket + \sqrt{2 r_1 r_3} |101\ket +  \sqrt{2 r r_3} |011\ket\,.
\label{qubit2}
\eeqa
The trace over region $C$ gives
\beqa
\rho_{A \bbc B}&=& r_1^2 |20\ket \bra 20|+ r^2|00\ket \bra 00| + r_3^2|02\ket \bra 02| + 2 r_1 r  |10\ket \bra 10|+ 2 r_3 r |01\ket \bra 01|  + 2 r_1 r_3 |11\ket \bra 11| \nonumber\\
&& + r_1 r_3(|2 0\ket \bra 02| + |02\ket \bra 2 0|)  + 2r \sqrt{r_1 r_3}(|10\ket \bra 01|+ |01\ket \bra 10|) \nonumber\\
&& + r_1\sqrt{2 r_1 r_3}(|20\ket \bra 11| + |11\ket \bra 20|) + r_3 \sqrt{2r_1 r_3} (|02\ket \bra 11| + |11\ket \bra 02|)\,,
\eeqa
and 
\beqa
\rho_{A \bbc B}^{T_{B}}&=& r_1^2 |20\ket \bra 20|+ r^2|00\ket \bra 00| + r_3^2|02\ket \bra 02| + 2 r_1 r  |10\ket \bra 10| + 2 r_3 r |01\ket \bra 01| + 2 r_1 r_3 |11\ket \bra 11| \nonumber\\
&& + r_1 r_3(|2 2\ket \bra 00| + |00\ket \bra 2 2|) + 2r \sqrt{r_1 r_3}(|11\ket \bra 00|+ |00\ket \bra 11|)  \nonumber\\
&&+ r_1\sqrt{2 r_1 r_3}(|21\ket \bra 10| + |10\ket \bra 21|) + r_3 \sqrt{2r_1 r_3} (|01\ket \bra 12| + |12\ket \bra 01|)\,.
\eeqa
In this case the non-vanishing contributions to $\rho_{A \bbc B}$ can be organized into a $6\times 6$ matrix, \beq
\rho_{A \bbc B}=\left(
\begin{array}{c|cccccc}
&00 &01 &10& 11&02&20 \\
\hline
00 &r^2&0&0& 0 &0&0\\
01 &0&2 r r_3&2r\sqrt{r_1 r_3}&0 &0&0\\
10 &0&2r \sqrt{r_1 r_3}&2 r r_1&0 &0&0\\
11  &0&0&0&2r_1 r_3 &r_3 \sqrt{2 r_1 r_3}& r_1 \sqrt{2 r_1 r_3}\\
02 &0&0& 0& r_3 \sqrt{2 r_1 r_3}&r_3^2&r_1r_3\\
20&0&0&0&r_1\sqrt{2 r_1 r_3}& r_1 r_3&r_1^2\\
\end{array}
\right)\,.
\eeq
The partially transposed matrix involves non-vanishing contributions from all the states and is instead a $9\times 9$ matrix given by
\beq
\left(
\begin{array}{c|ccccccccc}
&00 &01 &10& 11&02&20&21&12&22 \\
\hline
00 &r^2&0&0& 2r \sqrt{r_1 r_3} &0&0&0&0&r_1 r_3\\
01 &0&2 r r_3&0&0 &0&0&0&r_3 \sqrt{2 r_1 r_3}&0\\
10 &0&0&2 r r_1&0 &0&0&r_1 \sqrt{2 r_1 r_3}&0&0\\
11  &2r \sqrt{r_1 r_3}&0&0&2r_1 r_3 &0& 0&0&0&0\\
02 &0&0& 0& 0&r_3^2&0&0&0&0\\
20&0&0&0&0& 0&r_1^2&0&0&0\\
21&0&0&r_1\sqrt{2 r_1r_3}&0&0&0&0&0&0\\
12&0&r_3 \sqrt{2 r_1 r_3}&0&0&0&0&0&0&0\\
22&r_1r_3&0&0&0&0&0&0&0&0
\end{array}
\right)\,.
\eeq
The matrix $\rho_{A \bbc B}$ has non-vanishing eigenvalues
\beq
\lambda_1=r^2, \quad \lambda_2=2r(1-r)\quad \mathrm{and} \quad \lambda_3=(1-r)^2.
\eeq
Thus, the R\'enyi entropies are given by
\beq
S_n^2 (r_1,r_3)=\frac{\log(r^{2n} + (1-r)^{2n}+ (2r(1-r))^n)}{1-n}\,,
\eeq
and this is the expected formula found in \cite{the4ofus1,the4ofus2}. The matrix $\rho_{A \bigcup B}^{T_{\rm B}}$ has non-vanishing eigenvalues given by
\beqa
&& \lambda_1^t=r_1^2\,,\quad \lambda_2^t=r_3^2\,, \quad \lambda_{3}^t=r r_1 + r_1 \sqrt{r^2+2 r_1 r_3}\,, \quad \lambda_{4}^t=r r_1 - r_1 \sqrt{r^2+2 r_1 r_3}\,,\nonumber\\
&& \lambda_5^t= r r_3 + r_3 \sqrt{r^2+2 r_1 r_3}\,,\quad \lambda_6^t= r r_3 - r_3 \sqrt{r^2+2 r_1 r_3}\,,\quad \lambda_{7}^t\,,\quad \lambda_8^t\,, \quad \lambda_9^t\,, 
\eeqa
where $\lambda_{7,8,9}^t$ are the  roots of the cubic equation:
\beq
2 r_1^3 r_3^3 -r_1 r_3 (2 r^2  +r_1 r_3) x - (r^2 +
      2 r_1 r_3)x^2 +x^3=0\,.
      \label{pol}
\eeq
Let us consider a generic equation of the form $x^3 + b x^2 + c x+ d=0$. The roots of such an equation may be written in closed form as:
\beq
x_k= -\frac{1}{3}\left(b+\omega^k \tilde{\Delta} + \frac{\Delta_0}{\omega^k \tilde{\Delta}}\right), \quad \mathrm{with}\quad \omega=e^{\frac{2\pi i}{3}} \quad \mathrm{and} \quad k=0,1,2.
\eeq
with
\beq
\tilde{\Delta}=\sqrt[3]{\frac{\Delta_1 + \sqrt{\Delta_1^2-4\Delta_0^3}}{2}},
\eeq
and 
\beq
\Delta_0=b^2-3c, \qquad \Delta_1=2b^3-9bc+27 d.
\eeq
The roots of the polynomial (\ref{pol}) correspond to the identifications $\lambda_7^t:=x_0$, $\lambda_8^t:=x_1$ and $\lambda_9^t:=x_2$ as well as
\beq
b=- r^2 -
      2 r_1 r_3\,, \quad c= -r_1 r_3 (2 r^2  +r_1 r_3)\,,\quad \mathrm{and} \quad d=2 r_1^3 r_3^3\,.
\eeq
With these identifications we obtain the values of $\Delta_0$ and $\Delta_1$ reported in (\ref{d1}) and in (\ref{delta}).

Although there are now 9 non-vanishing eigenvalues and the expressions are much more complicated than for $\rho_{A \bbc B}$, it is easy to deduce that the sum of the $n$-th powers of the first six non-vanishing eigenvalues is the polynomial:
\beqa
\sigma_1(n):=\sum_{p=1}^6 (\lambda_p^t )^n=r_1^{2n}+r_3^{2n}+(r_1^n+r_3^n) \sum_{p=0}^{[\frac{n}{2}]} \frac{n}{n-p} \left(\begin{array}{c}
 n-p\\
 p\\
 \end{array}\right) 2^{n-p} r^{n-2p} r_1^p r_3^p\,.
 \label{sigma1}
\eeqa
Comparing to the general formula (\ref{de2}) we have that the term $r_1^{2n}$  corresponds to taking $p=2$, in which case, for $k=2$, the only possible value of $q$ is 0 with coefficient $\mathcal{A}_{2,0}=1$.
The term $r_3^{2n}$ is generated from (\ref{de2}) as well for $p=-2$, in which case the only possible value of $q$ is $q=2n$. This gives coefficient $\mathcal{A}_{-2,2n}=1$.  The remaining terms above correspond to the values $p=\pm 1$ in (\ref{de2}) in which case we get two sums, one proportional to $(r_1 r)^n$ and the other to $(r_3 r)^n$ as above. 
In this contribution the coefficients $\mathcal{A}_{\pm 1, q}$ once again are related to counting partitions into 0s and 1s with no consecutive 1s. 

The sum $(\lambda_7^t)^n+(\lambda_8^t)^n+(\lambda_9^t)^n$ is found (by inspection) to have the following structure:
\beq
\sigma_2(n):=\sum_{p=7}^9 (\lambda_p^t )^n=\sum_{q=0}^n \mathcal{A}_{0,q} \, r^{2(n-q)} r_1^q r_3^q\,,
\eeq
which corresponds to the $p=0$ contribution in (\ref{de2}). The coefficients are positive and integer and they can be systematically computed from the explicit formulae for the eigenvalues $\lambda_7^t, \lambda_8^t$ and $\lambda_9^t$, and/or by employing some useful properties of roots of third order polynomials as described for instance in \cite{sums}.  The first few coefficients are given by
\beqa
\mathcal{A}_{0,0}&=&1\,,\qquad \mathcal{A}_{0,1}=4n\,, \qquad 
 \mathcal{A}_{0,2}=n(8n-15)\,, \qquad \mathcal{A}_{0,3}= \frac{4n}{3}  (8n^2-45n+67)\,, \nonumber\\ 
\mathcal{A}_{0,4}&=&\frac{n}{6}  (64 n^3-720 n^2+2819 n-3849)\,, \nonumber\\ 
\mathcal{A}_{0,5}&=&\frac{2n}{15}  \left(64 n^4-1200 n^3+8735 n^2-29295 n+38196\right)\,,\nonumber\\ 
\mathcal{A}_{0,6}&=&\frac{n}{90}  \left(512 n^5-14400 n^4+166760 n^3-994905 n^2+3058553 n-3873780\right)\,,\nonumber\\ 
\mathcal{A}_{0,7} &=&\frac{2n}{315}  \left(512 n^6-20160 n^5+339080 n^4-3120495 n^3+16575818 n^2\right.\nonumber\\
&& \qquad \left.-48174105 n+59800950\right)\,,
\label{cobs}\eeqa
and, finally,
\beqa
\mathcal{A}_{0,n}=\left\{\begin{array}{cc}
2^n & \mathrm{for}\,\, n \,\, \mathrm{odd}\\
2+2^n & \mathrm{for}\,\, n \,\, \mathrm{even}
\end{array}\right.\,.
\label{cob2s}
\eeqa
Interestingly, all the coefficients above are positive integers for $n$ integer and they can be shown to agree with the formula (\ref{pro}), even though showing this analytically is relatively involved as there is a sum over partitions to perform. 
The replica logarithmic negativities are then given by 
\beq
  \mathcal{E}_n^2(r_1, r_3)={\log(\sigma_1(n)+\sigma_2(n))},
 \label{ds2}
 \eeq 
As usual, we are interested in the analytic continuation of this result to $n=1$. In this case, as we know all roots explicitly we can simply compute the logarithmic negativity using its formal definition as the sum of the absolute values of the eigenvalues of the partially transposed, reduced density matrix. We have that 
\beqa
\lim_{n\rightarrow 1} \sigma_1(n)=\sum_{p=1}^6 |\lambda_p^t|
=r_1^2+r_3^2 + 2 (r_1+r_3) \sqrt{r^2+2r_1r_3}\,.
\eeqa
For the remaining eigenvalues it is possible to use properties of cubic roots to show that in our case,
\beqa
\lim_{n\rightarrow 1} \sigma_2(n)=\sum_{p=7}^9 |\lambda_p^t| =-\lambda_7^t+\lambda_8^t+\lambda_9^t=\frac{1}{3} \left(r^2+r_1r_3+2^{2/3} \sqrt[3]{\Delta}+\frac{2 \sqrt[3]{2} \Delta_0}{\sqrt[3]{\Delta}}\right)\,,
\label{lims2}
\eeqa
namely, it can be shown that $\lambda_7^t<0$ whereas $\lambda_8^t$ and $\lambda_9^t$ are positive.  $\Delta$ and $\Delta_0$ were defined in (\ref{delta}), (\ref{d1}).
The logarithmic negativity is then
\beq
\mathcal{E}_1(r_1,r_3)=\log\left(\lim_{n\rightarrow 1} \sigma_1(n)+\lim_{n\rightarrow 1} \sigma_2(n)\right),
\eeq
as given by (\ref{de2}).

\subsection{Qubit Computation for three Identical Particles}
The qubit state for three particles takes the form:
\beqa
&&\!\!\! \!\!\! |\Psi^{(3)}_{\rm qb}\ket={r_1} \sqrt{r_1}|300\ket + r\sqrt{r} |030\ket + r_3 \sqrt{r_3} |003\ket + \sqrt{3 r_1 r} (\sqrt{r_1}|210\ket + \sqrt{r}|120\ket ) \nonumber\\
&& \!\!\! \!\!\! + \sqrt{3 r_1 r_3}(\sqrt{r_1} |201\ket + \sqrt{r_3} |102\ket )+  \sqrt{3 r r_3} (\sqrt{r} |021\ket+ \sqrt{r_3}|012\ket) + \sqrt{6 r r_1 r_3}|111\ket\,.
\label{qubit3}
\eeqa
The non-vanishing eigenvalues of the reduced density matrix are:
\beq
\lambda_1=r^3, \quad \lambda_2=(1-r)^3, \quad \lambda_3=3 r (1-r)^2, \quad \lambda_4=3 r^2 (1-r).
\eeq
The partially transposed reduced density matrix is a $16\times 16$ matrix with eigenvalues: 
\beqa
&&\lambda_1^t=r_1^3\,, \quad \lambda_2^t=r_3^3\,, \quad \lambda_3^{t}=\frac{r_1^2(3 r + \sqrt{3(3r^2+4 r_1 r_3)})}{2}\,, \quad \lambda_4^{t}=\frac{r_1^2(3 r - \sqrt{3(3r^2+4 r_1 r_3)})}{2}\,,\nonumber\\
&&
 \lambda_5^t=\frac{r_3^2(3 r + \sqrt{3(3r^2+4 r_1 r_3)})}{2}\,, \quad \lambda_6^t=\frac{r_3^2(3 r - \sqrt{3(3r^2+4 r_1 r_3)})}{2}\,, 
\eeqa
and the roots of the cubic and quartic equations,
\beqa
&& \!\!\!\! \! \!\!  \!\!\!\! \! \!\!  \!\!\!\! \! \!\! 9 r_1^6 r_3^3 - (9 r^2 r_1^3 r_3 +3 r_1^4 r_3^2) x - (3 r^2 r_1 +
    3 r_1^2 r_3) x^2 + x^3=0\,,\\
 &&   \!\!\!\! \! \!\!  \!\!\!\! \! \!\!  \!\!\!\! \! \!\!9 r_3^6 r_1^3 - (9 r^2 r_3^3 r_1 + 3 r_3^4 r_1^2) x - (3 r^2 r_3 +
    3 r_3^2 r_1) x^2 + x^3=0\,,\nonumber\\
    && \!\!\!\! \! \!\!  \!\!\!\! \! \!\!  \!\!\!\! \! \!\!9 r_1^6 r_3^6 + (9 r^3 r_1^3 r_3^3 + 6 r r_1^4 r_3^4) x - (3 r^4 r_1 r_3 +
    9 r^2 r_1^2 r_3^2 + 10 r_1^3 r_3^3) x^2 - (r^3 +
    6 r r_1 r_3) x^3 + x^4=0\,.\nonumber
    \eeqa
Similar to the two-particle case, it is easy to find a general formula for the sum of powers of the simpler eigenvalues
\beqa
\sum_{p=1}^6 (\lambda_p^t)^n
= r_1^{3n} + r_3^{3n}+ (r_1^{2n}+r_3^{2n})\sum_{p=0}^{[\frac{n}{2}]} \frac{n}{n-p} \left(\begin{array}{c}
 n-p\\
 p\\
 \end{array}\right) 3^{n-p} r^{n-2p} r_1^p r_3^p\,. 
 \label{abit}
 \eeqa
 However, once more, closed formulae for the sum of powers of the remaining eigenvalues are less straightforward to obtain. 
It is however possible to obtain an expression for the logarithmic negativity by employing general properties of the roots of cubic and quartic equations. We find that the contribution to the negativity from the first six eigenvalues is simply:
\beq
r_1^3+r_3^3 +  (r_1^2+r_3^2)\sqrt{3(3r^2+4 r_1 r_3)}.
\eeq 
The contribution from the roots of the two cubic equations is 
\beq
(r_1+r_3)( r^2 +
    r_1 r_3) +\frac{2^{2/3}}{3} \sqrt[3]{\Delta^1}+\frac{2 \sqrt[3]{2} \Delta_0^1}{3\sqrt[3]{\Delta^1}}+\frac{2^{2/3}}{3} \sqrt[3]{\Delta^2}+\frac{2 \sqrt[3]{2} \Delta_0^2}{3\sqrt[3]{\Delta^2}}\,,
\eeq
 where 
 \beq
 \Delta^{1,2} :=\Delta_1^{1,2}+\sqrt{(\Delta_1^{1,2})^2-4
   (\Delta_0^{1,2})^3}\,,
   \label{delta12}
   \eeq
   and
\beq
 \Delta_0^{1}=9 r_3^2 \left(r^4+5 r^2 {r_1} {r_3}+2 r_1^2 r_3^2\right)\,, \quad
\Delta_1^1=-27 r_3^3 \left(2 r^6+15 r_1 r_3 r^4+18 r_1^2 r_3^2 r^2-4 r_1^3 r_3^3\right)\,, \label{dd1}
\eeq
and $\Delta_0^2, \Delta_1^2$ are the same as above with $r_1$ and $r_3$ exchanged.
Finally, the contribution from the roots of the quartic equation takes the form,
\beq
\sqrt{\Omega_0+\frac{\Omega_1}{\sqrt[3]{\Omega_2+\sqrt{\Omega_3}}}+ \frac{2^{5/3}}{3}\sqrt[3]{\Omega_2+\sqrt{\Omega_3}}}\,,
\eeq
with
\beqa
\Omega_0&:=& r^6+20 r_1 r_3 r^4+60 r_1^2 r_3^2 r^2+\frac{80}{3} r_1^3 r_3^3\,,\nonumber\\
\Omega_1&:=&\frac{4 \sqrt[3]{2}}{3} r_1^2 r_3^2 \left(9 r^8+81 r_1 r_3 r^6+321 r_1^2 r_3^2 r^4+288 r_1^3 r_3^3 r^2+208 r_1^4 r_3^4\right)\,,\nonumber\\
\Omega_2&:=&r_1^3 r_3^3 \left(-54 r^{12}-729 r_1 r_3 r^{10}-4347 r_1^2 r_3^2 r^8-8910 r_1^3 r_3^3 r^6-7200 r_1^4 r_3^4 r^4\right.\nonumber\\
&& \left.+6912 r_1^5 r_3^5 r^2+4480 r_1^6 r_3^6\right)\,,\\
\Omega_3&:=& -243 r_1^8 r_3^8 \left(81 r^{20}+2970 r_1 r_3 r^{18}+36945 r_1^2 r_3^2 r^{16}+243720 r_1^3 r_3^3 r^{14}+906816 r_1^4 r_3^4 r^{12}\right.\nonumber\\
&& \left.+2130816 r_1^5 r_3^5 r^{10}+3325504 r_1^6
   r_3^6 r^8+3203584 r_1^7 r_3^7 r^6+1606656 r_1^8 r_3^8 r^4\right.\nonumber\\
   &&\left. +360448 r_1^9 r_3^9 r^2+65536 r_1^{10} r_3^{10}\right)\,,
\eeqa
In summary,
\beqa
&&\mathcal{E}^3(r_1,r_3)=\log\left(r_1^3+r_3^3 +  (r_1^2+r_3^2)\sqrt{3(3r^2+4 r_1 r_3)}+(r_1+r_3)( r^2 +
    r_1 r_3) \right.   \label{E3par} \\
    && \left.  +\frac{2^{2/3}}{3} \sqrt[3]{\Delta^1}+\frac{2 \sqrt[3]{2} \Delta_0^1}{3\sqrt[3]{\Delta^1}}+\frac{2^{2/3}}{3} \sqrt[3]{\Delta^2}+\frac{2 \sqrt[3]{2} \Delta_0^2}{3\sqrt[3]{\Delta^2}}+ \sqrt{\Omega_0+\frac{\Omega_1}{\sqrt[3]{\Omega_2+\sqrt{\Omega_3}}}+ \frac{2^{5/3}}{3}\sqrt[3]{\Omega_2+\sqrt{\Omega_3}}} \right)\nonumber\,.
\eeqa
\section{Entanglement Measures for an Arbitrary Number of Disconnected Regions from Qubits}
\label{appendixC}
It is possible to extend the calculations of subsection \ref{entropy2r} and \ref{324} to a situation where regions $A$ and $B$ are not simply connected. Here we will present the detailed computation for the R\'enyi entropies and later comment on how a similar generalization may work for the replica logarithmic negativities. 
\subsection{R\'enyi Entropies of $\alpha$ Disconnected Regions}
 Let us consider the case when one of our subsystems is composed of $\alpha$ disconnected regions  $R_m$ with  $m=1,\dots, \alpha$. Let $R_0$ be the rest of the system and $R':=\bbc_m R_m $. In this case the whole bipartite system is composed by $R_0 \bbc R^\prime$ and  described by a Hilbert space  $\mathcal{H} = \mathcal{H}_0 \otimes \mathcal{H}^\prime$ where now $\mathcal{H}^\prime = \mathcal{H}_1 \otimes \dots \otimes \mathcal{H}_\alpha $. Again, we can define an orthonormal basis  $\{ |{\underline{k}}\ket \, \in \mathcal{H} \; : \; \underline{k} = (k_0, \dots,  k_{\alpha}) \in \mathbb{N}_0^{\alpha+1} \, \} $ such that $k_m$ is the number of excitations in region $R_m$ and $\sum_{m=1}^\alpha k_m=k$ for a state of $k$ identical excitations. Let
\begin{equation}
\label{rrelation}
    r^\prime = \sum_{m=1}^\alpha r_m = 1 - r_0
\end{equation}
where $r_m$ is the scaled length of region $R_m$. We define the qubit state: 
\begin{equation}
\label{aqbitstate}
    |{\Psi^{(k)}_\alpha}\ket = \sum_{\underline{k}= \lbrace k_0, \dots, k_\alpha \rbrace \in \, \sigma_0^{\alpha+1} (k) } \,\left[ k! \prod_{\ell=0}^\alpha \frac{r_\ell^{k_\ell}}{k_\ell!}  \right]^{1/2}\, \delta_{\sum_{i=0}^\alpha k_i,k}\,   |{\underline{k}}\ket \, , 
\end{equation}
where $\sigma_0^{\alpha+1} (k)$ represents the set of integer partitions of $k$ into $\alpha + 1$ non-negative parts. It is easy to extend \eqref{trace} to the case of $\alpha$ disconnected regions. Indeed, if we now introduce $\mathcal{S}_n^k (r_1,\ldots,r_\alpha):= \exp{ \lbrace (1-n) S_n^k(r_1,\ldots,r_\alpha) \rbrace} $ where $S_n^k(r_1,\ldots,r_\alpha) $ is the R\'enyi entropy  for the state \eqref{aqbitstate} it is easy to see that this can be written as:
\begin{equation}
\label{asqbit}
    \mathcal{S}_n^k(r_1,\ldots,r_\alpha)=  \sum_{\lbrace k^{j}_p \, \in \,\mathbb{N}_0 \, ; \,j \in I_n \, ; \,p \in I^0_{\alpha} \rbrace } \, \prod_{i=1}^n k! \left(\prod_{\ell=0}^\alpha \frac{r_\ell^{k^{i}_\ell}}{k_\ell^{i}!}   \right) \delta_{k^{i}_0 + \sum_{m = 1}^\alpha k^{i}_m \;, k} \; \delta_{k^{i}_0 + \sum_{m = 1}^\alpha k^{i+1}_m \;, k} \; ,
\end{equation}
where $I_n = \lbrace 1, \dots, n \rbrace $ and $I^0_{\alpha} = \lbrace 0, \dots, \alpha  \rbrace $. The delta-functions introduce the contraints
\begin{equation}
\label{system}
   \sum_{m=1}^\alpha k^{i}_m + k^{i}_0  = k\, \quad \mathrm{and} \quad 
      \sum_{m=1}^\alpha k^{i+1}_m + k^{i}_0  = k \,, \quad \mathrm{for} \quad i=1,\ldots,n\,,
\end{equation}
with the identifications $k_m^0 \equiv k_m^n$ and $k_m^1\equiv k_m^{n+1}$. 
These constraints are equivalent to
\begin{equation}
    \sum_{m=1}^\alpha   k^{i}_m  =  \sum_{m=1}^\alpha   k^{i+1}_m:=\gamma \, ,
\end{equation}
where $\gamma$ is an arbitrary constant. As a consequence
\beq
\label{variables}
    k^{i}_\alpha = \gamma - \sum_{m=1}^{\alpha-1}   k^{i}_m \quad  \mathrm{and} \quad 
    k^{i}_0 = k -  \gamma=:q\, \quad \mathrm{for} \quad i=1,\ldots, n.
    \eeq
As in the two region case  $k^{i}_0$ does not depend on any $k^{i}_m$s. Substituting \eqref{variables} into \eqref{asqbit} we have:
\begin{equation}
\label{asqbit2}
    \mathcal{S}_n^k(r_1,\ldots,r_\alpha)= \sum_{q \in \mathbb{Z}} \,   \sum_{\lbrace k^{j}_p \, \in \,\mathbb{N}_0 \, ; \,j \in I_n \, ; \,p \in I_{\alpha-1} \rbrace } \,  \left[{k\choose q} \, r_0^{q} \right]^n \,\prod_{i=1}^n  \left(  \frac{(k-q)! \, r_\alpha^{k-q -\sum_{m=1}^{\alpha-1}    k^{i}_m }}{(k-q -\sum_{m=1}^{\alpha-1}   k^{i}_m)!} \,  \prod_{m=1}^{\alpha-1} \frac{r_m^{   k^{i}_m }}{(k^{i}_m )!}   \right) \; .
\end{equation}
Again the multinomial coefficients constrain the sums. The presence of $q!$ in the denominator means that $q\geq 0$. We know also that $k^{i}_m$ must be non-negative for all $i=1,\ldots, n$ and $m=1,\ldots, \alpha-1$. Furthermore the only non-zero terms in the sums are given by $k-q -\sum_{m=1}^{\alpha-1}   k^{i}_m \geq 0$ and thus $q\leq k$. In summary, for the same reasons as in the two region case $0 \leq q \leq k$. \\
We can re-write \eqref{asqbit2} as: \begin{small}
\begin{equation}
\label{asqbit3}
    \mathcal{S}^k_n(r_1,\ldots,r_\alpha)= \sum_{q=0}^{k}  \, \left[ \, {k\choose q} \, \right]^n \, r_0^{nq}  \,   \prod_{i=1}^n \, \sum_{s_i =0}^{k-q} \,
    \sum_{\{ k^{i}_1, \dots, k^{i}_{\alpha -1} \} \, \in \sigma_0^{\alpha -1} ( s_i )   }
  \,  \left[ {k- q\choose s_i} \,  r_\alpha^{k-q -s_i }   \;   \; s_i ! \,  \prod_{m=1}^{\alpha-1}\frac{r_m^{   k^{i}_m }}{(k^{i}_m )!}   \right] \;  .
\end{equation}
\end{small}
It is easy to see that
\begin{equation}
  \sum_{\{ k^{i}_1, \dots, k^{i}_{\alpha -1} \} \, \in \sigma_0^{\alpha -1} ( s_i )  }   s_i ! \,  \prod_{m=1}^{\alpha -1} \frac{r_m^{   k^{i}_m }}{(k^{i}_m )!} = \left( \, \sum_{m=1}^{\alpha-1 }\, r_m \,  \right)^{s_i} \; , \quad \forall \, i=1,\ldots,n\,,
\end{equation}
and thus \eqref{asqbit3} becomes:
\begin{small}
\begin{equation}
\label{asqbit4}
     \mathcal{S}^k_n (r_1,\ldots,r_\alpha)=\sum_{q=0}^{k}   \, \left[ \, {k\choose q} \, \right]^n \, r_0^{nq}  \,\prod_{i=1}^n  \left[  \sum_{s_i = 1}^{k-q} \, {k- q\choose s_i} \,  r_\alpha^{k-q -s_i } \,  \left( \, \sum_{m=1}^{\alpha-1 }\, r_m \,  \right)^{s_i} \,  \right] \; .
\end{equation}
\end{small}
Furthermore we can notice that
\begin{align}
    \prod_{i=1}^n  \left[  \sum_{s_i = 1}^{k-q} \, {k- q\choose s_i} \,  r_\alpha^{k-q -s_i } \,  \left( \, \sum_{m=1}^{\alpha-1 }\, r_m  \,  \right)^{s_i} \,  \right] \; \,& = \left[ \sum_{s= 0}^{k-q} {k- q\choose s} \,  r_\alpha^{k-q -s} \,  \left( \, \sum_{m=1}^{\alpha-1 }\, r_m \,  \right)^{s}  \right]^n \nonumber\\
     & = \left[ \, \left( \, \sum_{m = 1}^{ \alpha }\, r_m \,  \right)^{k-q}  \right]^n \nonumber\\
     & = \left( 1 - r_0 \right)^{n(k-q)}\,,
\end{align}
where in the last line we used \eqref{rrelation}. By recalling the R\'enyi entropy, we finally have:
\begin{equation}
\label{asqbit5final}
    S^k_n(r_1,\ldots,r_\alpha)=:S^k_n(r_0)= \frac{1}{1-n} \, \log \left( \, \sum_{q=0}^k \, \left[ \, {k\choose q} \,   r_0^{q}  \, (1 - r_0 )^{k-q} \right]^n \, \right) .
\end{equation}
notice it takes the same form as \eqref{ren2} (with $r$ replaced by $r_0$). Therefore, the R\'enyi (and related) entropies of this particular class of qubit states depend only on the relative size of the two parts in the bipartition and not on whether or not they are connected. 
\subsection{Replica Logarithmic Negativities of Disconnected Regions}
A very similar computation can be performed for the replica logarithmic negativities. The starting point is the assumption that regions $A$ and $B$ are now disconnected, namely 
\beq
 A=\bbc\limits_{i=1}^{\alpha} A_i \qquad \mathrm{and} \qquad B=\bbc\limits_{i=1}^{\beta} B_i\,,
 \eeq
 so that $A$ consists of a number $\alpha$ and $B$ of a number $\beta$ of disconnected regions. Let regions $A_i$ and $B_i$ have scaled lengths given by $r_1^i$ and $r_3^i$, respectively. 
 Then our results for the (replica) logarithmic negativities will still hold up to the identifications:
 \beq
 r_1= \sum\limits_{i=1}^\alpha r_1^i\,, \qquad r_3= \sum\limits_{i=1}^\beta r_3^i,\, \qquad \mathrm{and}\qquad r=1-r_1-r_3\,.
 \eeq
In the qubit picture this can be shown in a very similar way as for the R\'enyi entropies in the previous section, so we do not present the computation here. 

\section{Form Factors and Residue Formulae}
\label{Ape}
Let $u(\beta)$ and $v(\beta)$ be generic functions of $\beta$ such that $v(\beta)$ has a single zero at $\beta=\theta$. The corresponding residua of the following first, second, and third order poles are given by
\begin{eqnarray}
\mathrm{Res}_{\beta\sim\theta}\frac{u\left(\beta\right)}{v\left(\beta\right)} & = & \frac{u\left(\theta\right)}{v'\left(\theta\right)}\,,\\
\mathrm{Res}_{\beta\sim\theta}\frac{u\left(\beta\right)}{\left(\beta-\theta\right)v\left(\beta\right)} & = & \frac{u'\left(\theta\right)}{v'\left(\theta\right)}-\frac{u\left(\theta\right)v''\left(\theta\right)}{2\left[v'\left(\theta\right)\right]^{2}}\,,\\
\mathrm{Res}_{\beta\sim\theta}\frac{u\left(\theta\right)}{\left(\beta-\theta\right)^{2}v\left(\beta\right)} & = & \frac{1}{12\left[v'\left(\theta\right)\right]^{3}}\left[3u\left(\theta\right)\left[v''\left(\theta\right)\right]^{2}-2u\left(\theta\right)v'\left(\theta\right)v'''\left(\theta\right)\right. \nonumber\\
&  &  \left.-6u'\left(\theta\right)v'\left(\theta\right)v''\left(\theta\right)+6u''\left(\theta\right)\left[v'\left(\theta\right)\right]^{2}\right]\,.
\end{eqnarray}
These generic formulae have been employed in order to obtain (\ref{residuos}). In particular, the function $\tilde{g}(r)$ defined in (\ref{gtilde}) results from the evaluation of residua of third order poles of the type
\beq
\mathrm{Res}_{\beta\sim\theta} \frac{e^{i x P(\beta)}}{(e^{i L P(\beta)}-1)(\beta-\theta)^2}= i e^{i x P(\theta)} \left[L E(\theta) \tilde{g}(r) + O(L^{0})\right]\,,
\eeq
with $r:=\frac{x}{L}$ and where $P(\theta), E(\theta)$ are the momentum and energy. Lower order poles of a similar type give the following residua
\beq
\mathrm{Res}_{\beta\sim\theta} \frac{e^{i x P(\beta)}}{(e^{i L P(\beta)}-1)(\beta-\theta)}= e^{i x P(\theta)} \left[r-\frac{1}{2} + O(L^{-1})\right]\,,
\eeq
and
\beq
\mathrm{Res}_{\beta\sim\theta} \frac{e^{i x P(\beta)}}{e^{i L P(\beta)}-1}= -\frac{ie^{i x P(\theta)} }{LE(\theta)}\,.
\eeq
These results justify the claim that only third order poles produce contributions that are proportional to the volume.

The function \eqref{g_function} is the result of the second order residue
\beq
\mathrm{Res}_{\beta\sim\theta}\frac{e^{i x P(\beta)}}{(e^{i L P(\beta)-i\frac{2\pi \epsilon p}{n}}-1)\left(\beta-\theta\right)^2} =  \frac{i g^n_{\epsilon p}(r)}{(1-e^{i\frac{2\pi \epsilon p}{n}})(1-e^{-i\frac{2\pi \epsilon p}{n}})}\,,
\eeq
and the denominator factors are cancelled by the form factor residues in the calculation.

\end{document}